%% file: 000-MainManuscript.tex
\documentclass[acmsmall]{acmart}

\setcopyright{cc}
\setcctype{by-nc-sa}
\acmJournal{PACMHCI}
\acmYear{2025} \acmVolume{9} \acmNumber{7} \acmArticle{CSCW431} \acmMonth{11} \acmDOI{10.1145/3757612}

\input{99-Commands}

\begin{document}

\title[Privacy Threats from VR Deceptive Design]{Immersive Invaders: Privacy Threats from Deceptive Design in Virtual Reality Games and Applications}
%

\author{Hilda Hadan}
\email{hhadan@uwaterloo.ca}
\orcid{https://orcid.org/0000-0002-5911-1405}
\affiliation{
    \institution{Stratford School of Interaction Design and Business, University of Waterloo}
    \city{Waterloo}
    \country{Canada}
}

\author{Michaela Valiquette}
\email{michaelavaliquette@cmail.carleton.ca}
\orcid{https://orcid.org/0009-0009-5335-3014}
\affiliation{
    \institution{Human-Computer Interaction, Carleton University}
    \city{Ottawa}
    \country{Canada}
}

\author{Lennart E. Nacke}
\email{lennart.nacke@acm.org}
\orcid{https://orcid.org/0000-0003-4290-8829}
\affiliation{
    \institution{Stratford School of Interaction Design and Business, University of Waterloo}
    \city{Waterloo}
    \country{Canada}
}

\author{Leah Zhang-Kennedy}
\email{lzhangke@uwaterloo.ca}
\orcid{https://orcid.org/0000-0002-0756-0022}
\affiliation{
    \institution{Stratford School of Interaction Design and Business, University of Waterloo}
    \city{Waterloo}
    \country{Canada}
}

\renewcommand{\shortauthors}{Hilda Hadan, Michaela Valiquette, Lennart E. Nacke, and Leah Zhang-Kennedy}

\begin{abstract}
\input{00-Abstract}

\end{abstract}

\ccsdesc[500]{Security and privacy~Human and societal aspects of security and privacy}
\ccsdesc[500]{Human-centered computing~Virtual reality}
\ccsdesc[500]{Human-centered computing~Empirical studies in HCI}
\ccsdesc[500]{Applied computing~Computer games}

\keywords{User Privacy, Dark Patterns, Deceptive Design, Virtual Reality, Games User Research}

\received{October 2024}
\received[revised]{April 2025}
\received[accepted]{August 2025}

\maketitle

\section{Introduction}
\label{sec:introduction}

\input{01-Introduction}

\section{Related Work}
\label{sec:related-work}
\input{02-RelatedWork}

\section{Methodology}
\label{sec:methodology}
\input{03-Methodology}

\section{Findings}
\label{sec:findings}
\input{04-Findings}

\section{Discussion}
\label{sec:discussion}
\input{05-Discussion}

\section{Conclusion}
\label{sec:conclusion}
\input{06-Conclusion}

\begin{acks}
\input{99-Acknowledge}
\end{acks}

\bibliographystyle{ACM-Reference-Format}
\bibliography{01-References}

\appendix
\input{98-Appendix}

\end{document}

%% file: 99-Commands.tex
\usepackage{float}

\usepackage[shortlabels]{enumitem}

\newenvironment{quot}[2]{
\begin{quote}
\begin{center}
\vspace{1.5mm}
\hspace*{-1mm}\textit{``#1''}\hspace*{-1mm}
\end{center}

\begin{flushright}
\textit{---~\faFilePowerpoint[regular]~#2}
\end{flushright}
\vspace{1.5mm}
\end{quote}
}

\usepackage{xspace} 
\newcommand{\guide}{\textit{Assessment Guide}\xspace}
\newcommand{\cif}{CI framework\xspace}
\newcommand{\VRapps}{VR games and apps\xspace}
\newcommand{\brightpatterns}{``bright patterns''\xspace}

\usepackage{microtype}  
\newcommand{\pattern}[1]{\texttt{\textls[-40]{#1}}}
\newcommand{\name}[1]{\textit{#1}}
\newcommand{\code}[1]{``#1''}

\usepackage{multirow}
\usepackage{fontawesome5}

\usepackage{svg}

\usepackage{xcolor,colortbl}
\definecolor{paleblue}{HTML}{A3D2FF}
\definecolor{darkblue}{HTML}{64b3ff}
\definecolor{palepink}{HTML}{ff84b1}  
\definecolor{palepurple}{HTML}{eabee8}
\definecolor{darkpurple}{HTML}{d57ed1}
\definecolor{paleorange}{HTML}{FFB703}
\definecolor{darkorange}{HTML}{ff7f47}
\definecolor{palegreen}{HTML}{CAFFBF}
\definecolor{darkgreen}{HTML}{0ca688}

\definecolor{palered}{HTML}{FFADAD}
\definecolor{palebrown}{HTML}{99582A}
\definecolor{paleyellow}{HTML}{FDFFB6}
\definecolor{palecyan}{HTML}{9BF6FF}
\definecolor{palewhite}{HTML}{fffffc}
\definecolor{palegray}{HTML}{C8C7D6}

\newcommand{\BeatSaber}[1][1.5em]{\raisebox{-0.8em}{\includegraphics[height=#1]{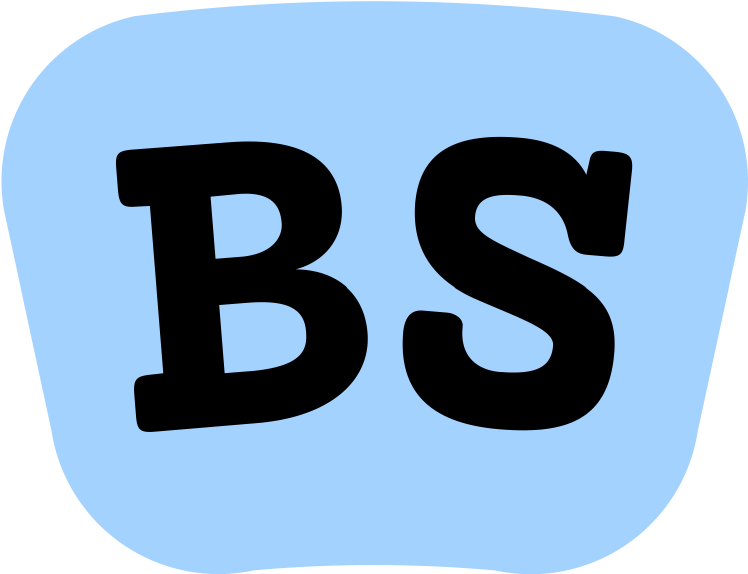}}}
\newcommand{\Moss}[1][1.5em]{\raisebox{-0.8em}{\includegraphics[height=#1]{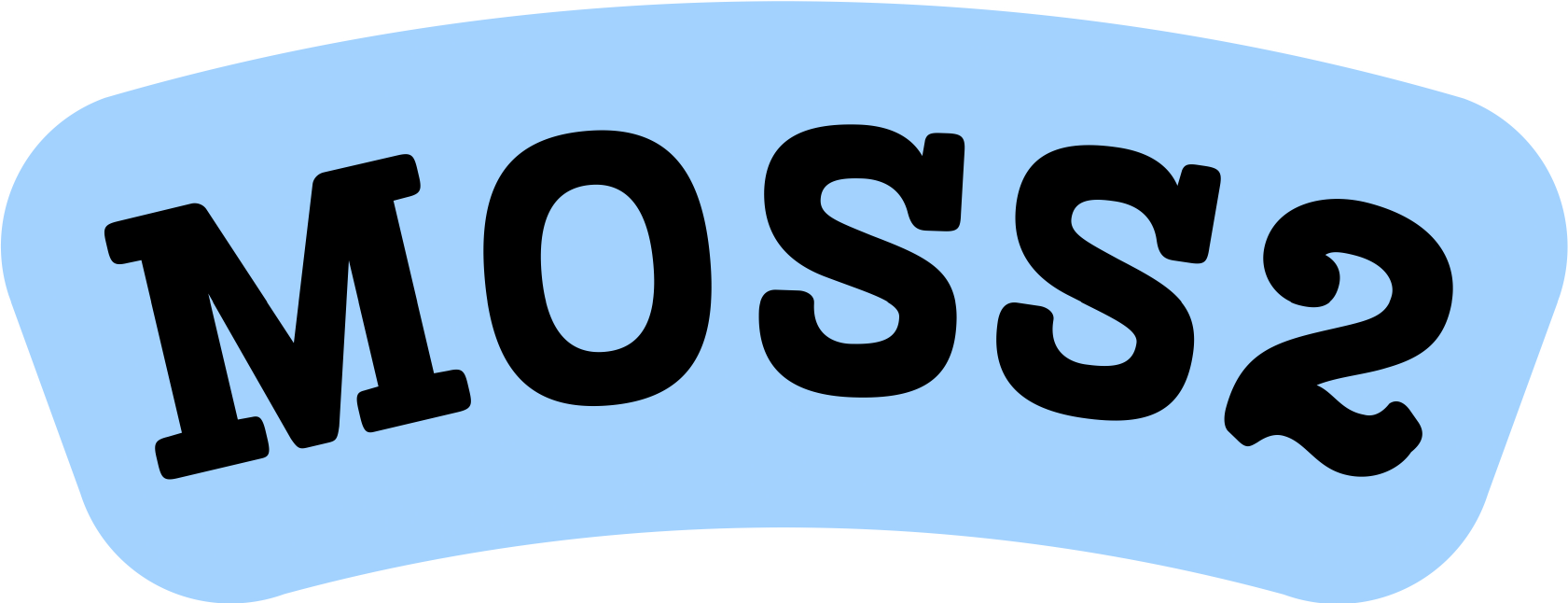}}}
\newcommand{\Room}[1][1.5em]{\raisebox{-0.8em}{\includegraphics[height=#1]{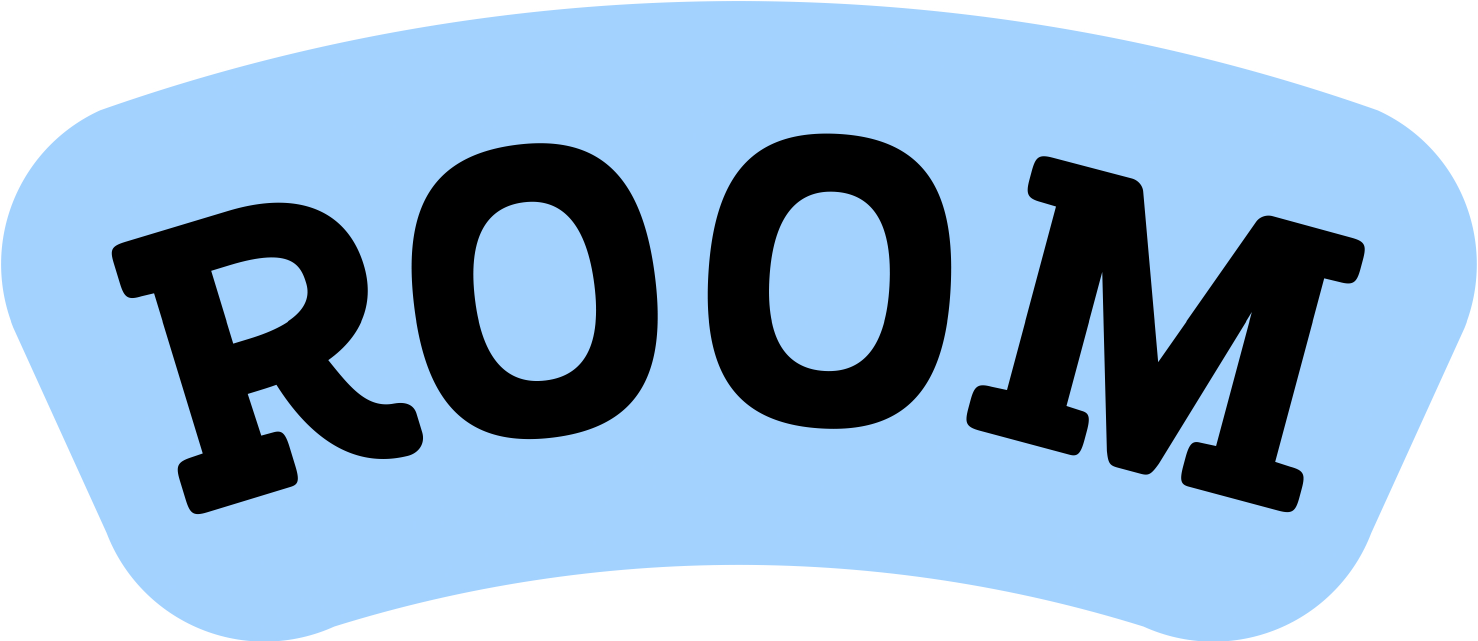}}}
\newcommand{\AFT}[1][1.5em]{\raisebox{-0.8em}{\includegraphics[height=#1]{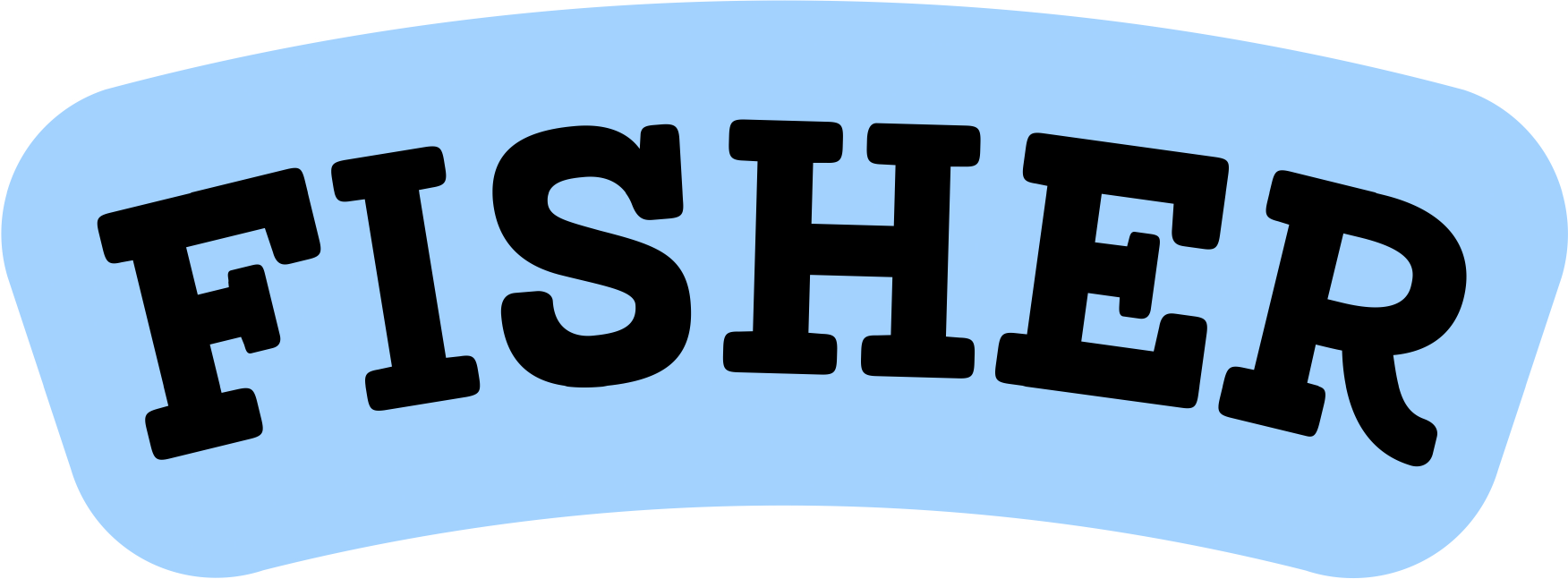}}}
\newcommand{\Rabbit}[1][1.5em]{\raisebox{-0.8em}{\includegraphics[height=#1]{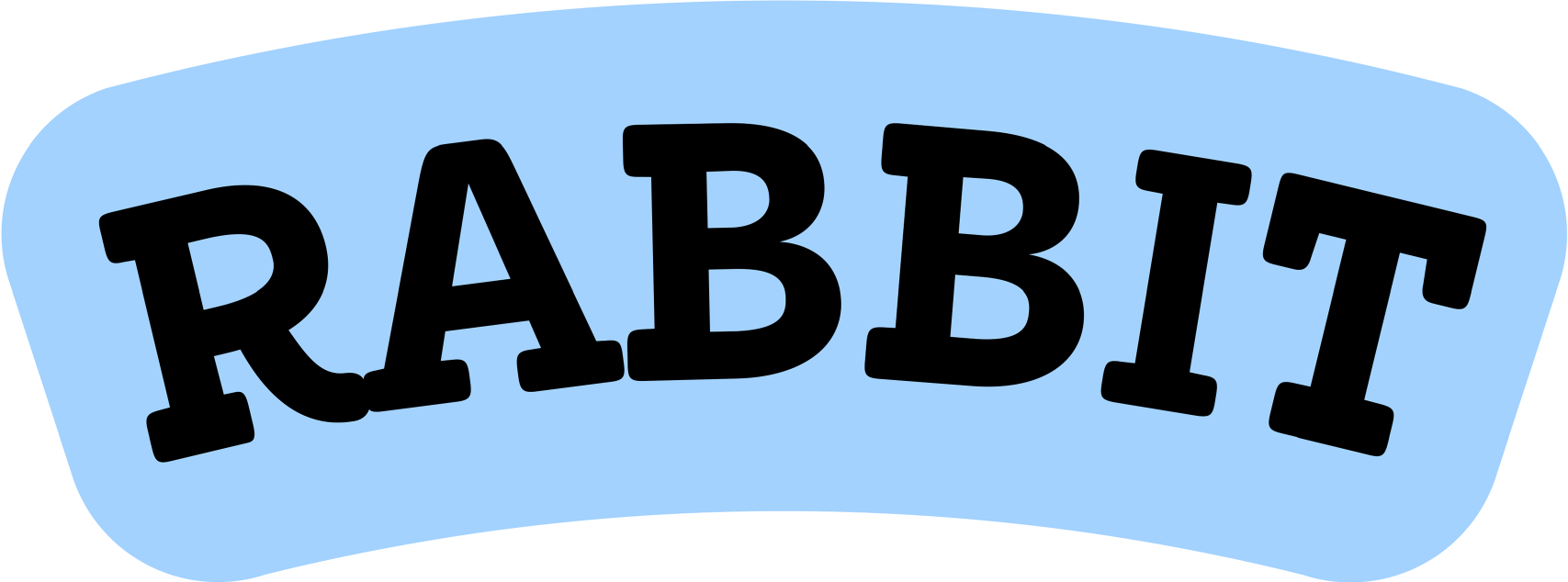}}}
\newcommand{\LEGO}[1][1.5em]{\raisebox{-0.8em}{\includegraphics[height=#1]{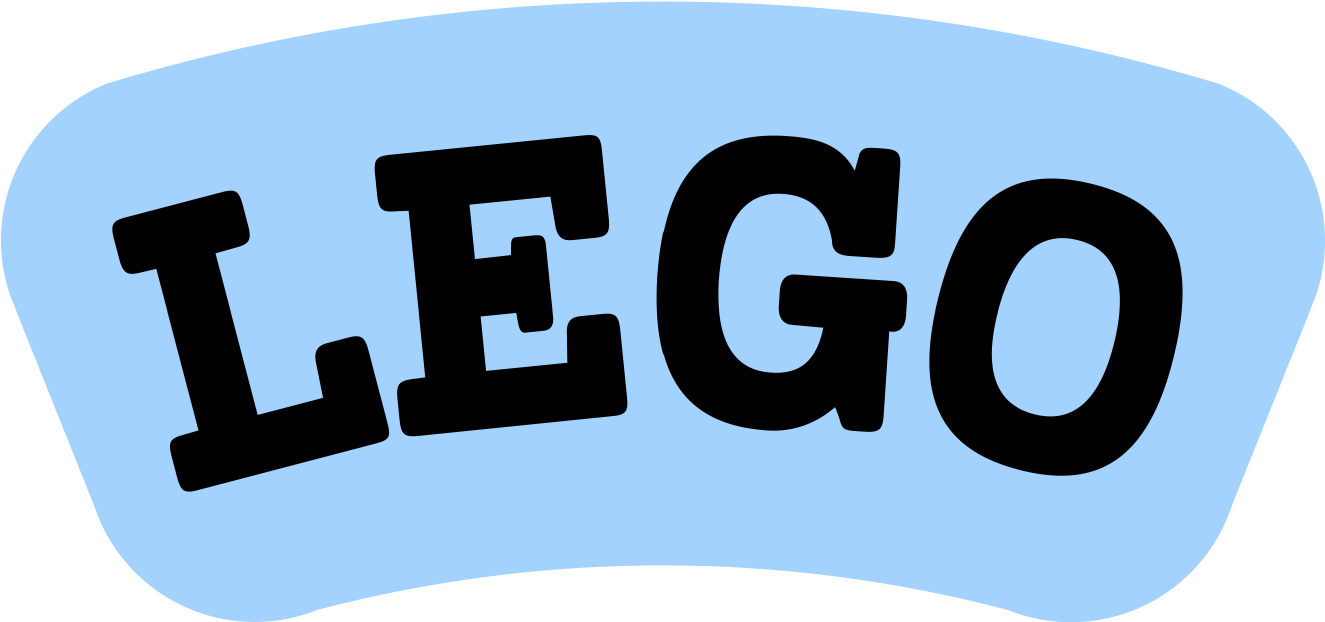}}}
\newcommand{\Climb}[1][1.5em]{\raisebox{-0.8em}{\includegraphics[height=#1]{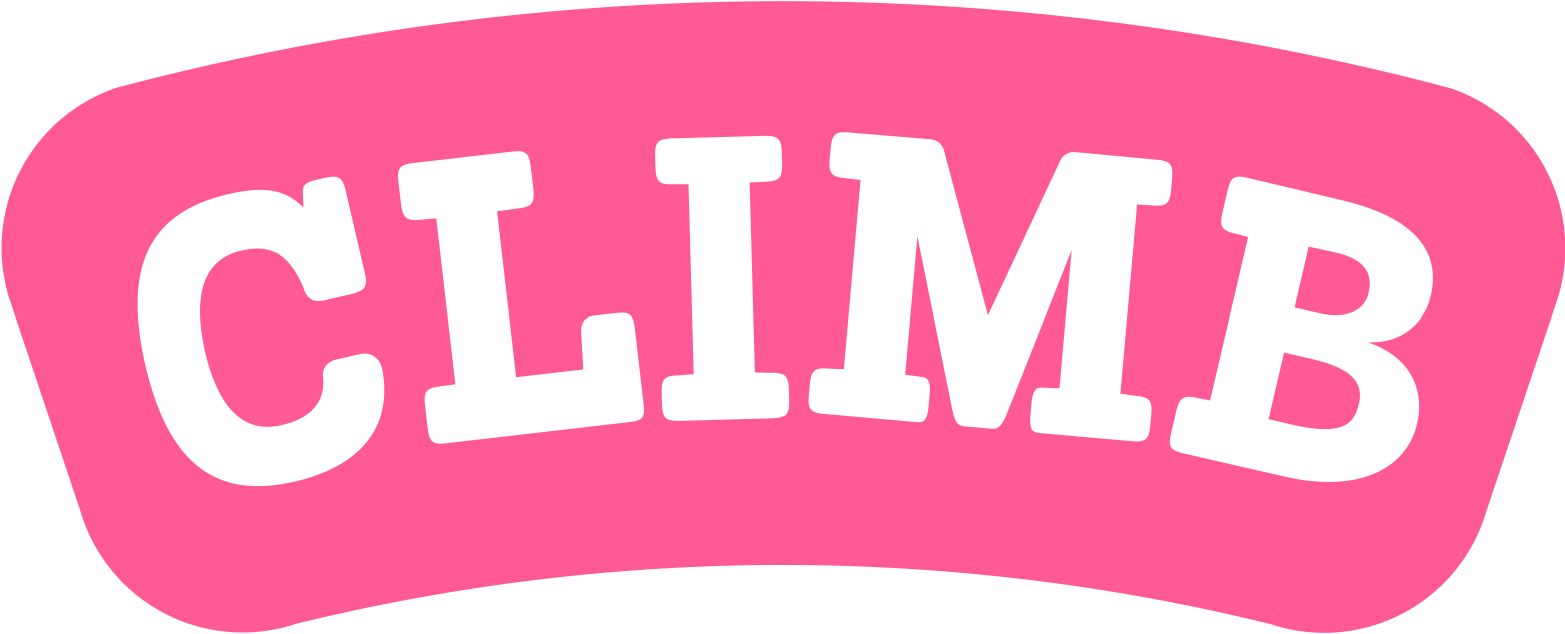}}}
\newcommand{\VRChat}[1][1.5em]{\raisebox{-0.8em}{\includegraphics[height=#1]{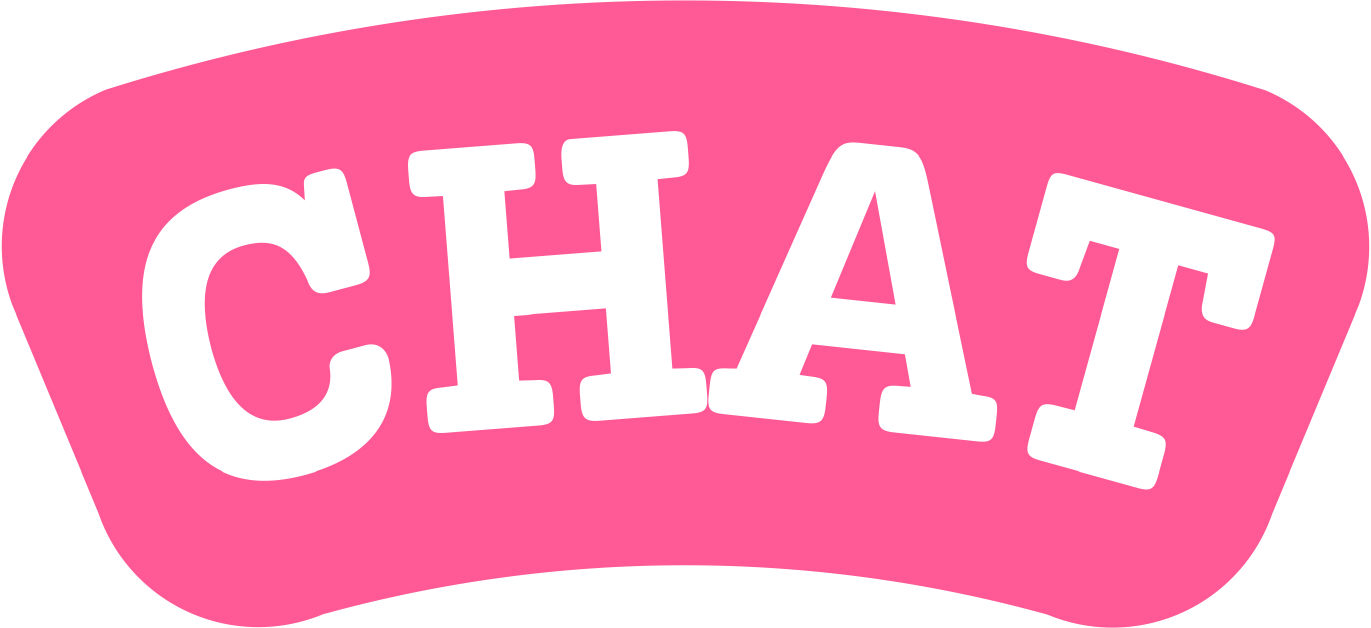}}}
\newcommand{\Horizon}[1][1.5em]{\raisebox{-0.8em}{\includegraphics[height=#1]{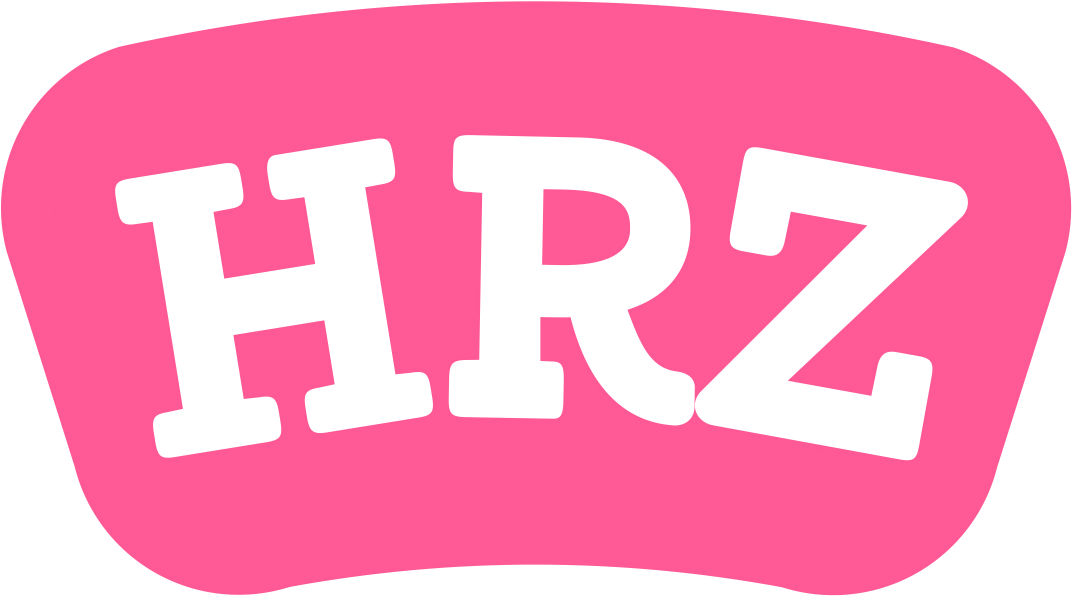}}}
\newcommand{\TRIPP}[1][1.5em]{\raisebox{-0.8em}{\includegraphics[height=#1]{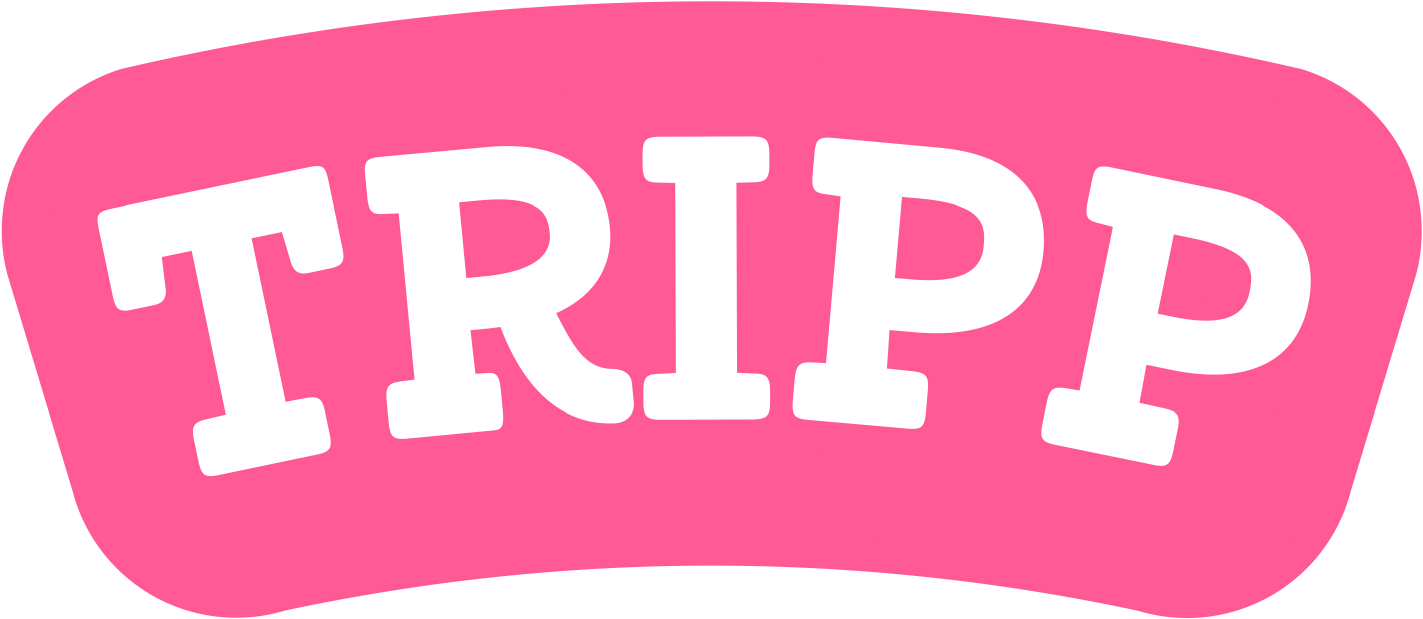}}}
\newcommand{\Supernatural}[1][1.5em]{\raisebox{-0.8em}{\includegraphics[height=#1]{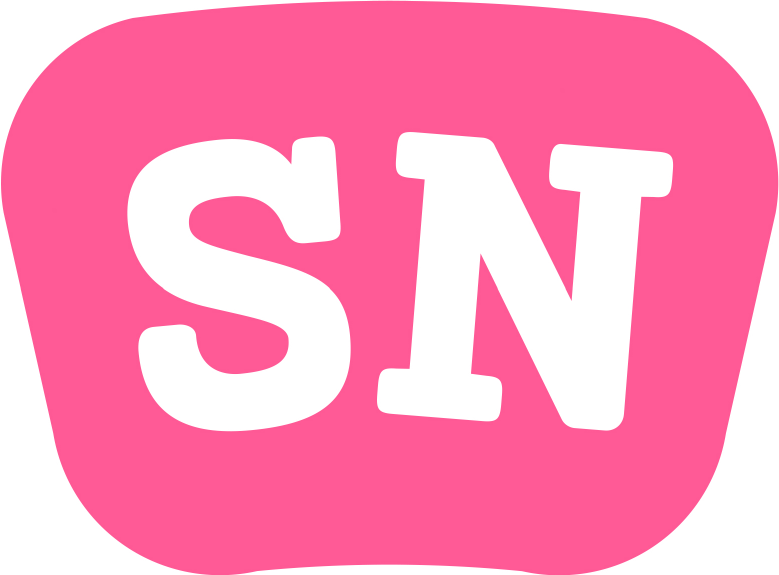}}}
\newcommand{\Immersed}[1][1.5em]{\raisebox{-0.8em}{\includegraphics[height=#1]{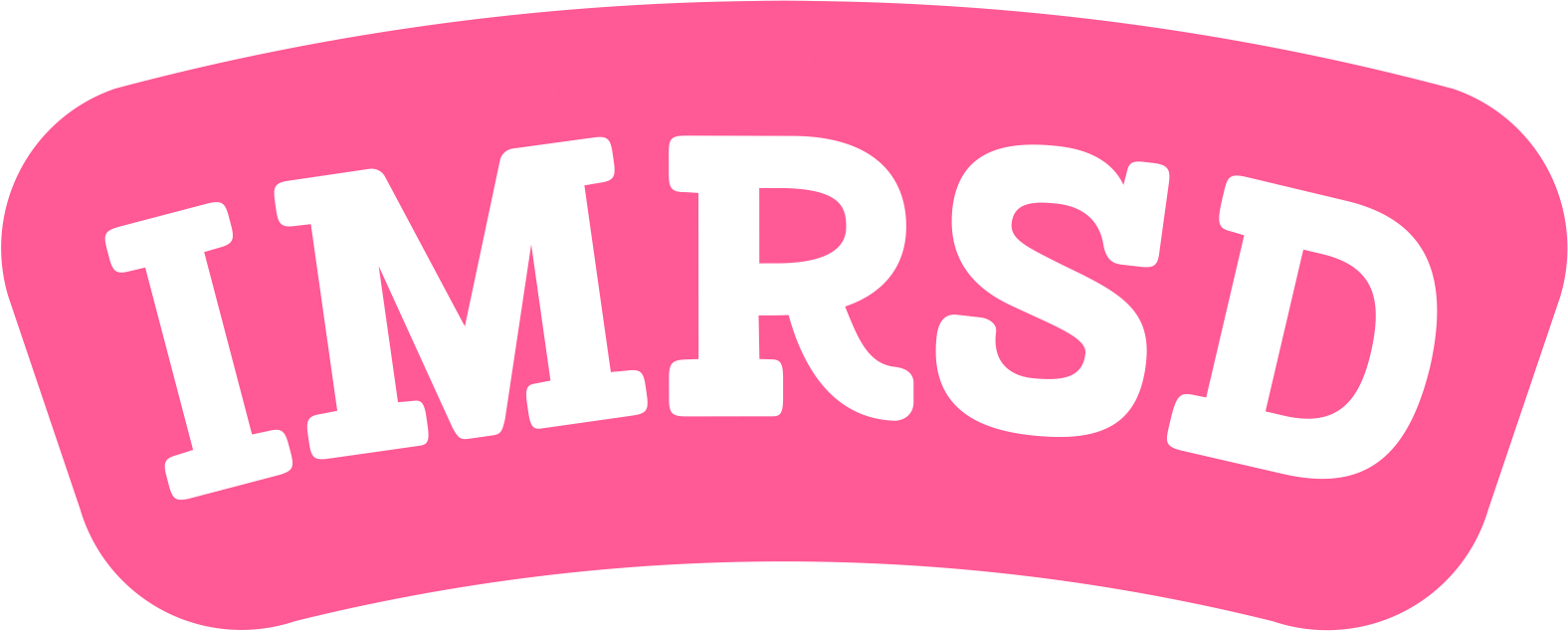}}}

%% file: 00-Abstract.tex
Virtual Reality (VR) technologies offer immersive experiences but collect substantial user data. While deceptive design is well-studied in 2D platforms, little is known about its manifestation in VR environments and its impact on user privacy. This research investigates deceptive designs in privacy communication and interaction mechanisms of 12 top-rated VR games and applications through autoethnographic evaluation of the applications and thematic analysis of privacy policies. We found that while many deceptive designs rely on 2D interfaces, some VR-unique features, while not directly enabling deception, amplified data disclosure behaviors, and obscured actual data practices. Convoluted privacy policies and manipulative consent practices further hinder comprehension and increase privacy risks. We also observed privacy-preserving design strategies and protective considerations in VR privacy policies. We offer recommendations for ethical VR design that balance immersive experiences with strong privacy protections, guiding researchers, designers, and policymakers to improve privacy in VR environments.

%% file: 01-Introduction.tex
Deceptive Design (formerly ``dark patterns'')\footnote{We follow the ACM Diversity and Inclusion Council's guideline for inclusive language and adopt the term ``deceptive design'' instead of ``dark patterns'' in our study. See: \url{https://www.acm.org/diversity-inclusion/words-matter}} refers to design practices that distort or impair users' ability to make informed decisions, regardless of the design intent~\cite{luguri2021shining,bosch2016tales,Brignull2023book}. Users face privacy harms when deceptive tactics, such as bad defaults, mandatory registrations, and social pyramid, impair their decision-making and lead to unintended data disclosures~\cite{opc2024sweep,bosch2016tales,gunawan2022redress,gray2018dark}. Deceptive designs are particularly problematic in social computing, where issues such as data consent interactions and social engineering techniques raise concerns about data ethics and privacy~\cite{gray2021enduser,bosch2016tales,gunawan2021comparative}. 

As a new medium for gaming and social interaction, Virtual Reality (VR) opens a new frontier for manipulative tactics through immersive visual, audio, and haptic feedback that disconnects users from the real world~\cite{speicher2019mixed}. VR's spatial and realistic displays can enhance the subtlety of deceptive designs~\cite{hadan2024deceived,krauss2024what}, making it harder for users to critically assess content~\cite{cummings2022all}. VR simulations using bodily sensor data enable hyper-personalized deceptive tactics~\cite{buck2022security,mhaidli2021identifying}. Additionally, VR-enabled social experiences also introduce challenges to maintaining user privacy~\cite{maloney2020anonymity}.

Although academic researchers and government agencies are actively regulating deceptive designs~\cite{gray2024ontology,opc2024sweep,EUDSA2022,California2020CPRA} and sanctioning companies for invading user privacy with deceptive practices~\cite{FTC2022Epics,AZAGSettlement2022,CNIL2021google,BEUC2024monetization}, these efforts primarily address issues in the 2D mobile and computer platforms (e.g.,~\cite{Brignull2023book,gray2024ontology,mildner2023defending,mildner2023engaging}). Research on deceptive practices in VR and their privacy implications remains limited (e.g.,~\cite{hadan2024deceived}), and deceptive game design studies have largely focused on temporal and monetary manipulations, neglecting privacy issues (e.g.,~\cite{hhadan2024ow2,king20233d,petrovskaya2021predatory}). As commercial games and social applications increasingly shift to VR, it becomes crucial to understand how deceptive practices evolve in immersive technology and the implications for user privacy.

Our research examines the manifestation of deceptive designs in the privacy communication and interaction mechanisms of top-rated commercial VR applications. This is particularly important since users often lack awareness of VR's data collection capabilities~\cite{hadan2024privacy}. Our research objective is to explore how these designs obscure or distort users' understanding of privacy information and impair their privacy decisions. We also explore privacy-enhancing VR designs, which we referred to as \brightpatterns\footnote{We adopt the term \brightpatterns from the literature~\cite{grassl2021dark,truong2022bright,sandhaus2023promoting} to prioritize clarity for this specific class of that literature frequently referred to as the user empowerment alternative to deceptive design. We follow this terminology rather than introducing a new term to prevent confusion, as there is currently no standardized terms in any legislation to describe this kind of design patterns.} that support users' ability to make informed privacy decisions and better understand privacy information~\cite{grassl2021dark,truong2022bright,sandhaus2023promoting}. Such insights are essential for developing VR safeguards that contribute to safer and privacy-respecting VR gaming and social interaction environments. With this objective in mind, we focus on two primary research questions (RQs):

\begin{enumerate}[label= \textbf{RQ\arabic*:}]
    \item To what extent do commercial VR games and applications implement deceptive and ``bright patterns'' that might influence users' privacy decisions?
    \item What specific data practices and privacy risks are concealed by or made more manageable through design strategies in VR games and applications?
\end{enumerate}

To answer these RQs, we examined the privacy communication and interaction mechanisms of 12 top-rated \VRapps, along with their data practices described in their privacy policies (see ~\autoref{subsec:research-context}). These titles were selected for their popularity, market impact, and diverse genres and functionalities. Therefore, they represent the most highly recommended\footnote{Recommended Games---PCVR, Quest \& PSVR. ~\url{https://www.reddit.com/r/VRGaming/comments/ymsewg/recommended_games_pcvr_quest_psvr/}}, top-rated\footnote{Meta Quest Store.~\url{https://www.meta.com/experiences/}}, and widely downloaded products on major VR platforms\footnote{VR Space. (2024, February 14). \textit{The 50 most downloaded VR games of all time.} VR Space.~\url{https://vr.space/news/the-50-most-downloaded-vr-games-of-all-time/}}. The genres of the 12 games and applications selected range from puzzle and adventure games to applications for productivity, fitness, and social interaction. We refer to the former as ``games'' and the latter as ``apps'', which are applications that offer playful experiences but do not necessarily rely on core game elements or demonstrate game characteristics~\cite{deterding2011game,lucero2013playful}. This diverse selection, which we hereby refer to as ``VR games and apps,'' provides a broad overview of privacy practices in various VR use cases and allows us to identify deceptive practices affecting a large number of users in various VR contexts.

Building on prior deceptive design and privacy research~\cite{hadan2024computer,bourdoucen2023privacy,gray2018dark,king2019unfair}, we used a two-phased methodology. First, we conducted an autoethnographic evaluation of 12 \VRapps. Then, we contrasted these findings with the data practices outlined in the applications' privacy policies derived through thematic analysis. This two-phased approach enabled us to identify discrepancies between the privacy information presented to users, the ways user-facing privacy mechanisms influence decision-making, and the actual privacy issues disclosed in the publisher's data practices. 

Our research findings make three main \textbf{contributions}: 
\textit{First}, we show that while deceptive design patterns in the registration and subscription processes, consent interactions, privacy settings, quality-of-life (QoL) features, and notification mechanisms in \VRapps still heavily rely on 2D interfaces in VR, their presence in the privacy communication and interaction mechanisms of \VRapps demonstrates how traditional manipulative tactics are being adapted in immersive environments. These tactics are further amplified by VR-unique features, with increased harm to users' privacy. \textit{Second}, our analysis of privacy policies shows that convoluted and complex policies, already a known usability issue on mobile and web platforms, continue to plague users in VR, obscure their understanding, and hinder their privacy decision-making. We found that substantial data transmissions are not transparently communicated through VR design mechanisms, with many relying on submissive consent practices. Some lacked privacy settings entirely, which further increases users' confusion and unawareness of privacy risks. These findings emphasize the need for transparent and usable privacy communication, both in design mechanisms and policy construction. \textit{Third}, we identified privacy-enhancing VR design strategies and protective considerations in privacy policy that can serve as a foundation for more effective privacy communication and interaction in \VRapps. Based on these results, we make recommendations for the ethical design of \VRapps that balance immersive experiences with strong privacy protections to guide VR researchers, designers, and policymakers in creating privacy-preserving \VRapps and crafting clearer privacy policies for this evolving technology.

%% file: 02-RelatedWork.tex
\subsection{Deceptive Design and Its Privacy Implications}
\label{subsec:literature-deceptivedesign}

Deceptive Designs exploit both people's internal cognitive biases and external social pressures that constitute complex human decision-making~\cite{simon1997models,waldman2020cognitive,thaler2009nudge}. They manipulate users into making choices that are not in their best interest and lead to various negative consequences~\cite{mathur2019dark,Brignull2023book,mathur2021what}. Literature has identified four issues associated with deceptive design, including loss of autonomy and decision-making capability~\cite{mathur2021what}, financial loss~\cite{mathur2019dark,luguri2021shining,lewis2014irresistible,bongard2021manupulated}, labor and cognitive burdens (e.g., emotional distress)~\cite{nouwens2020dark,maier2019dark,bongard2021manupulated,gray2021enduser}, and invasion of privacy~\cite{bosch2016tales,mathur2021what,gunawan2022redress}. Within the context of video games, research has found deceptive designs for psychological and social manipulation~\cite{zagal2013dark,hhadan2024ow2}, attention capture~\cite{roffarello2023defining}, and mechanisms that prolong gameplay or encourage impulsive spending~\cite{geronimo2020UI,karlsen2019exploited,zagal2013dark,fitton2019F2P}. Building on early deceptive design research~\cite{Brignull2010deceptive,zagal2013dark,gray2018dark,mathur2019dark}, academic researchers and government agencies have further investigated and categorized deceptive design across various digital platforms, such as social media~\cite{mildner2023defending,mildner2023engaging}, websites~\cite{Brignull2023book,oecd2022dark,opc2024sweep,GPEN2024}, mobile apps~\cite{geronimo2020UI,gunawan2021comparative}, games~\cite{hhadan2024ow2,king20233d,BEUC2024monetization}, and 3D immersive environments~\cite{Greenberg2014proxemic,hadan2024deceived,krauss2024what}. Research has also proposed user empowerment design patterns---``bright patterns''---to counteract deceptive design~\cite{grassl2021dark,truong2022bright,sandhaus2023promoting}. In recent work, ~\citet{gray2024ontology} compiled an ontology of 64 deceptive designs based on regulatory and academic sources, and \citet{hadan2024computer} further expanded it into a VR-specific assessment guide. 

\textit{Relevance.} 
While deceptive design literature provided a foundation for our analysis, we primarily relied on \citet{hadan2024computer}'s \textit{VR Deceptive Game Design Assessment Guide} as our deductive codebook due to its integration of foundational works in deceptive design research (e.g.,~\cite{gray2024ontology,Brignull2010deceptive,gray2018dark,mathur2021makes}) and regulatory works that focus on how deceptive designs affect user privacy (e.g.,~\cite{bosch2016tales,oecd2022dark,cma2022UK,ftc2022bringing}). This \guide further includes frameworks specifically targeting VR and game-related deceptive designs~\cite{hadan2024deceived,king20233d,zagal2013dark}. In ~\autoref{subsubsec:diary-analysis-method}, we detail how this \guide shaped our exploration of privacy-impacting deceptive design in VR games and apps included descriptions of deceptive patterns from literature~\cite{hadan2024computer,gray2024ontology} in~\autoref{tab:diary-theme}.

\subsubsection{Privacy Implications of Deceptive Design}

As research on deceptive design evolves, it has expanded beyond issues of autonomy, financial harm, and cognitive burden to address privacy implications (e.g.,~\cite{bosch2016tales,zhang2024navigating,gunawan2022redress}). In contrast to back-end privacy issues such as data breaches or compromised network traffic, users' immediate perception of privacy predominantly hinges on privacy interfaces. Regulatory efforts have sought to regulate the deceptive design issue~\cite{UKICO,CAADCA}, and impose sanctions on companies that manipulate children and invade their privacy~\cite{FTC2022Epics,FTC_genshin2025}. Building upon established strategies for privacy preservation~\cite{hoepman2014privacy}, ~\citet{bosch2016tales} formulated a taxonomy of eight privacy-related deceptive design strategies and documented their prevalence across popular online platforms. \citet{zhang2024navigating} investigated the reasons behind implementing these strategies from the perspective of design practitioners. \citet{gunawan2022redress} summarized three privacy-related contexts where deceptive designs frequently appear on websites and mobile apps, such as user settings, exit requests, and ``entry'' requests and consent interactions. Deceptive practices in consent mechanisms have also been identified, such as missing terms of service, bundled consent, pre-checked opt-in boxes, and nagging~\cite{gunawan2021comparative,soe2020circumvention,nouwens2020dark}. These strategies create the illusion of informed and free choice~\cite{waldman2020cognitive}. Additionally, privacy policies can employ deceptive tactics, such as hiding hyperlinks~\cite{jensen2004privacy}, using positive framing, exploiting time gaps between notice and choice~\cite{adjerid2013sleights}, and omitting privacy information~\cite{linden2018privacy,zimmeck2019maps}. Legalistic language further discourages reading and hinders user understanding~\cite{jensen2004privacy,fabian2017large,opc2024sweep}.

\textit{Relevance.} 
Studies have shown that users are less aware of privacy harms caused by deceptive designs compared to other harms~\cite{bongard2021manupulated,bourdoucen2023privacy}. However, existing academic and regulatory efforts have primarily focused on 2D websites and mobile applications (e.g.,~\cite{gunawan2021comparative,opc2024sweep,gunawan2022redress,geronimo2020UI}), and overlooked their impact in gaming and immersive environments. Given users' unawareness and lack of transparency in online games' data practices~\cite{kroger2023surveilling,seif2020data,bourdoucen2023privacy} and VR technology's data collection capabilities~\cite{hadan2024privacy}, our study explores how deceptive design practices manifest in gaming and VR environments and potentially compromise user privacy in these contexts.

\subsubsection{Importance of Considering User Perspectives in Deceptive Design Research}
\label{subsubsec:user-perspectives-literature}

The impact of deceptive design largely depends on users' ability to recognize such patterns, which varies depending on their literacy in this area~\cite{zagal2013dark,geronimo2020UI,luguri2021shining}. For instance, \citet{geronimo2020UI} revealed that users often remain unaware of deceptive design in mobile apps, while \citet{luguri2021shining} found that subtle deceptive practices are more likely to go unnoticed by less-educated users. \citet{m2020towards} identified five key factors that influence users' susceptibility to deceptive design: frequency of exposure, perceived trustworthiness, level of frustration, misleading behavior, and the physical appearance of the user interface. Research also highlights the mixed reactions of players to game mechanisms that involve deceptive designs. For example, some players view daily tasks as rewarding and motivating, but others see them as pressure-inducing due to the ``Fear of Missing Out'' and the obligation to participate~\cite{frommel2022daily,hhadan2024ow2}. Similarly, grinding may be perceived positively when it involves variations and achievements, but it can also be seen as tedious~\cite{karlsen2019exploited,zagal2013dark,hhadan2024ow2}. Given the complexity of identifying deceptive design, the literature advocates for user-centric approaches to discern practices that may not be overtly unethical~\cite{gray2021enduser,gray2023mapping,hhadan2024ow2} and derive ethical design solutions~\cite{petrovskaya2024ask,hhadan2024ow2}. However, non-expert participants are often unable to correctly identify malicious designs, let alone specific deceptive design patterns~\cite{geronimo2020UI}. 

\textit{Relevance.} 
To address the challenge while maintaining a user-centric focus, we adopted an autoethnographic approach~\cite{rapp2018autoethnography,Kaltenhauser2024autoethnographic} to enable researchers to engage with VR applications as users, systematically document their frustrations, concerns, and observations over time
~\cite{rapp2018autoethnography,Kaltenhauser2024autoethnographic}, and refine their interpretations through an introspective process~\cite{rapp2018autoethnography,adams2016handbook}. Autoethnography has been used to explore privacy risks~\cite{laato2022balancing}, usability issues with privacy controls~\cite{fassl2023can,turner2022hard}, and deceptive design in both computer~\cite{karlsen2019exploited,rapp2017games} and VR and AR games~\cite{hadan2024computer,laato2022balancing}. Given the subtle nature of deceptive designs in VR~\cite{krauss2024makes,hadan2024deceived}, the autoethnography approach that paired our researchers' expertise with a first-hand user experience is ideal for studying deceptive design in VR~\cite{hadan2024computer}. 

\subsection{Privacy Concerns of VR and Games and Their Interplay with Deceptive Design}

\subsubsection{Privacy Implications of VR Technology and Its Deceptive Design}

The rise of VR technology has introduced new avenues for user profiling~\cite{buck2022security,hadan2024privacy}, as VR sensors track user movements, generate feedback, and enable inferences about users' physical and mental conditions~\cite{miller2020personal}, habitual movements~\cite{miller2020personal,pfeuffer2019behavioural}, cognition, emotions, and personality~\cite{buck2022security, mhaidli2021identifying}. Enhanced user experiences in VR often encourages the sharing of private information, with companies often promising improved services in exchange~\cite{bonnail2022exploring}. These data can be exploited to manipulate user behavior~\cite{mhaidli2021identifying,buck2022security,BEUC2024monetization} and drive profits for companies~\cite{mhaidli2021identifying,egliston2021examining,BEUC2024monetization}. However, studies show that even experienced users are often unaware of the types, uses, and management of the data collected by immersive technology~\cite{hadan2024privacy,abraham2022implications,gallardo2023speculative}, which can lead to unclear expectations and increased privacy risks. Research suggests that deceptive designs that invade user privacy from non-VR environments are likely to migrate into VR~\cite{hadan2024deceived,krauss2024what}. For example, XR\footnote{Extended Reality (XR) describes immersive technologies such as Virtual Reality (VR), Augmented Reality (AR), and Mixed Reality (MR).} variations of ``toying with emotion'' such as hyper-personalized digital characters mimicking friends or family, can manipulate users into revealing private information~\cite{hadan2024deceived,gray2018dark}. Similarly, perception hacking in VR might trick users into interacting with undesired virtual objects, leading to accidental information disclosure or unintended consent~\cite{gray2018dark,su2022perception}. Moreover, VR's immersive nature may introduce novel deceptive designs~\cite{hadan2024deceived}, such as immersion in realistic and multi-sensory environments~\cite{speicher2019mixed}, which can obscure users' sense of reality~\cite{hadan2024deceived,krauss2024what,cummings2022all}. These XR features can lead to new deceptive practices, such as false memory implantation, blind-spot tracking, and reality distortion, all of which exploit human cognitive and perceptual limitations~\cite{schlembach2021forced,su2022perception,bonnail2022exploring}. 

\textit{Relevance.}  
While recent studies have examined the potential for deceptive design in immersive environments, they have primarily relied on hypothetical scenarios, expert predictions, and controlled experiments (e.g.,~\cite{hadan2024deceived,krauss2024what,su2022perception}) that often focus broadly on XR technology but overlooked the contextual factors of VR. To address this gap, our research investigates how deceptive practices manifest in real-world \VRapps, with special attention to their impact on user privacy.

\subsubsection{Privacy Implications of Video Game and Its Deceptive Design}

As a reason behind VR adoption~\cite{nair2023exploring}, video games collect extensive data without sufficient disclosure~\cite{bourdoucen2023privacy,kroger2023surveilling}. During account registration, users often provide personal information such as full name, date of birth, email address, postal address, and payment details~\cite{kroger2023surveilling}. Beyond these manual inputs, gaming devices can capture a wide array of data, including voice, gestures, heart rate, facial expressions, location, and hardware configurations~\cite{russell2018privacy,kroger2023surveilling}. Games also track users' in-game actions and may request access to other applications or social media profiles~\cite{canossa2009patterns,russell2018privacy}. The granularity of data collection varies, with more advanced games gathering a broader range of information~\cite{drachen2014comparison}. Such detailed data collection enables companies to profile users and infer their demographic traits, interests, personality, emotional states, skills, and financial habits~\cite{martinovic2014you,russell2018privacy,kroger2023surveilling}. Users may adopt fictional identities but their virtual behavior often reflects their real-world personalities~\cite{spronck2012player,el2013game}. The entertainment value of games and the complexity of data practices make it difficult and demotivating for users to stay informed~\cite{kroger2023surveilling,russell2018privacy}. In fact, many users remain unaware of or misunderstand how their data is handled in games~\cite{bourdoucen2023privacy}. In addition, privacy policies for video games are often vague and omit critical details~\cite{russell2018privacy}, which leaves opportunities for deceptive designs. Furthermore, proposals to use player-NPC (non-player character) communication, such as emotional states communicated through NPC dialogues, can lead to privacy intrusions~\cite{frommel2021gathering} and facilitate emotional manipulation~\cite{hadan2024computer}. Therefore, games' entertainment value, NPC intimacy, and vague privacy policies~\cite{frommel2021gathering,russell2018privacy,kroger2023surveilling} all increase the risk of users being subtly deceived into sharing private information.

\textit{Relevance.} 
While prior studies have examined privacy in gaming through information flow analysis~\cite{nair2023exploring} and data-driven prediction~\cite{martinovic2014you}, research on user attitudes and awareness of data-sharing practices in online gaming remains limited (e.g.,~\cite{bourdoucen2023privacy}). Regulatory guidelines primarily target privacy in computer and mobile games~\cite{OPC2019gaming,opc2024sweep}, with minimal attention to those in VR. Although some studies have touched on deceptive designs that invade user privacy in VR, they often rely on practitioners' and literature's hypotheses (e.g.,~\cite{krauss2024what,hadan2024deceived}), a narrow selection of commercial VR games (e.g.,~\cite{hadan2024computer}), and a limited focus on consent interactions and concerns in computer and mobile games (e.g.,~\cite{gunawan2022redress,bourdoucen2023privacy,drachen2014comparison}). Deceptive design in broader privacy contexts and a wider range of VR applications remains a gap in the literature. In this study, we aim to understand privacy-invasive data practices in VR games and the role of deceptive design in these processes.

%% file: 03-Methodology.tex

\begin{figure}[!t] 
  \includegraphics[width=\textwidth]{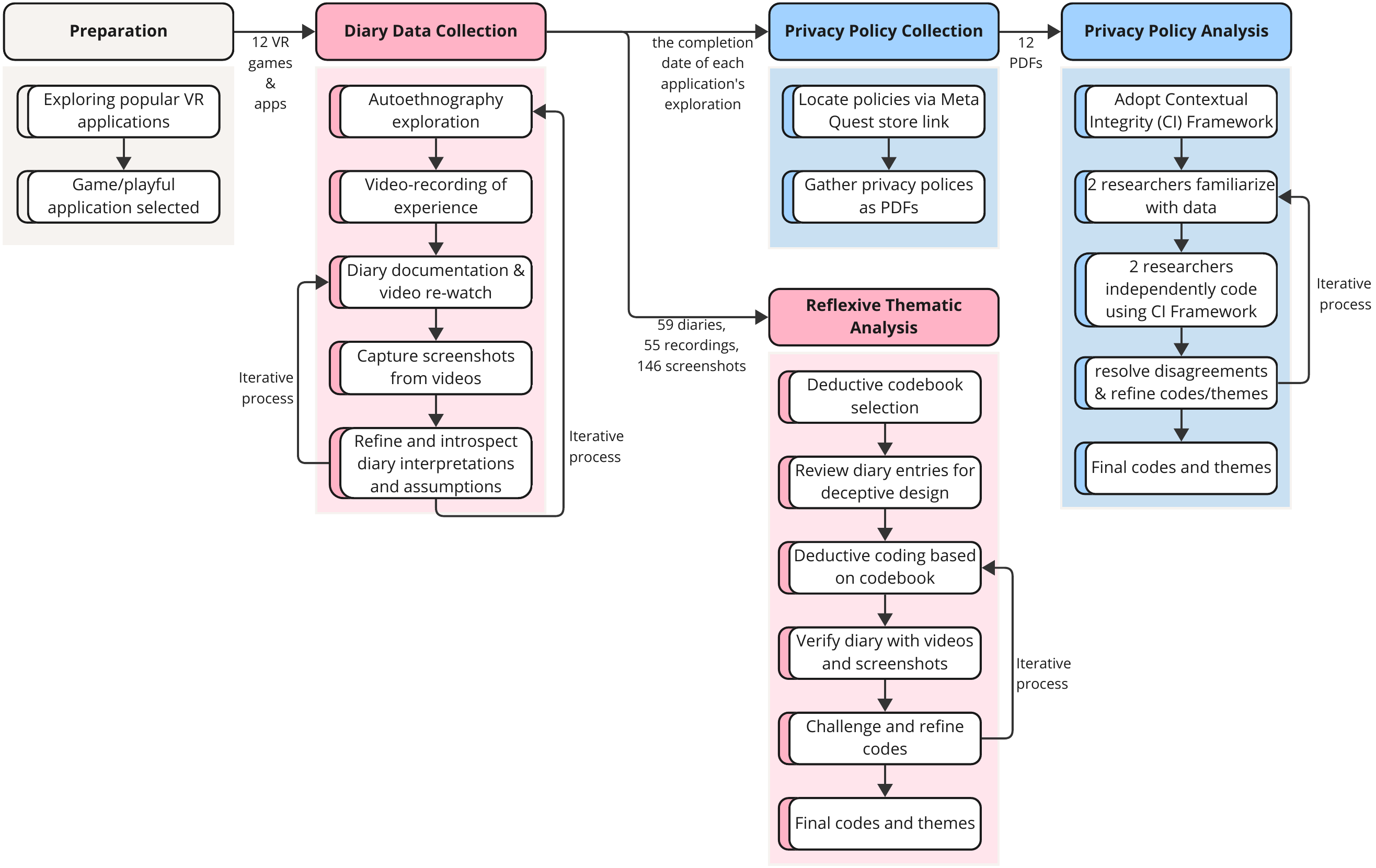}
  \vspace{-7mm}
  \caption{Flowchart of our two-phased methodology, with \textcolor{palepink!160}{\textbf{pink}} denotes the process of our autoethnographic evaluation of VR games' and apps' privacy communication and interaction design mechanisms, and \textcolor{paleblue!160}{\textbf{blue}} denotes the process of our privacy policy analysis. The reflexive thematic analysis and privacy policy data collection were conducted in parallel. We completed each application's autoethnographic exploration before collecting its privacy policy to ensure the policy was up-to-date during our analysis.}
  \Description{Flowchart of our two-phased methodology, with pink denotes the process of our autoethnographic evaluation of VR games' and apps' privacy communication and interaction design mechanisms, and blue denotes the process of our privacy policy analysis.}
  \label{fig:diary-flowchart}
\end{figure}

Building upon methodology in previous deceptive design and privacy literature~\cite{hadan2024computer,bourdoucen2023privacy,gray2018dark,king2019unfair}, in this research, we adopted a two-phased methodology (see~\autoref{fig:diary-flowchart}): (1) an autoethnographic evaluation of VR games' and apps' privacy communication and interaction design mechanisms with our researchers with expertise in deceptive design, user privacy, and games research, and (2) a thematic analysis of privacy policies of the evaluated \VRapps. 

\input{Table-AppSummary}

\subsection{Research Context}
\label{subsec:research-context}

We evaluated 12 popular VR games ($n=6$) and apps ($n=6$) that were top-rated on the Meta Quest Store\footnote{See footnote 4.} and highly recommended by VR users on Reddit\footnote{See footnote 3.} to cover diverse game genres and application categories to represent various gameplay styles and user experiences. \autoref{tab:app-summary} summarizes the selected VR games' and apps' publisher, initial release year, main gameplay mechanics, availability of a privacy policy, and available platforms. Although our analysis was performed on the Meta Quest 3 headset, the selected apps are available on multiple VR platforms and are widely accessed by millions of users worldwide. This wide user base 
enables us to identify privacy-related deceptive design patterns that are likely to manifest in the mainstream VR user experience.

\subsection{Design Mechanism Evaluation}

We employed an autoethnography method, which situates researchers as both the ``protagonist'' and observer, and enables a first-person engagement and understanding of the VR environment~\cite{rapp2018autoethnography,Kaltenhauser2024autoethnographic}. Our decision to adopt this methodology is informed by literature on HCI, games, and security and privacy research (e.g.,~\cite{rapp2017designing,rapp2017games,laato2022balancing,fassl2023can,turner2022hard}), as discussed in ~\autoref{subsubsec:user-perspectives-literature}. 

\subsubsection{Ethical Considerations and Limitations}
\label{subsubsec:ethical_limitations}

Following the research ethics guidelines~\cite{tpcs2022}, we carefully designed our methodology to prioritize ethical considerations. Before initiating the autoethnographic diary data collection, we consulted our institution's Research Ethics Board (REB) to confirm that our approach did not require formal ethical review. To protect the privacy of other users, we focused our investigation on single-player modes in the selected \VRapps to avoid other users' involvement without consent. For apps (e.g., VRChat, Meta Horizon Worlds, and Immersed) that inherently include social interactions, we centered our analysis on the researcher's journey and avoided video recordings that involved other users. We also consulted our university library on the appropriate use of screenshots for research purposes in this manuscript, under Fair Dealing guidelines in the Copyright Act of our country.

Our autoethnographic study, while providing valuable insights, has limitations. We have provided a detailed reflexivity statement (see~\autoref{para:reflexivity}) to address potential biases from the lead researcher's background. However, we recognize that our findings may not fully represent the experiences of a broader user base with varying levels of deceptive design literacy and perceptions~\cite{zagal2013dark,geronimo2020UI,luguri2021shining,frommel2022daily}. Nonetheless, we are confident that our autoethnographic study on the 12 cross-platform \VRapps allowed us to experience a wide variety of deceptive tactics and privacy issues in the privacy communication and interaction mechanisms in VR environments.

\subsubsection{Data Collection}

The lead researcher with expertise in VR, deceptive design, gaming, and usable privacy engaged and fully immersed in all 12 \VRapps during a six-month timeframe. For games and apps with a clear endpoint (e.g., storyline completion), exploration continued until the endpoint was reached to ensure a comprehensive understanding of the privacy mechanisms. For genres without a defined endpoint (e.g., Beat Saber), the exploration followed the principle of saturation, ending when no new privacy issues and deceptive designs were observed and no new insights emerged from the privacy communication and interaction mechanisms. This autoethnography process resulted in 59 diary entries, 55 video recordings, and 146 screenshots. 

\paragraph{Diary Entries}

Dairy entries focused on the design mechanisms that allow users to (1) obtain privacy information, and (2) make privacy choices and express privacy preferences. Following an established method in literature~\cite{hadan2024computer}, the lead researcher systematically documented the experience in two segments in each diary entry: (1) detailed observations of privacy communication and interaction design mechanics that incorporated deceptive designs, and (2) reflections on experiences, reactions, thoughts, and emotional responses when the design mechanics were encountered. Using this strategy, the lead researcher personally experienced and immersed in the \VRapps. This process resulted in 55 immediate reflections after each session and 4 cross-session entries, identifying common trends across the games and apps. These diary entries ranged from 67 to 340 words in length, averaging approximately 146 words each. ~\autoref{tab:app-summary}~details hours spent and diary entries for each VR game and app, with an example diary entry in~\autoref{app-sec:example-diary}.

\paragraph{Video Recordings and Screenshots}

Video recordings were captured throughout gameplay and application sessions to document instances of deceptive designs and user interactions in VR. Each session was experienced and screen-recorded using a Meta Quest 3 headset, with the recording paused after key privacy interactions, such as setting adjustments or consent agreements, to allow the lead researcher to immediately document reflections in a diary entry. The researcher repeatedly reviewed the video recordings during the diary entries to pinpoint moments of interaction with VR design mechanisms involving deceptive designs. This re-watching process enabled deeper reflection on how these designs might subtly influence user behavior and decision-making, which may not be immediately apparent during real-time interactions. Screenshots were also taken to visually support the text-based observational diary entries, with multiple screenshots taken for each design mechanism to provide clear insight into its design elements and functionality. The most illustrative screenshots and explanations of these design mechanisms are included in~\autoref{app-sec:screenshots-deceptive}. The diary entries, along with the corresponding video recordings and screenshots, were organized on Miro\footnote{Miro---the Virtual Workspace for Innovation.~\url{https://miro.com/}} for easier visual interpretation. Upon completing the autoethnography evaluation of all 12 \VRapps, these data were then transferred to Dovetail\footnote{Dovetail---Thematic Analysis Software.\url{https://dovetail.com/}} for thematic analysis. 

\subsubsection{Data Analysis}
\label{subsubsec:diary-analysis-method}

The lead researcher analyzed data using the \textit{deductive reflective thematic analysis} method~\cite{clarke2021thematic} to preserve the personal and reflexive nature of the autoethnography study~\cite{rapp2018autoethnography,fassl2023can,hadan2024computer} while enabling a ``theoretically-informed exploration of qualitative data''~\cite[p.~263]{clarke2021thematic}. Prior studies have extensively used this method to analyze autoethnographic data (e.g.,~\cite{laato2022balancing,fassl2023can,hadan2024computer}). To offer transparency in our findings and interpretations~\cite{rapp2018autoethnography}, we provide a \textit{Reflexivity Statement} that follows the recommendations for qualitative research~\cite{may2014reflexivity,saldana2021coding,clarke2021thematic}. 

\paragraph{Reflexivity Statement}
\label{para:reflexivity}

To enhance the rigor of our qualitative research, we provide a \textit{Reflexivity Statement} for our autoethnographic evaluation of \VRapps to acknowledge potential biases in data coding and theme development, following guidance from the literature~\cite{may2014reflexivity,saldana2021coding,clarke2021thematic}. 

Our reflexive thematic analysis was led by our lead researcher who has a background in deceptive design and user privacy research, and has previously published on these topics in the context of VR and traditional computer environments (e.g.,~\cite{hadan2024computer,hadan2024deceived,hadan2024privacy,hhadan2024ow2}). Our research team's collective expertise in interaction design, usable privacy, and games user research also ensures the comprehensiveness and quality of our study.

The lead researcher self-identifies as empathetic and sensitive, with a strong emphasis on protecting personal boundaries. This personality sharpened the researcher's attention to subtle privacy intrusions and design patterns that exploit users' engagement with lifelike, responsive VR elements, but  may have also influenced the focus and interpretation of the data compared to a typical VR user. We also acknowledge that the lead researcher's positive and negative personal experience with video games, VR technology, and deceptive design may have influenced the research.

\paragraph{Thematic Analysis}

Our analysis primarily relied on an evolving \textit{VR Deceptive Game Design Assessment Guide}~\cite{hadan2024computer} as our deductive codebook, given its integration of insights from foundational deceptive design works~\cite{gray2024ontology,Brignull2010deceptive,gray2018dark,mathur2021makes} and VR and game-related~\cite{hadan2024deceived,king20233d,zagal2013dark} and privacy deceptive design frameworks~\cite{bosch2016tales,oecd2022dark,cma2022UK}. We also draw insights from broader literature (as discussed in ~\autoref{subsec:literature-deceptivedesign}) to capture any deceptive design patterns beyond those in the codebook. We conducted the analysis in three stages. First, the lead researcher reviewed the data based on the deductive codebook, with a focus on identifying potential deceptive designs in the privacy communication and interaction design mechanisms. In the second stage, the diary entries from the recordings were deductively coded. The video recordings and screenshots were used to verify the observations and reflections documented in the diaries. This involved an iterative coding process, where the analysis was continuously challenged and refined based on the researcher's critical insights and expertise in deceptive design within VR experiences. In the third stage, our research team collectively discussed the findings to ensure no key aspects were overlooked in the data. In the end, we identified and synthesized seven deceptive design patterns from privacy communication and interaction mechanisms of six VR games, and 14 patterns from six VR apps (see~\autoref{tab:diary-theme}). 

\subsection{Privacy Policy Analysis}

We analyzed the privacy policies of the selected \VRapps, and compared them with findings from our design mechanics evaluation. This approach reveals the data collection and sharing practices of games~\cite{bourdoucen2023privacy,rauti2020location} and apps~\cite{bui2021consistency}, and those in VR~\cite{trimananda2022OVR,brehm2023understanding}. Our goal was to uncover privacy issues and hidden data practices distorted by deceptive designs, which leave users vulnerable to unwanted data practices and unrecognized privacy risks.

\subsubsection{Data Collection}

We collected privacy policies for the selected \VRapps from their official websites, accessed through the Meta Quest Store. Since these apps operate on the Meta Quest platform and rely on its device storage and functionality, we gathered this platform's privacy policy and found it shares the same privacy policy as the VR app \name{Meta Horizon Worlds}. Meta provides two privacy policies: the ``Supplemental Meta Platforms Technologies Privacy Policy''\footnote{Supplemental Meta Platforms Technologies Privacy policy (June, 20, 2024).~\url{https://www.meta.com/ca/legal/privacy-policy/}. Last accessed on August 2, 2024.} specifically for Meta's VR products and the broader ``Meta Privacy Policy''\footnote{Meta Privacy Policy (June 26, 2024).~\url{https://www.facebook.com/privacy/policy/}. Last accessed on July 28, 2024.} covering all Meta services, including Facebook and Messenger. For this research, we focused on the VR-specific policy and its related explanations about data practices, as the broader policy covers information about unrelated products. In total, the 12 privacy policies we collected cover both the data practices of the selected \VRapps as well as their operating platform.

Since privacy policies often change with software updates, we collected them on the completion day of each autoethnography exploration. The Meta Quest Store privacy policy was collected in August 2024, after the final VR app evaluation. This approach ensured that our privacy policy analysis corresponded to the exact versions of the \VRapps at the time of exploration. We manually downloaded the policies as PDF files. For multi-page policies, we compiled all content into a single PDF. This data collection process resulted in 12 PDF files, which were then uploaded to Dovetail\footnote{See footnote 10.} for analysis. \autoref{app-sec:readability} presents the retrieval dates of each privacy policy.   

\subsubsection{Data Analysis}
\label{subsubsec:policy-analysis-method}

\begin{figure}[t!]
  \includegraphics[width=\textwidth]{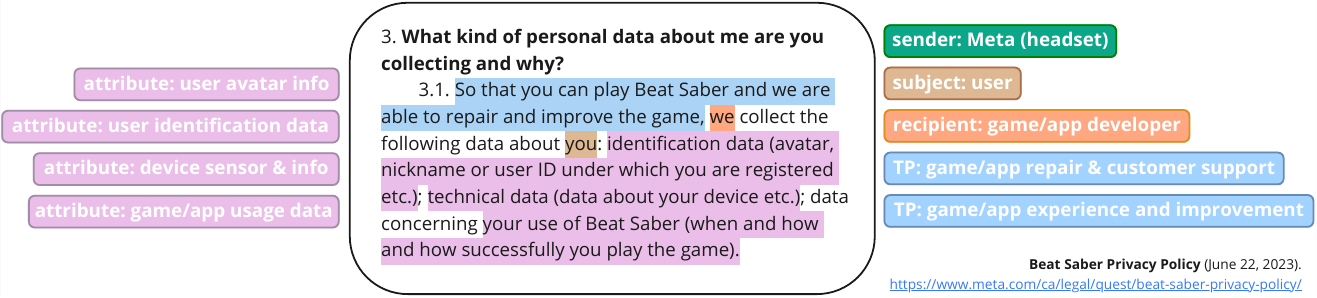}
  \vspace{-7mm}
  \caption{TP=Transmission Principle. Example of our coding process using the contextual integrity framework (\cif)~\cite{Shvartzshnaider2019going} on a snippet of the privacy policy of VR game \name{Beat Saber}. Middle: The snippet; left and right: The respective thematic coding. The color of the codes correspond to the respective \cif themes in our methodology description in ~\autoref{subsubsec:policy-analysis-method}.}
  \Description{Example of our coding process using the contextual integrity framework.}
  \label{fig:CI-coding-example}
\end{figure}

Our analysis used a hybrid deductive-inductive approach~\cite{swain2018hybrid,proudfoot2023inductive}, which
allows us to deductively analyze privacy policies using a theoretical framework from the literature~\cite{nissenbaum2004privacy,Shvartzshnaider2019going}, while inductively developing codes directly from our observations and identifying themes specific to our data. Specifically, to capture a comprehensive picture of different elements such as personal information collected, sharing practices, purposes of use, and potential risks described in privacy policies~\cite{OPC2018consent}, we adopted Nissenbaum's privacy as \textit{contextual integrity (CI) framework}~\cite{nissenbaum2004privacy} as the starting point in our analysis. We were further inspired by~\citet{Shvartzshnaider2019going}'s work on applying CI to privacy policies~\cite{FTC_PrivacyCon2019}. This deductive approach enabled us to systematically annotate, assess, and compare the \textit{information flows} (i.e., a self-contained description of an information transfer) disclosed in the privacy policies~\cite{Shvartzshnaider2019going}. In particular, we focused on codes relevant to the following five themes (colors correspond to the examples in~\autoref{fig:CI-coding-example}):

\begin{itemize}
    \item \textcolor{darkgreen}{\textbf{Sender.}} Entities that transfer or share information, such as a person, company, website, or device. Examples include the ``user'' themselves and the ``Meta'' headset.
    \item \textcolor{darkorange}{\textbf{Recipient.}} Entities that ultimately receive information, such as a person, company, website, or device. Examples include ``app/game publisher'' and ``third-parties''.
    \item \textcolor{darkblue}{\textbf{Transmission principle.}} Conditions under which information transfers occur, such as in the ``registration \& subscription'' process.
    \item \textcolor{darkpurple}{\textbf{Attribute.}} Type of information in the transfer, such as ``game/app activity data'' or ``body, fitness, and health data''. 
    \item \textcolor{palebrown}{\textbf{Subject.}} The individual whose information is transferred, either explicitly or implicitly, through pronouns or possessives. Examples include users (i.e., ``you'') and their friends (i.e., ``your friends'').
\end{itemize}

The lead researcher and a research assistant independently analyzed the 12 privacy policies uploaded to Dovetail\footnote{See footnote 10.}. The researchers began by familiarizing themselves with the data and taking rough notes of trends they observed. Then, three policies (25\% of data) were randomly selected from the data and were read in detail and coded using inductive line-by-line coding based on the five themes from the \cif~\cite{Shvartzshnaider2019going} and any potential privacy issues and protection measures that appeared in the privacy policies. The researchers then met to address disagreements, refine codes, and collaboratively revisited the three policies to ensure insights were fully captured. In the second iteration, the two researchers independently analyzed five more policies (42\% of data), held a discussion to compare coding results, refine and add to the codes, and collaboratively revisit the same five policies to ensure no neglected insights. In the final iteration, the remaining four privacy policies (33\% of data) were independently analyzed, a discussion meeting was held, the codes were further refined and added to, and four policies were collaboratively revisited. Upon completing the coding process, the two researchers collaboratively discussed the larger patterns across the codes and grouped them into themes. \autoref{fig:CI-coding-example} presents an illustrative example of our line-by-line coding process. ~\autoref{fig:CI-sankey} presents the user information flow revealed in privacy policies based on the \cif themes~\cite{Shvartzshnaider2019going}. Beyond the data practices, our researchers also noticed additional privacy issues and protective measures mentioned in the privacy policies, which we present in~\autoref{tab:policy-codebook}.

%% file: Table-AppSummary.tex
\begin{table}[!ht]
\renewcommand{\arraystretch}{1.2}  
\centering
\caption{Summary of Virtual Reality (VR) Games and Apps in Our Research.}
\label{tab:app-summary}
\resizebox{\textwidth}{!}{%
\begin{tabular}{@{}llp{0.155\textwidth}lp{0.43\textwidth}cccc@{}}
\toprule
\textbf{Title (acronym$^a$)} & \textbf{Genre$^b$} & \textbf{Publisher} & \textbf{Year} & \textbf{Description} & \textbf{Privacy Policy$^c$} & \textbf{Platforms$^d$} & \textbf{Hours} & \textbf{Diaries} \\ \midrule
Beat Saber & Rhythm & Beat Games & 2019 & Players slash incoming beat cubes to the & In-game, & PCVR, CVR,& 18 & 9\\ 
(\BeatSaber)& & & & rhythm of the music. & On-website & SAVR& & \\ \cmidrule(l){1-9} 
Moss: Book II & Action & Polyarc & 2022 & Players guide the mouse Quill through fan- & On-website & PCVR, CVR, & 8 & 9 \\  
(\Moss) & & & & tasy environments, solve puzzles and battle enemies on her journey. & & SAVR& & \\ \cmidrule(l){1-9}
The Room VR & Puzzle & Fireproof & 2020 & Players investigate mysterious environme- & On-website & PCVR, CVR, & 8 & 3 \\ 
(\Room) & & Games & & nts, solve intricate puzzles, and uncover a secret within an ancient Egyptian artifact. & & SAVR& & \\ \cmidrule(l){1-9} 
A Fisherman's Tale & Interactive &Vertigo Games  & 2019 & Players guide a tiny fisherman through a  & On-website &  PCVR, CVR, & 3 & 2 \\ 
(\AFT) & Story & & & series of dollhouse-like worlds to uncover a strange story. & &SAVR & & \\ \cmidrule(l){1-9}  
Down the Rabbit Hole & Adventure & Beyond Frames & 2020 & Players guide a lost girl through a whimsi-  & On-website &  PCVR, CVR, & 3 & 2 \\  
(\Rabbit) & & Entertainment  & & cal prequel world to Alice in Wonderland.& & SAVR& & \\ \cmidrule(l){1-9}   
LEGO\textregistered Bricktales $^e$ & Adventure & Thunderful & 2023 & Players build solutions to help LEGO char- & On-website & PCVR, CVR,  & 31 & 8 \\
(\LEGO) & & Publishing AB & & acters navigate various themed environments and overcome challenges.  & & SAVR& & \\ \cmidrule(l){1-9} 
The Climb 2 & Sports & Crytek & 2021 & Users climb mountains, skyscrapers, and  & In-app, & PCVR, SAVR& 9 & 4 \\  
(\Climb)& & & & caves, and find their way to the top and enjoying the views. & On-website & & & \\ \cmidrule(l){1-9} 
VRChat & Social & VRChat Inc. & 2019 &  Users create avatars, explore diverse vir- & In-app, & PCVR, SAVR& 4 & 3 \\
(\VRChat) & & & & tual worlds, and interact with others. & On-website & & & \\ \cmidrule(l){1-9} 
Meta Horizon Worlds$^f$ & Social & Facebook & 2020 & Users explore, build, and interact in imme- & In-app,  & SAVR& 4 & 3 \\ 
(\Horizon) & & & & rsive, user-created worlds. & On-website & & & \\ \cmidrule(l){1-9}  
TRIPP & Meditation & TRIPP, Inc. & 2019 & Users go on guided mindfulness journeys & In-app, & PCVR, CVR, & 5 & 7 \\  
(\TRIPP)& & & & with calming visuals and sounds to help them relax and focus. & On-website &SAVR & & \\ \cmidrule(l){1-9}   
Supernatural & Fitness & Within & 2020 & Users go on guided fitness sessions in bea- & In-app, & SAVR & 8 & 4 \\  
(\Supernatural) & & Unlimited, Inc. & & utiful natural virtual locations with motivating music.  &On-website  & & & \\ \cmidrule(l){1-9}   
Immersed & Productivity & Immersed Inc. & 2020 & Users join virtual workspaces, set up vir- & On-website & SAVR & 40 & 5 \\  
(\Immersed) & & & & tual monitors, and stay focused in a distraction-free environment.& & & & \\ \bottomrule
\multicolumn{9}{l}{\begin{tabular}[c]{@{}l@{}}\textit{Note.} Information in this table was retrieved from the Meta Quest Store website.~\url{https://www.meta.com/experiences/} \\
$a.$ VR game acronyms are colored \textcolor{paleblue!160}{\textbf{Blue}} and VR app acronyms are colored \textcolor{palepink!160}{\textbf{Pink}}. \\
$b.$ For apps listed under multiple genres, the first listed genre was chosen.\\ 
$c.$ A link to each game's and application's developer privacy policy is available on their Meta Quest Store page.   \\
$d.$ ``PCVR'', ``CVR'', and ``SAVR'' refer to computer-based VR (e.g., Valve Index, Meta Oculus Rift, HTC Vive),  console-based VR (e.g., PlayStation VR), and stand-\\ alone VR(e.g., Meta Quest VR, PICO VR) systems, respectively.\\
$e.$ We included a second adventure game---LEGO\textregistered Bricktales---due to its appeal to children and child data privacy concerns raised by the Office of the Privacy \\ Commissioner (OPC)~\cite{opc2024sweep} in its analysis of the game's website. \\
$f.$ We included a second VR social platform---VRChat---given its popularity among children around 15 years old~\cite{ISDVRChat}. \\
\end{tabular}}
\end{tabular}%
}
\end{table}

%% file: 04-Findings.tex
We first present the findings from our autoethnography evaluation of the design mechanics, then ground these findings in our thematic analysis of their privacy policies and the operation platform---Meta Oculus Quest 3 headset. We go beyond \VRapps' privacy-impacting design mechanics and provide insights into the types of user information collected and the purposes for which it is used. We discuss our findings in~\autoref{sec:discussion} within recent privacy deceptive design research (e.g.,~\cite{gunawan2022redress,bourdoucen2023privacy,gunawan2021comparative,krauss2024what}) and regulatory works (e.g.,~\cite{GPEN2024,oecd2022dark,opc2024sweep}), and draw implications for game and application designers, researchers, and policymakers in the VR field.

\subsection{Deceptive Design Patterns in VR Privacy Communication Design Mechanisms}
\label{subsec:privacy-mechanics}

To address RQ1, our analysis of 59 diary entries identified five categories (themes) of 14 VR-adapted and VR-amplified deceptive design patterns that influence user privacy decisions. Although these patterns are previously observed from the mobile and web deceptive design literature~\cite{hadan2024computer,gray2024ontology,oecd2022dark}, we focus our observations on the specific issues in the privacy communication and interaction design mechanisms in VR and how the effects of deceptive design are amplified by the technology.

This section describes the 14 deceptive design patterns in the sequence we experienced them within VR design mechanisms to illustrate how these patterns influence users' privacy decisions. For clarity, we denote VR games' and apps' titles in italics text (e.g., \name{Beat Saber}), use \pattern{monospace} font for deceptive design patterns (e.g.,~\pattern{Privacy Mazes}), and use quoted italics text for the descriptions directly extracted from the VR design mechanisms. To ensure a clear context for the findings presented in~\autoref{tab:diary-theme}, we include each VR game and app's special icon (e.g., \raisebox{0.4em}{\scriptsize\BeatSaber} for \name{Beat Saber}) upon its first mention. Screenshots and short video clips of our researcher's interactions are available at OSF public repository:~\url{https://osf.io/axzve/}.

\input{Table-DiaryThemes}
\input{Table-DiaryThemes-part2}

\subsubsection{Registration and Subscription Design Mechanisms}
\label{subsubsec:subscription-deceptive}

To access the core features, the VR apps \name{TRIPP} (\raisebox{0.4em}{\scriptsize\TRIPP}) and \name{Supernatural} (\raisebox{0.4em}{\scriptsize\Supernatural}) require users to create independent accounts separate from their VR platform accounts, exemplifying the characteristics of \pattern{Forced Registration}. Furthermore, forcing users to disclose personal information through the registration process (i.e., \pattern{Privacy Zuckering}) is often not essential for the apps' functionality.  For example, \name{TRIPP} asks users to provide their email address to start demo sessions, and unlocking all functionalities, including detailed privacy settings, requires users to register an account and pay a subscription fee with personal information like name and credit card details. In contrast, other VR games and apps operate without forcing registration, such as \name{Supernatural} which allows users to use their VR platform account during the first entry but requires account registration with their name and email address upon re-entry. While such design mechanisms may support cross-device experiences, they are unnecessary for users who only intend to use the applications in VR platforms. 

Both \name{Supernatural} and \name{TRIPP} require users to pay a subscription fee. Our evaluation revealed that \name{Supernatural} employs \pattern{Endorsements and Testimonials} to entice potential subscribers to provide credit card information by promising positive experiences during the subscription process. However, the authenticity of these endorsements is questionable and likely comes from paid influencers rather than genuine users, as the highly polished, professional-looking photos far exceed the quality of typical user uploads (see~\autoref{app-sec:screenshots-deceptive}). At the same time, users are immersed in a realistic, vibrant, and lifelike environment that mirrors the beauty of natural surroundings. This combination creates a persuasive atmosphere, where users might feel compelled to provide payment information, believing the endorsements reflect authentic experiences.

Both VR apps' registration and subscription mechanisms also use \pattern{Visual Prominence} and \pattern{False Hierarchy}. \name{Supernatural}'s \textit{``sign up''} and \textit{``pairing additional user account''} options, which request users' personal information and details about others, are brightly colored and prominently positioned at the top. Similarly, \name{TRIPP}'s \textit{``start demo''} option is visually emphasized when asking for users' email. These mechanisms use a bright green color for acceptance buttons for options that require financial commitment and careful consideration exemplifying \pattern{Positive or Negative Framing} by evoking emotional safety and subtly encouraging users to accept privacy-invasive options without fully understanding the implications of their choices.

\subsubsection{Consent Interaction Design Mechanisms}
\label{subsubsec:consent-interactions}

Upon launching, VR game \name{The Room VR} (\raisebox{0.4em}{\scriptsize\Room}) and the VR application \name{VRChat} (\raisebox{0.4em}{\scriptsize\VRChat}) prompt users to grant access to their \textit{``photo, media, and files''} on the VR device, and the VR application \name{Immersed} (\raisebox{0.4em}{\scriptsize\Immersed}) prompts users to grant access to their microphone. These permission request interfaces exhibit \pattern{Bad Defaults} by having \textit{``Save my preference''} option checked by default and they use \pattern{Visual Prominence} by making the \textit{``Allow''} option more visually bright and noticeable than other options. This combination of visual emphasis and default selection may cause users to grant access without fully considering or remembering their decision, potentially forgetting to revoke it later. Additionally, requesting such permissions in \name{The Room VR} which lacks related functions and in \name{VRChat} before users use relevant functions, can also be considered as \pattern{Privacy Zuckering}, as it may expose user data to unwanted usage.

Users are often prompted to review privacy policies before accessing core functions~\cite{bourdoucen2023privacy} in web and mobile games, but this is less common in VR. VR games like \name{Moss: Book II} (\raisebox{0.4em}{\scriptsize\Moss}), \name{The Room VR}, \name{A Fisherman's Tale} (\raisebox{0.4em}{\scriptsize\AFT}), \name{Down the Rabbit Hole} (\raisebox{0.4em}{\scriptsize\Rabbit}), and \name{LEGO\textregistered Bricktales} (\raisebox{0.4em}{\scriptsize\LEGO}), and VR application \name{Immersed} lack a formal in-app privacy policy explaining what data is collected, by whom, and for what purpose. This lack of crucial privacy information demonstrates \pattern{Hidden Information} and leads to \pattern{Submissive Acceptance} of data practices without awareness. For instance, in \name{The Room VR}, users grant access to photos, media, and files without a clear reason. In \name{VRChat}, the privacy policy is presented only after permission requests. While the Meta Quest Store offers a link to the publisher's privacy policy at the bottom of the page, it is overshadowed by large video previews on top (see~\autoref{fig:screenshot16} in Appendix), which can cause the link to go unnoticed.

While the VR game \name{Beat Saber} (\raisebox{0.4em}{\scriptsize\BeatSaber}) and apps \name{The Climb 2} (\raisebox{0.4em}{\scriptsize\Climb}), \name{VRChat}, \name{Meta Horizon Words} (\raisebox{0.4em}{\scriptsize\Horizon}), \name{TRIPP}, and \name{Supernatural} provide in-app privacy policies to inform users about data practices upon first entry, these documents are often lengthy and difficult to read. This complexity discourages users from fully reading the terms and hinders their comprehension of potential privacy risks. Since content written above an 8th-grade reading level can be difficult for many people to understand~\cite{opc2024sweep,UnitedForLiteracy2022}, they exhibit characteristics of \pattern{Complex \& Lengthy Language}. Our findings from privacy policy analysis further support this observation (see ~\autoref{subsubsec:policy-ambiguity} and~\autoref{app-sec:readability}). 

We further identified \pattern{Positive or Negative Framing} in the consent mechanisms in the VR application \name{TRIPP}, where the \textit{``accept''} button is bright green. Despite being grayed-out until the \textit{``I agree''} box is checked, the vibrant green color suggests emotional safety and may prompt users to accept the policy without fully reading the lengthy statements (see~\autoref{app-sec:screenshots-deceptive}).

\subsubsection{Privacy Settings and Resource Design Mechanisms}

During users' interaction with \VRapps, having accessible privacy settings and resources is essential to their privacy. Our evaluation revealed the presence of \pattern{Privacy Mazes} in the VR game \name{A Fisherman's Tale} and the VR application \name{The Climb 2}. In \name{The Climb 2}, the privacy policy is buried three layers deep within the settings menu, requiring users to first navigate to \textit{``Options''} among six menu items in the first layer, then find the \textit{``Privacy Policy''} option in the second layer to open up the detailed privacy information on the third layer. Similarly, in \name{A Fisherman's Tale}, the game defaults to a \textit{``standing''} gameplay posture without clear information about height calibration, which is only available under the settings menu. This convoluted design makes privacy information hard to find, and potentially discourages users from understanding data practices and privacy risks. 

We also identified the use of \pattern{Bad Defaults} in the settings menu of VR apps \name{Meta Horizon Words}, including automatically turning on the microphone (\textit{``people nearby can hear you talk''}), auto-joining the \textit{``everyone''} voice channel, and automatic leaderboard participation (\textit{``allow leaderboard to show your name and score in any world''}). While these features encourage social interaction, they may leave users who prefer to keep certain information private feeling exposed, such as automatically having their name displayed on a leaderboard without consenting. Although \name{VRChat} also auto-joins users to the world voice channel by default, we do not consider this privacy-invasive since the microphone is in ``toggle'' mode and remains off until manually activated.

Additionally, in \name{A Fisherman's Tale}, \name{TRIPP}, and \name{Supernatural}, users are subjected to \pattern{Mandatory Acceptance}, as they do not provide available settings for users to customize their privacy preferences without registration. Therefore, users are forced to consent to all publisher-stated data practices, even if these practices are irrelevant to the game's and application's functions.

\subsubsection{Quality-of-Life Feature Design Mechanisms}
\label{subsubsec:qol-deceptive}

Several quality-of-life (QoL) features---which are designs that make interaction smoother, more customizable, and enjoyable in \VRapps---also employ privacy deceptive design patterns. In \name{The Climb 2}, users are initially asked to select their gender and skin color, framed as \textit{``select your race and gender''}. Although this question is meant for avatar customization, the wording demonstrates the characteristic of \pattern{Trick Questions}. As users see the realistic rendering of their ``hands'' in the avatar creation process from a first-person perspective (see~\autoref{app-sec:screenshots-deceptive}), they may feel compelled to create avatars that resemble their real-life appearance. As a result, data from users' avatars can be used to infer their personal characteristics. Moreover, \name{Beat Saber} uses \pattern{Privacy Zuckering} when users must enter their credit card information before they can view and compare the prices of individual music and bundles.

\subsubsection{Notification Design Mechanisms}

The use of excessive permission notifications is common. In \name{VRChat}, we identified the use of \pattern{Nagging}, where users who select \textit{``Not allow''} during the entry permission request are repeatedly interrupted by a \textit{``permission error''} message each time they transfer to a different virtual interaction space (i.e., \textit{``world''}). Users must press \textit{``okay''} each time, which can become irritating and may ultimately grant permission to stop the interruptions.

\subsection{Privacy-Enhancing ``Bright Patterns'' in VR
}
\label{subsec:bright-patterns}

Through the lead researcher's autoethnographic documentation, we also noted several \brightpatterns~\cite{grassl2021dark,truong2022bright}, as privacy-enhancing countermeasures of deceptive patterns. Similar to the previous section, game and application titles are italicized, \brightpatterns are set in \pattern{monospace} font, and verbatim descriptions from VR design mechanisms are set in quoted italics.

\input{Table-BrightPatterns}

\subsubsection{Consent Interaction Design Mechanisms}
\label{subsubsec:consent-bright}

Our diary entries revealed that the VR game \name{Beat Saber} and VR apps \name{The Climb 2}, \name{VRChat}, \name{Meta Horizon Worlds}, \name{TRIPP}, and \name{Supernatural} all present an \pattern{accessible privacy policy} upon users' first entry and keep it accessible in the menu for later revisits. Of these, \name{Beat Saber}, \name{The Climb 2}, \name{VRChat}, \name{Meta Horizon Worlds} and \name{Supernatural} include the full privacy policy content within the menu alongside a hyperlink to an external page. In contrast, \name{TRIPP} only provides a hyperlink to an external privacy policy webpage. Although the language of these policies remains lengthy and complex (as noted in ~\autoref{subsubsec:consent-interactions}), making them available within core applications is an important first step toward improving the accessibility of privacy communication. Additionally, \name{Meta Horizon Worlds} avoided the use of visual prominence in its permission request interfaces. Options in its user agreement form and  the settings menu use balanced positions and consistent colors, exemplifying an \pattern{Unbiased presentation of choices}. 

\subsubsection{Privacy Settings and Resource Design Mechanisms}

Both \name{VRChat} and \name{TRIPP} prioritize \pattern{privacy-friendly defaults} and allow users to opt-in based on their preferences. For instance, the acceptance box to user agreements is unchecked by default in both apps and the microphone is set to \textit{``toggle''} mode to prevent unintended listening by others in \name{VRChat} (see~\autoref{app-sec:screenshots-bright}). 

We found \pattern{clear control mechanisms} from \name{Meta Horizon Worlds} and \name{VRChat}, where illustrative images were used to guide users to navigate to the settings menu and manage specific controls (see~\autoref{app-sec:screenshots-bright}). The diary entries highlight a consistent desire for {balanced privacy and functionality}. For example, in \name{Meta Horizon Worlds}, privacy-conscious users are entirely excluded from leaderboard competition rather than being offered more inclusive options, such as hiding their nicknames on the leaderboard, allowing participation without compromising privacy.

\subsubsection{Quality-of-Life Feature Design Mechanisms}
\label{subsubsec:qol-bright}

\name{Meta Horizon Worlds} and \name{VRChat} offer private onboarding spaces and \name{Immersed} allows users to learn privacy controls in a single-user workspace before joining public areas (\pattern{private privacy configuration space}). Furthermore, \name{Immersed} clearly labels virtual workspaces as \textit{``private''} or \textit{``public''}, and \name{VRChat} implemented a microphone icon on the screen to keep users constantly aware of their microphone status (see~\autoref{app-sec:screenshots-bright}).

\name{VRChat} and \name{Supernatural} also maintained \pattern{balanced privacy and functionality} in their QoL features. While \name{VRChat} asks users to provide information such as lifestyle and social preferences to help find matching friends, it is optional. In \name{Supernatural}, users are asked to choose their fitness goal and preferred music genre for fitness sessions, with the options to skip these steps.

\subsubsection{Notification Design Mechanisms}
\label{subsubsec:notification-bright}

Unlike \name{VRChat}'s nagging tactics, \name{The Room VR} does not repetitively prompt users after denying permission for photos, media, and files (\pattern{Unbiased presentation of choices}). We further found \pattern{transparent communication of data practices} from \name{The Room VR}, \name{VRChat}, and \name{Meta Horizon Worlds}. These apps provide brief descriptions of data practices upon entry and during data collection that clearly state what is collected and its purpose. For example, \name{The Room VR} notifies users when their height is calibrated for gameplay.

Overall, the VR \brightpatterns demonstrate that fair and balanced design mechanisms can be implemented in privacy communication and interaction design mechanisms. Unlike deceptive designs, \brightpatterns promote transparent privacy communication, preserve user autonomy in privacy decisions, and allow users to manage their privacy while preserving user experience.

\subsection{VR Information Flows from Privacy Policy Statements}
\label{subsec:policy-analysis}

To identify data practices and potential privacy risks users may unknowingly face, we addressed RQ2 by conducting a thematic analysis of the privacy policies of the 12 \VRapps, and their operating VR platform, Meta Oculus Quest, using the Contextual Integrity framework (CI)~\cite{Shvartzshnaider2019going}.

\begin{figure}[t!]  
  \includegraphics[width=\textwidth]{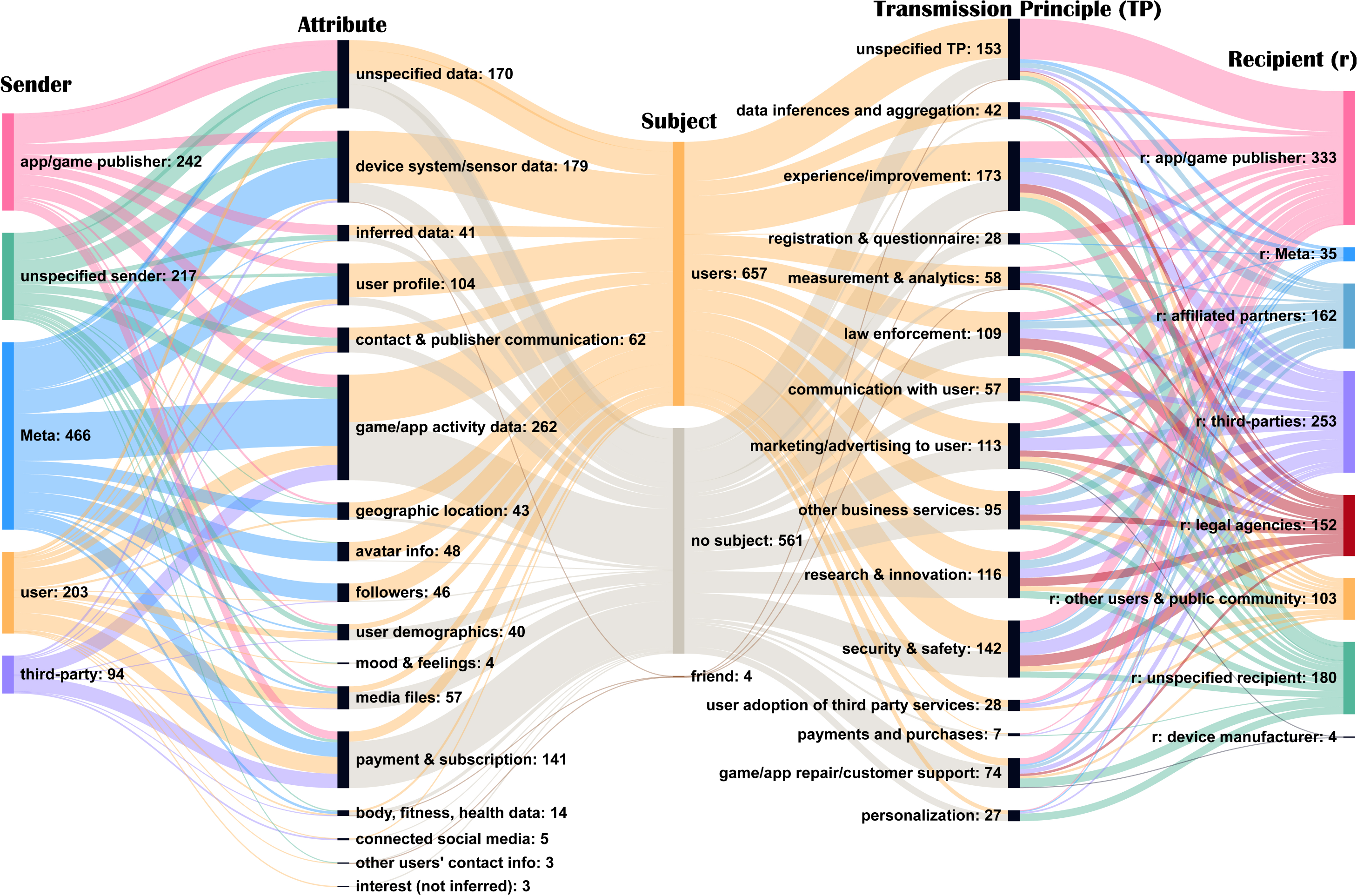}
  \vspace{-7mm}
  \caption{This Sankey Diagram visualizes the 1222 information flows from 209 privacy policy statements through our thematic analysis of 12 privacy policies. The diagram moves from left to right, showing the codes in information flows across the five \cif themes. For example, \code{app/game publisher} appears as the sender in 242 flows. Colors distinguish different information flows. For example, colors flowing from \textit{sender} to \textit{attribute} identify the sender of an \textit{attribute}. A detailed breakdown of the layers is available in~\autoref{app-sec:flow-by-sender}.}
  \Description{Sankey Diagram that demonstrates the user information flow based on our analysis of privacy policies.}
  \label{fig:CI-sankey}
\end{figure}

We identified 1,222 information flows from 209 policy statements across 12 privacy policies (see~\autoref{fig:CI-sankey}). The sender layers in the diagram are superimposed in the paper to provide a high-level overview showing the complexity of information flows across the five \cif themes: (1) sender, (2) attribute, (3) subject, (4) transmission principle, and (5) recipient. ~\autoref{subsubsec:policy-analysis-method}~includes a detailed explanation of each CI framework theme in our analysis, and~\autoref{app-sec:flow-by-sender} provides a layer-by-layer breakdown of the information flows specific to each sender.

We acknowledge that despite our efforts to accurately classify and differentiate information flows, ambiguous language in privacy policies sometimes limited our precision. This occurred when multiple CI factors were grouped together in policy statements. In these cases where a single statement involves multiple instances of the same CI factor, we derived multiple information flows from the statement. For example, if an \textit{attribute} is transmitted to various \textit{recipients}, the transmission to each \textit{recipient} represents a separate information flow. Despite this limitation inherited from the ambiguity in privacy policies, we believe the CI classification still offers valuable insights into the broader patterns of data collection and sharing in VR apps. 

In reporting our analysis, game and application titles are italicized (e.g., \name{Beat Saber}), codes from thematic analysis themes are placed in quotations (\code{game/app publisher}), and direct descriptions from VR design mechanisms are in quoted italics. ``Information flows'' is abbreviated as ``i-flows''.

\subsubsection{Sender, Recipient, and Subject in the Information Flows} As summarized in~\autoref{fig:CI-sankey}, our analysis identified five primary senders in the data transmissions: \code{Meta}, \code{app/game publisher}, \code{user}, \code{third-party}, and \code{unspecified sender}. \code{Meta,} the operational platform for \VRapps and the publisher of \name{Meta Horizon Worlds}, acted as the sender in 466 (38\%) i-flows. This reflects \code{Meta}'s role as the data handler and intermediary between users and other entities in the data flow. In less than a quarter of i-flows ($n=242$, 20\%), the \code{app/game publisher} acted as the sender by outsourcing data processing tasks, such as analytics, marketing, and user support, to third-party service providers. In these cases, they transmitted the collected information about users and their devices to external parties. In 203 (17\%) i-flows, the \code{user} acted as the sender by actively providing or entering data, such as during the registration process or interacting with game or application features. In a few cases ($n=94$, 8\%), \code{third-party services} acted as the sender when \VRapps allowed users to sync data from external devices and services, such as health information from wearable devices. We also identified 217 (18\%) i-flows where no clear sender (\code{unspecified sender}) was identified due to the fragmented presentation of information within policy sections, which leaves the point of origin of the data ambiguous. 

Our analysis identified eight primary recipients of user data. The \code{app/game publisher} was the recipient of 333 (27\%) i-flows, and their \code{affiliated partner companies}, such as parent companies, subsidiaries, and joint ventures, were involved in some ($n=162$, 13\%). Contractual \code{third-party} service providers, such as marketing, service, and analytics vendors, served as the recipient of 253 (21\%) i-flows. \code{Legal agencies}, including government authorities and professional advisors (e.g., lawyers, auditors, insurers), were recipients in 152 (12\%) i-flows. \code{Other users \& public community} were less involved as recipients ($n=103$, 8\%) in games and apps with public interaction spaces. The VR operational platform \code{Meta} was the recipient in 35 (3\%) i-flows, often in collaboration with the publisher for legal requests, protection activities, customer support, and other user-related activities. The VR \code{device manufacturer} was a minor recipient ($n=4$, <1\%), primarily for device improvements and system-related needs. 180 (15\%) i-flows had \code{unspecified recipient}.  

We identified two primary subjects indirectly involved in the data transmission. Around half are \code{users} ($n=657$, 54\%) and only a few are \code{friends} ($n=4$, <1\%). We applied \code{no subject} to 561 (46\%) i-flows when they are directly involved in data transmission as the sender or recipient. 

The wide range of senders and recipients involved in the i-flows shows that user data can be distributed across various entities beyond the user's knowledge. Many of these entities, such as \code{affiliated partners} and \code{third-parties}, are not directly facilitating core functions or promised services. Additionally, many policy statements have vague unspecified senders and recipients. Given the deceptive design practices identified in ~\autoref{subsec:privacy-mechanics}, the lack of transparent communication and diminished user agency in user agreements becomes even more concerning. 

\subsubsection{Attribute and Transmission Principle}
\label{subsubsec:flow-attribute-transmission}

As demonstrated in~\autoref{fig:CI-sankey}, we identified 17 data attributes transmitted across 15 transmission principles. Only a small portion supported core functionality \code{experience/improvements} ($n=173$, 14\%), promised services, or tailored \code{personalization} ($n=27$, 2\%). For example, fitness and meditation apps rely on \code{body, fitness, health data} ($n=14$, 1\%), \code{mood \& feelings} ($n=4$, <1\%), and users' \code{interest} ($n=3$, <1\%). Social VR platforms collect \code{user profile} ($n=104$, 8\%), \code{avatar information} ($n=48$, 4\%), and \code{media files} ($n=57$, 5\%). VR games monitor users' \code{game/app activity data} ($n=262$, 21\%) and \code{device system/sensor data} ($n=179$, 15\%). Apps require a registration and subscription process users' \code{payment \& subscription} details ($n=141, 12\%$) and for \code{payments and purchases} ($n=7$, <1\%). Upon \code{user adoption of third party services}, only 2\% ($n=28$) of the apps transmit their data to external services.

Some data transmissions, while not essential to core functionality, are apparent in user interactions. These include \code{user demographics} ($n=40$, 3\%) and \code{connected social media} ($n=5$, <1\%) collected during \code{registration \& questionnaire} ($n=28$, 2\%) for account creation and social media synchronization. Additionally, when users request \code{game/app repair/customer support} ($n=74$, 6\%), their \code{contact \& publisher communication} ($n=62$, 5\%) are necessarily transmitted. 

Beyond these apparent uses, many data attributes are transmitted under less apparent principles but still contribute to enhancing UX and protection. These transmission principles include \code{measurement \& analytics} ($n=58$, 5\%), \code{research \& innovation} ($n=116$, 9\%), \code{security \& safety} ($n=142$, 12\%), \code{law enforcement} ($n=109$, 9\%), and \code{communication with user} ($n=57$, 5\%).

In contrast, many data attributes are transmitted covertly under principles that do not align with user interests or core VR functionality, primarily benefiting publishers while potentially compromising user privacy. For instance, we found i-flows involving \code{inferred data} ($n=41$, 3\%) from \code{data inferences and aggregation} ($n=42$, 3\%) that can reveal more information about users than they intend to disclose~\cite{miller2020personal,pfeuffer2019behavioural}. Users' \code{geographic location} ($n=43$, 4\%) and \code{followers} ($n=46$, 4\%) were often shared for unclear purposes, such as (\code{other business services} ($n=95$, 8\%) or entirely \code{unspecified TP} ($n=153$, 13\%). While not encountered in our diary evaluation, the privacy policies of \name{TRIPP} and \name{Supernatural} revealed their referral features that require \code{other users' contact info} ($n=3$, <1\%), which may result in unwanted contact. Additionally, we identified 170 ($14\%$) i-flows with vague and \code{unspecified data}. All these data attributes can be used for \code{marketing/advertising to users}, as shown in 113 ($9\%$) i-flows.

The diverse data attributes and transmission principles reveal the complexity of information flows. The deceptive design patterns in privacy communication and interaction mechanisms (as discussed in ~\autoref{subsec:privacy-mechanics}) further exacerbate the difficulty in user understanding and privacy decision-making. Even when users share personal data for enhanced gameplay purposes, they may unintentionally expose themselves to undisclosed and subtly disguised data practices.

\subsubsection{Complexity, Ambiguity, Omission, and Deception in the Policy Statements Introduce Further Privacy Barriers}
\label{subsubsec:policy-ambiguity}

\input{Table-PolicyCodebook}
\input{Table-PolicyCodebook-part2}

Beyond complex information flows, our thematic analysis uncovered additional privacy issues within the 12 privacy policies. As demonstrated in~\autoref{tab:policy-codebook}, three VR games and four VR apps presented a \code{plethora of policies}, where multiple distinct policies are linked within a single main policy. This forces users to navigate several documents to fully understand their privacy rights and data practices. We also identified \code{blurred accountability} in four VR games and four apps, where responsibility for data management, obligations, and liability is unclear.

The privacy policies of three VR games and four apps also exhibited \code{no clear retention period}, as they do not specify time frames or conditions for data removal. Additionally, we found  \code{missing/unclear information} in three VR games and four apps, where statements failed to differentiate between VR services and other platforms, or lacked information on VR data entirely, leaving users confused about how their data is handled in the VR context.

We also identified \code{submissive consent} in the privacy policies of two VR games, where user consent is implied through the act of downloading or using the games. Some policies even suggested that consent was unnecessary for \textit{``legitimate reasons.''} Additionally, we found \code{false rights} in the privacy policies of three VR games and two VR apps to create an illusion of user rights without actual provision. This included requiring users to agree to terms without explicit consent mechanisms in the game or application, and offering revocable rights subject to the company's interpretation of \textit{``legitimate interests.''} Finally, we identified \code{user-burdening policies} in the VR game \name{Down the Rabbit Hole} and the VR application \name{Supernatural}. These policies impose restrictions or obligations on users, such as requiring users to forfeit security and privacy expectations for their data and irrevocably granting the publisher all rights to user data exchanged within the game.

\quot{We do not need your consent as for all forms of processing, we have other legitimate reasons because this is necessary for fulfillment of agreement or it is in our legitimate interest.}{Beat Saber Privacy Policy (06/22/2023)}

To assess the challenge of interpreting convoluted privacy policy language, we evaluated the readability of the 12 selected policies using the Flesch Reading Ease Score in Microsoft 365, following the literature's approach~\cite{GPEN2024}. ~\autoref{app-sec:readability} shows the results, with lower scores indicating more difficult content requiring higher education levels to understand. We found that all 12 privacy policies scored between 30 and 50, which indicates they are \textit{``difficult to read''} and require at least a \textit{``college-level''} understanding\footnote{Flesch-Kincaid readability tests. (n.d.). In Wikipedia. Retrieved April 14, 2023, from \url{https://en.wikipedia.org/wiki/Flesch\%E2\%80\%93Kincaid_readability_tests}}. These findings align with our earlier identification of \pattern{Complex \& Lengthy Language} in VR design mechanisms, as discussed in ~\autoref{subsec:policy-analysis}.  

\subsubsection{Clarity, Proactiveness, and Guidance in Privacy Policies Ease User Concern on Privacy}
\label{subsubsec:policy-clarity}

Our analysis also uncovered various approaches by publishers to ensure transparent privacy communication and protect user privacy, although these measures vary across different \VRapps. As shown in~\autoref{tab:policy-codebook}, all 12 privacy policies reported implementing a range of \code{security \& privacy protections}, such as encryption, data minimization, disassociation from user accounts, restrictions on unauthorized activities, physical safeguards, privacy-friendly defaults, and adherence to third-party security practices. These measures reflect the publishers' commitment to safeguarding user data. Additionally, the privacy policies for all 12 \VRapps clearly present \code{information on post-collection rights \& notification}, including users' rights to access, correct, and delete their data, as well as updates on privacy policies. They also provide contact information for users to exercise these rights, which ultimately enhances user control over their personal information after collection. Moreover, the publishers of three VR apps claimed to obtain \code{proactive consent} from users before processing certain data and implemented an explicit consent mechanism, which is a genuine effort compared to \code{false rights} that merely claim consent without effective mechanisms.

Moreover, we found that the privacy policies of six VR games and four VR apps included \code{country-specific information} with tailored privacy details for users in specific regions such as the European Economic Area (EEA), United States, United Kingdom, Switzerland, and Korea. This information clarifies publishers' compliance with local regulations and provides users with relevant legal context. Additionally, the privacy policies of three VR games and four VR apps offered \code{children-specific instructions} that outline data collection practices for minors, including non-collection policies and guidance on requesting access, correction, or deletion of their data. These policies often advise children to seek parental involvement in privacy-related decisions.

Finally, three VR apps offered \code{direct guidance \& communication} by offering explicit guidelines on adjusting settings to prevent unwanted data collection. They are also transparent about how user data is processed, shared, and for what purposes. An exemplary privacy policy is from \name{Meta Horizon Worlds}, where representative icons were used to illustrate different types of data collection (e.g., motion tracking, eye tracking) to improve user understanding of data practices.

\quot{The software fits a generic hand and body model over the estimated [approximate] points on your hands and body,... to ensure hand and body tracking works properly; this data will be deleted within 90 days\dots}{Meta Privacy Policy (06/20/2024).}

%% file: Table-DiaryThemes.tex
\begin{table}[!t]
\renewcommand{\arraystretch}{1.2}  
\centering
\caption{Based on the VR Deceptive Game Design Assessment Guide from literature~\cite{gray2024ontology,hadan2024computer}, this table presents deceptive design we synthesized from our diary evaluation on the 12 VR games' and apps' privacy communication and interaction mechanisms.}
\label{tab:diary-theme}
\resizebox{\textwidth}{!}{%
\begin{tabular}{@{}lp{0.17\textwidth}p{0.17\textwidth}p{0.3\textwidth}p{0.6\textwidth}p{0.15\textwidth}@{}}
\toprule
\multicolumn{2}{l}{\textbf{\textit{Theme}}/Subtheme} & \textbf{Pattern} & \textbf{Description}~\cite{gray2024ontology,hadan2024computer} & \textbf{Synthesized diary entries$^a$} & \textbf{Mechanisms} \\ \midrule
\rowcolor{palegray!50}\multicolumn{6}{l}{\textit{\textbf{Sneaking}}} \\
& Hiding Information & Submissive Acceptance$^b$ & Trick users into accepting an agreement without prior access to the agreement content. & I feel tricked because agreement was assumed and I did not get to view the actual privacy policy. \raisebox{0.1em}{~\Moss ~\Room ~\AFT ~\Rabbit ~\LEGO} & Consent interactions \\
 & & & & \raisebox{0.1em}{  ~\Immersed ~\VRChat} & \\  \cmidrule(l){1-6}
\rowcolor{palegray!50}\multicolumn{6}{l}{\textbf{\textit{Obstruction}}} \\
 & Adding Steps & Privacy Mazes & Bury privacy controls under layers of confusing menus. & I feel irritated having to dig through various settings tabs to find data practices that affect my privacy.~\AFT~\Climb & Settings \& resources \\ \cmidrule(l){1-6}
\rowcolor{palegray!50}\multicolumn{6}{l}{\textbf{\textit{Social Engineering}}}  \\
 & Social Proof & Endorsements and Testimonials & Portray biased or fabricated testimonials as genuine to influence users' purchase decisions. & Seeing testimonials tempts me to enter my credit card for the free trial, even though I don't know if they are from genuine users or paid influencers.~\Supernatural & Registration \& subscription \\ \cmidrule(l){1-6}
\rowcolor{palegray!50}\multicolumn{6}{l}{\textbf{\textit{Interface Interference}}}  \\
 & Manipulating Choice Architecture & False Hierarchy & Prioritize certain options, make it difficult to compare choices. & I feel inclined to click on the ``sign up'' or ``start the free trial'', which grants access to my personal data, because they are higher positioned than other options.~\Supernatural & Registration \& subscription \\ \cmidrule(l){3-6}
 & & Visual Prominence & Highlight distracting elements, causing users to forget or lose focus on their initial intent. & I feel more likely to click the ``sign up'', ``allow'', and ``start demo'' options with bigger size and eye-catching colors, which grants access to my personal data, than other available options.~\Room ~\VRChat ~\TRIPP ~\Supernatural ~\Immersed & Registration \& subscription, Consent interactions \\ \cmidrule(l){3-6}
 & Hidden Information & Hidden Information & Obscure crucial details or disguising them as unimportant. & I feel confused about why certain data are collected due to the missing privacy policy and privacy settings. \raisebox{0.1em}{~\Moss ~\Room ~\AFT ~\Rabbit ~\LEGO ~\Immersed}  & Consent interactions, Settings \& resources \\ \cmidrule(l){5-6}
 & & & & I feel confused about why my media and photo library access is required since the permission was requested prior to viewing the privacy policy or using related function. ~\VRChat & Consent interactions \\  \cmidrule(l){3-6}
 & Bad Defaults & Bad Defaults & Set default options that benefit the company, forcing users to manually change settings to avoid privacy risks. & I feel concerned about accidental data-sharing because the option is checked by default.~\Room ~\VRChat ~\Horizon & Settings \& resources, Consent interactions \\ \cmidrule(l){3-6}
 & Emotional or Sensory Manipulation & Positive or Negative Framing & Hide or downplay critical information through visual cues. & I feel inclined to click on ``accept'', ``agree'', and ``start demo'' because the bright green buttons make me feel safe to do so. ~\TRIPP ~\Supernatural & Registration \& subscription, Consent interactions \\ \cmidrule(l){3-6}
 & Trick Questions & Trick Questions & Use confusing wording or double negatives to manipulate users' choices. & I feel tricked into revealing personal information because the game asked for my gender and skin color instead of asking ``create your avatar.''~\Climb & QoL features* \\ \cmidrule(l){3-6}
 & Language Inaccessibility & Complex \& Lengthy Language$^c$ & Use hard-to-understand words and sentence structures, hindering users' informed decisions. & While the game/application has a privacy policy, it is too lengthy and hard to read. \raisebox{0.1em}{~\BeatSaber ~\Climb} \raisebox{0.1em}{~\VRChat ~\Horizon ~\TRIPP ~\Supernatural} & Consent interactions \\ 
 \bottomrule
 \multicolumn{6}{l}{\begin{tabular}[c]{@{}l@{}}\textit{Note.} Themes and codes in this table follow the sequence in our deductive codebook~\cite{hadan2024computer}. *QoL features = Quality-of-Life features. \\
 $a.$ VR games and apps acronyms match those in~\autoref{tab:app-summary}. VR game acronyms are colored \textcolor{paleblue!160}{\textbf{Blue}} and VR application acronyms are colored \textcolor{palepink!160}{\textbf{Pink}}.\\
 $b.$ New deceptive design patterns synthesized from our diary entries, and are distinct from our initial codebook from literature~\cite{hadan2024computer}.\\
 $c.$ We renamed the theme to highlight the ``lengthy'' issue in privacy policy language, in addition to the complexity noted in the original codebook~\cite{hadan2024computer}. \\
\end{tabular}}
\end{tabular}%
}
\end{table}

%% file: Table-DiaryThemes-part2.tex
\begin{table}[!t]
\renewcommand{\arraystretch}{1.2}  
\centering
\caption*{Table 2 Continued. Based on the VR Deceptive Game Design Assessment Guide from literature~\cite{gray2024ontology,hadan2024computer}, this table presents deceptive design we synthesized from our diary evaluation on the 12 VR games' and apps' privacy communication and interaction mechanisms.}
\label{tab:diary-theme-part2}
\resizebox{\textwidth}{!}{%
\begin{tabular}{@{}lp{0.17\textwidth}p{0.17\textwidth}p{0.3\textwidth}p{0.6\textwidth}p{0.15\textwidth}@{}}
\toprule
\multicolumn{2}{l}{\textbf{\textit{Theme}}/Subtheme} & \textbf{Pattern} & \textbf{Description}~\cite{gray2024ontology,hadan2024computer} & \textbf{Synthesized diary entries$^a$} & \textbf{Mechanisms} \\ \midrule
\rowcolor{palegray!50} \multicolumn{6}{l}{\textbf{\textit{Forced Action}}} \\ 
 & Nagging & Nagging & Disrupt user focus with repeatedly unwanted interruptions to push actions they would rather avoid. & I feel annoyed by constant ``permission error'' messages because I didn't grant access to my photos and media.~\VRChat & Notification \\ \cmidrule(l){3-6}
 & Forced Registration & Forced Registration & Require account creation for tasks that shouldn't need it, potentially extracting unnecessary personal data. & I feel annoyed that nothing is accessible without registration, even after I paid for subscription. I have to sacrifice my privacy without knowing potential risks or if it's worth it.\raisebox{0.1em}{~\TRIPP ~\Supernatural} & Registration \& subscription \\ \cmidrule(l){3-6}
 & Forced Communication or Disclosure & Privacy Zuckering & Trick users into thinking it's essential for the service, leading to overshare personal data. & I feel pressured to give my credit card information to be able to see the item prices or demo session, and even after doing so I was asked for more personal details upon re-entry.\raisebox{0.1em}{~\BeatSaber ~\Supernatural} & Registration \& subscription, QoL features* \\ \cmidrule(l){5-6}
 & & & & I feel confused about granting media and photo library access as I do not see related functions. \raisebox{0.1em}{~\Room ~\VRChat} & Consent interactions \\ \cmidrule(l){3-6}
 & & Mandatory Acceptance$^b$ & Trick users into accepting all terms in the agreement without customization. & I feel coerced into agreeing to all data sharing practices because I cannot opt-out of those that are irrelevant to the functions. \raisebox{0.1em}{~\AFT ~\TRIPP ~\Supernatural} & Settings \& resources \\
 \bottomrule
 \multicolumn{6}{l}{\begin{tabular}[c]{@{}l@{}}\textit{Note.} Themes and codes in this table follow the sequence in our deductive codebook~\cite{hadan2024computer}. *QoL features = Quality-of-Life features. \\
 $a.$ VR games and apps acronyms match those in~\autoref{tab:app-summary}. VR game acronyms are colored \textcolor{paleblue!160}{\textbf{Blue}} and VR application acronyms are colored \textcolor{palepink!160}{\textbf{Pink}}.\\
 $b.$ New deceptive design patterns synthesized from our diary entries, and are distinct from our initial codebook from literature~\cite{hadan2024computer}.\\
 $c.$ We renamed the theme to highlight the ``lengthy'' issue in privacy policy language, in addition to the complexity noted in the original codebook~\cite{hadan2024computer}. \\
\end{tabular}}
\end{tabular}%
}
\end{table}

%% file: Table-BrightPatterns.tex
\begin{table}[!t]
\renewcommand{\arraystretch}{1.2}  
\centering
\caption{This table presents privacy-enhancing bright designs we identified from the 12 VR games and applications, and the desired ethical design strategies synthesized from our diary entries.}
\label{tab:bright-theme}
\resizebox{\textwidth}{!}{%
\begin{tabular}{@{}p{0.35\textwidth}p{0.8\textwidth}p{0.25\textwidth}@{}}
\toprule
\textbf{Bright Pattern$^a$} & \textbf{Synthesized diary entries$^b$} & \textbf{Interface} \\ \midrule
Accessible privacy policy & I feel satisfied that the privacy policy is accessible both at the start and in the menu for future checks. ~\BeatSaber ~\Climb ~\VRChat ~\Horizon ~\TRIPP ~\Supernatural & Consent interactions \\ \cmidrule(l){1-3}
Unbiased presentation of choices & I feel relieved that all options are presented fairly and without bias, as I was not repeatedly prompted for permission after my initial refusal, and there was no misleading visual prominence or false hierarchy in the user agreement form and settings menu. ~\Room ~\Horizon & Consent interactions, Notification \\ \cmidrule(l){1-3}
Privacy-friendly default & I feel relieved knowing that the default settings respect my privacy concerns by leaving privacy-related options unchecked. ~\VRChat ~\TRIPP & Settings \& resources \\ \cmidrule(l){1-3}
Private privacy configuration space &  I feel relieved knowing that I can fine-tune my privacy settings and avatar in a private environment before joining other users. ~\Horizon ~\VRChat ~\Immersed & QoL features* \\ \cmidrule(l){1-3}
Transparent communication of data practices & I feel satisfied that the message clearly communicates the purpose of data collection, despite lacking details on its external purpose of use. ~\Room ~\Horizon & Notification \\  \cmidrule(l){2-3}
& I feel relieved that the microphone icon at my screen clearly communicates when my microphone is on or off. ~\VRChat & QoL features* \\ \cmidrule(l){2-3}
& I feel relieved that the application specifies room privacy, which reduces my chance of accidentally entering a public space.~\Immersed & QoL features* \\ \cmidrule(l){1-3}
Clear control mechanisms & I feel empowered knowing where in the menu I can control what gets shared with other players in the shared worlds.~\Horizon ~\VRChat & Settings \& resources \\ \cmidrule(l){2-3}
& I wish I could control my data sharing preferences beyond just other players to include the application and game publisher and external entities. & Settings \& resources \\ \cmidrule(l){1-3}
Balanced privacy and functionality & I feel relieved that providing information like my pronouns, lifestyle, and VR preferences is not mandatory for my use of the functionality.~\VRChat ~\Supernatural & QoL features* \\ \cmidrule(l){2-3}
& I wish I could still participate in the leaderboard ranking without my name being publicly displayed. &  Settings \& resources \\  
 \bottomrule
 \multicolumn{3}{l}{\begin{tabular}[c]{@{}l@{}}\textit{Note.} *QoL features = Quality-of-Life features.\\
 $a.$ Patterns are organized by their frequency of appearance, with the most common among all VR games and apps at the top.\\
 $b.$ We included acronyms for VR games and apps, matching those in~\autoref{tab:app-summary}, from which the diary entries originated. VR game acronyms \\ are colored \textcolor{paleblue!160}{\textbf{Blue}} and VR app acronyms are colored \textcolor{palepink!160}{\textbf{Pink}}. Entries not linked to acronyms are desired ethical design strategies from our \\ diary entries and do not exist in the selected VR games and apps. \\
\end{tabular}}
\end{tabular}%
}
\end{table}

%% file: Table-PolicyCodebook.tex
\begin{table}[!t]
\renewcommand{\arraystretch}{1.2}  
\centering
\caption{Final themes from our thematic analysis of 12 VR privacy policies and the Meta Quest platform.}
\label{tab:policy-codebook}
\resizebox{\textwidth}{!}{%
\begin{tabular}{@{}lp{0.25\textwidth}p{0.6\textwidth}p{0.5\textwidth}@{}}
\toprule
\multicolumn{2}{l}{\textbf{\textit{Theme}}/Code} & \textbf{Description$^a$} & \textbf{Example policy statements}\\ \midrule
\rowcolor{palegray!40}\multicolumn{4}{l}{\textbf{\textit{Privacy Inhibitors}}} \\
& Plethora of policies & Multiple distinct privacy policies linked within a single main policy, requiring users to click on each link to fully understand their privacy rights and data practices. ~\Moss ~\AFT ~\Rabbit ~\VRChat ~\Horizon & ``Learn more about the Hand and Body Tracking feature... [URL].'' --- Meta Quest Platform \& Meta Horizon Worlds Privacy Policy (06/20/2024) \\ 
& & \raisebox{0.1em}{~\TRIPP ~\Supernatural} & \\ \cmidrule(l){3-4}
& Blurred accountability &  Unclear responsibility for user data management, obligations, and liability when data is transmitted with third-parties and entities in jurisdictions internationally. ~\Moss ~\AFT ~\Rabbit ~\LEGO ~\VRChat ~\Horizon & ``You will need to apply these opt-out settings... We cannot offer any assurance as to whether the companies we work with participate in the opt-out programs.'' --- TRIPP Privacy Policy (11/30/2023) \\ 
& & \raisebox{0.1em}{~\TRIPP ~\Supernatural}  & \\ \cmidrule(l){3-4} 
& No clear retention period & Ambiguous duration of user data retention without specifying timeframes or conditions for automatic data deletion. ~\Moss ~\AFT ~\Rabbit ~\VRChat ~\Horizon   & ``User data is stored indefinitely on Immersed servers.'' --- Immersed Privacy Policy (08/01/2024) \\ 
& & \raisebox{0.1em}{~\TRIPP ~\Supernatural} & \\ \cmidrule(l){3-4}
& Missing/unclear information & Vague statements that fail to distinguish between VR games and apps and other services, or lack details about VR games and apps entirely. ~\Moss ~\Room ~\Rabbit ~\VRChat ~\TRIPP ~\Supernatural & ``This privacy and data policy applies and has effect in respects of all games, related online services...'' --- The Room VR Privacy Policy (02/14/2022) \\ \cmidrule(l){3-4}
& Submissive consent & Assumed user consent to data practices based solely on their download or use of VR games and apps, or claims that user consent is unnecessary due to ``legitimate reasons.'' ~\Moss ~\Room & ``We do not need your consent as for all forms of processing, we have other legitimate reasons...'' --- Beat Saber Privacy Policy (06/22/2023)\\ \cmidrule(l){3-4}
& False rights & Statements create illusion of user rights without actual provision, such as claiming sharing user data based on their consent without having explicit consent design mechanisms, and offering limited, revocable rights that are contingent on the company's interpretation of ``legitimate interests.'' ~\AFT ~\Rabbit ~\LEGO ~\Immersed & ``... the processing is carried out with your consent...'' --- LEGO\textregistered Bricktales (11/30/2023) but we did not observe an explicit consent form in the game (see~\autoref{tab:diary-theme}) \\ \cmidrule(l){3-4}
& User-burdening policy & Statements that impose excessive restrictions or obligations on users, such as requiring users to forfeit security and privacy expectations for data, and irrevocably grant the publisher all rights to user data exchanged within the game. ~\Rabbit ~\Supernatural & ``We cannot and do not guarantee the security of your personal information and you transmit such information at your own risk.'' --- Supernatural Privacy Policy (07/01/2024) \\ 
\bottomrule
 \multicolumn{4}{l}{\begin{tabular}[c]{@{}l@{}}\textit{Note.} $a.$ We included acronyms for VR games and apps, matching those in~\autoref{tab:app-summary}, indicating the source of each privacy policy \\ statement. VR game acronyms are colored \textcolor{paleblue!160}{\textbf{Blue}} and VR application acronyms are colored \textcolor{palepink!160}{\textbf{Pink}}.\\
\end{tabular}}
\end{tabular}%
}
\end{table}

%% file: Table-PolicyCodebook-part2.tex
\begin{table}[!ht]
\renewcommand{\arraystretch}{1.2}  
\centering
\caption*{Table 4 Continued. Final themes from our thematic analysis of 12 VR privacy policies and the Meta Quest platform.}
\label{tab:policy-codebook-part2}
\resizebox{\textwidth}{!}{%
\begin{tabular}{@{}lp{0.25\textwidth}p{0.6\textwidth}p{0.5\textwidth}@{}}
\toprule
\multicolumn{2}{l}{\textbf{\textit{Theme}}/Code} & \textbf{Description$^a$} & \textbf{Example policy statements}\\ \midrule
\rowcolor{palegray!40}\multicolumn{4}{l}{\textbf{\textit{Protective Measures}}} \\
& Security \& privacy protections & Protective measures like encryption, data minimization, disassociation from user accounts, restrictions on unauthorized activities, physical safeguards, privacy-friendly defaults, and ensuring third-party security practices. ~\BeatSaber ~\Moss ~\Room ~\AFT ~\Rabbit ~\LEGO & ``This [user data] is made available to us in aggregated form and we are not able to single out any one individual from this data...'' --- The Room VR Privacy Policy (02/14/2022) \\ 
& & \raisebox{0.1em}{~\Climb ~\VRChat ~\Horizon ~\TRIPP ~\Supernatural ~\Immersed} & \\ \cmidrule(l){3-4}
& Information on post-collection rights \& notification & Clear notification about privacy policy updates and guidelines on users' rights to their data, including access, correction, and deletion, along with the publisher's contact information for related requests. ~\BeatSaber ~\Moss ~\Room ~\AFT ~\Rabbit ~\LEGO  & ``Users can request the deletion of their data at any time by clicking on the `Delete User Data' option or by emailing us at ... '' --- Immersed Privacy Policy (08/01/2024) \\ 
& & \raisebox{0.1em}{~\Climb ~\VRChat ~\Horizon ~\TRIPP ~\Supernatural ~\Immersed} & \\ \cmidrule(l){3-4}
& Proactive consent & Obtain explicit user consent before sharing personal data. Those that claimed to obtain user consent but do not have a consent interface in the VR games and apps are categorized as ``false rights.'' ~\Horizon ~\TRIPP ~\Supernatural & ``For sensitive personal information that we collect, we will use or disclose it with your specific consent...'' --- Supernatural Privacy Policy (07/01/2024)  \\ \cmidrule(l){3-4}
& Country-specific information & Privacy details tailored for users in specific regions, such as the European Economic Area (EEA), United States, United Kingdom, Switzerland, and Korea. ~\BeatSaber ~\Moss ~\Room ~\AFT ~\Rabbit ~\LEGO   & ``As a European resident, you also have the right to...'' --- The Climb 2 Privacy Policy (n.d.)  \\ 
& & \raisebox{0.1em}{~\Climb ~\VRChat~\TRIPP ~\Supernatural} & \\ \cmidrule(l){3-4}
& Children-specific instructions & Guidelines for children on data collection, non-collection, and how to request access, correction, or deletion of their data, with advice on involving parents. ~\Moss ~\Room ~\Rabbit ~\VRChat ~\Horizon ~\TRIPP & ``To provide social features your child's Meta Horizon profile will be set to private by default...'' --- Meta Quest Platform \& Meta Horizon Worlds Privacy Policy (06/20/2024) \\ 
& & \raisebox{0.1em}{~\Supernatural} & \\ \cmidrule(l){3-4}
& Direct guidance \& communication & Explicit guidelines on adjusting settings to prevent unwanted data collection, along with transparent information on how user data is processed, shared with third parties, for what purpose. ~\VRChat ~\Horizon ~\Immersed & ``If you pay for access to certain features... we may use your personal information to provide you access to them... If you request approval to... we may use your personal information to decide whether to grant your request...'' --- VRChat Privacy Policy (11/22/2023) \\ 
\bottomrule
 \multicolumn{4}{l}{\begin{tabular}[c]{@{}l@{}}\textit{Note.} $a.$ We included acronyms for VR games and apps, matching those in~\autoref{tab:app-summary}, indicating the source of each privacy policy \\ statement. VR game acronyms are colored \textcolor{paleblue!160}{\textbf{Blue}} and VR application acronyms are colored \textcolor{palepink!160}{\textbf{Pink}}.\\
\end{tabular}}
\end{tabular}%
}
\end{table}

%% file: 05-Discussion.tex
In this section, we contextualize our findings within existing research (e.g.,~\cite{gunawan2022redress,bourdoucen2023privacy,gunawan2021comparative,krauss2024what}) and regulatory efforts (e.g.,~\cite{GPEN2024,oecd2022dark,opc2024sweep}) on VR privacy in ~\autoref{subsec:discussion-contrasting}. In ~\autoref{subsec:implications}, we provide design recommendations for VR researchers, designers, and policymakers to mitigate deceptive practices, prioritize user privacy, and improve user awareness and control.

\subsection{Deceptive Design in Immersive VR Compared to Web and Mobile Platforms}
\label{subsec:discussion-contrasting}

Some deceptive design patterns from mobile and web are amplified by VR’s unique features, while others were absent due to VR design mechanisms that are distinct from other platforms.

\subsubsection{Migration of Deceptive Design Practices From Mobile and Web Platforms to VR}

Our analysis reveals a disconcerting migration of deceptive design patterns from mobile and web platforms into VR environments~\cite{gunawan2022redress,gunawan2021comparative,GPEN2024}. We observed \pattern{submissive acceptance} in VR design mechanisms and \pattern{false rights} in privacy policies that align with issues of assumed consent and illusory mandatory permissions from prior mobile and web research~\cite{gunawan2022redress,gunawan2021comparative}. This coercion forces users to accept all data practices without opt-out options (\pattern{mandatory acceptance}), which further resembles the bundled consent problem~\cite{gunawan2022redress,gunawan2021comparative,luguri2021shining,bongard2021manupulated}. The convoluted privacy settings in VR apps (\pattern{privacy maze}) reflect similar challenges in web environments~\cite{GPEN2024}, while the issue of missing privacy information and settings (\pattern{hidden information}) parallels issues with lacking configure buttons and privacy policies on websites~\cite{gunawan2022redress,gunawan2021comparative}. Similarly, \pattern{bad defaults} and \pattern{forced registration} in VR privacy settings and consent forms resemble issues found on the web and mobile environments ~\cite{GPEN2024,bosch2016tales,gray2018dark,gunawan2022redress}. 

Many \VRapps lacked privacy policies, and those that had one often used \pattern{complex \& lengthy language} that reflected similar \textit{``information overload''}~\cite{gunawan2022redress} challenges found in website and game privacy policies~\cite{GPEN2024,russell2018privacy}. We also identified \pattern{nagging} notifications in VR, similar to those seen in mobile and web contexts~\cite{gunawan2021comparative}. The presence of games and apps separate from the policy of their VR platforms further complicates users' comprehension, as they have to navigate and understand both without a clear connection between them~\cite{bourdoucen2023privacy}. Additionally, we found \pattern{false hierarchy} and \pattern{visual prominence} in VR, such as bright-colored buttons, echoing the grayed-out options that subtly steer user behavior in web and mobile environments~\cite{gunawan2022redress,gunawan2021comparative}. 

The migration of deceptive design patterns into VR represents a concerning threat, as the lack of sufficient disclosure of their data collection practices of VR applications can worsen with deceptive design practices~\cite{bourdoucen2023privacy,kroger2023surveilling}. Our identified patterns still rely heavily on 2D interface elements in VR, such as menu screens and pop-up notifications, rather than fully capitalizing VR's 3D features (e.g., unbounded displays, whole-body interaction, and motion sensing)~\cite{krauss2024what}. In other words, we found that immersive features, although enhance the effectiveness of deceptive patterns, do not directly enable them in the applications. Thus, these patterns may also appear in other immersive technologies, such as Augmented Reality and Mixed Reality, and in other interactive digital media that use 2D interfaces for privacy communication and interaction. The combination of these designs and the ambiguous privacy policies can significantly disadvantage users. This challenge is exacerbated by immersive technologies' extensive, continuous, and granular data collection, which often is not recognized by users~\cite{hadan2024privacy,buck2022security,miller2020personal}.   

\subsubsection{VR Amplifies the Impact of Deceptive Design on User Privacy}

We found that several deceptive design patterns documented in mobile and web literature (e.g.,~\cite{gray2024ontology,mathur2019dark,Brignull2023book,oecd2022dark}) are amplified by features unique to VR~\cite{krauss2024what,hadan2024deceived}. One example is that the artificial \pattern{endorsement and testimonials} discussed in ~\autoref{subsubsec:subscription-deceptive} used as social proof to encourage payment information disclosure~\cite{gray2024ontology,mathur2019dark} can become more problematic in VR. This is because, unlike mobile or web contexts where users can easily pause and seek external information on a secondary device, the occlusion of reality by wearing a VR headset~\cite{krauss2024what} and the disruption of closing the application to switch to a browser in VR discourage users from doing so. This barrier, combined with the lack of demo or trial options, forces users to rely solely on in-app endorsements to assess the value of services (detailed in ~\autoref{subsubsec:subscription-deceptive}). 
Similarly, we observed that the VR application \name{TRIPP} provided hyperlinks to its privacy policy and terms of service during user registration, as discussed in the \pattern{accessible privacy policy} pattern in ~\autoref{subsubsec:consent-bright}). While this is common in web and mobile environments (e.g., on Steam~\cite{bourdoucen2023privacy}),  users can quickly open a new tab to skim the content. In VR, accessing these external links can be disruptive because users must exit the running application. This disruption makes accessing external resources in VR more difficult. 

The realistic rendering of virtual elements in VR can also enhance the effectiveness of \pattern{trick questions}~\cite{Brignull2023book,mathur2019dark,oecd2022dark}. One example is the avatar creation mechanisms we observed in \name{The Climb 2}, where users are prompted to select their gender and race while viewing realistic virtual ``hands'' from a first-person perspective (see~\autoref{app-sec:screenshots-deceptive}). This avatar creation process is designed to resemble the user but blurs the line between voluntary choices and subtle manipulation~\cite{krauss2024what,hadan2024deceived}. ~\autoref{subsubsec:flow-attribute-transmission} shows that the fusion of virtual representation and physical identity may cause users to disclose personal characteristics without fully realizing how their data is shared with third parties.

We did not observe deceptive patterns that relied on VR's ''device sensing'' features, nor evidence of hyper-personalized manipulations as discussed in the literature~\cite{krauss2024what,hadan2024deceived}. This may be due to most current VR apps needing to focus on functionality over advanced data-driven manipulation. Although our privacy policy analysis revealed that VR data are used for marketing (see ~\autoref{subsubsec:flow-attribute-transmission}), these can be legitimate business strategies when executed correctly~\cite{Brignull2023book}. Therefore, we argue that the current VR data collection primarily aims to ensure safety and facilitate VR interaction rather than deceive users.

\subsubsection{Absence of Some Common 2D Deceptive Design Patterns in VR Due to Modality Differences}

Our analysis in ~\autoref{subsec:privacy-mechanics} reveals differences in privacy communication and interaction mechanisms between VR and traditional web or mobile platforms that have led to the absence of many common deceptive design patterns. For instance, we did not encounter problems with settings that cannot be saved~\cite{gunawan2021comparative}, as privacy settings were often missing entirely in VR (\pattern{hidden information}). Similarly, deceptive patterns used in account deletion and logout, common on mobile and web platforms~\cite{gunawan2021comparative,opc2024sweep,GPEN2024}, were mostly absent because most VR apps do not require separate accounts beyond the one tied to the VR headset. However, the challenge of data deletion remains. ~\autoref{subsubsec:policy-clarity} shows that while privacy policies outline users' post-collection data rights and provide contact information for data deletion requests, requesting data removal needs significant user effort such as filling out lengthy forms or sending personalized email requests. This can be viewed as a form of ``obstruction'' that discourages users from initiating the process~\cite{gray2024ontology,Brignull2023book}.

Cookie banners common in web platforms~\cite{gunawan2022redress,GPEN2024} are noticeably absent in VR contexts. We found consent interactions are mostly submissive and paired with deceptive patterns, and privacy policies are often missing in VR environments. While deceptive consent practices are ethically concerning, the complete absence of explicit consent mechanisms presents an even greater issue.

\subsection{Implications for Privacy-Enhancing Designs in VR Games and Applications}
\label{subsec:implications}

Based on our findings on privacy-enhancing \brightpatterns in VR communication and interaction mechanisms (see~\autoref{tab:bright-theme}), and good practices in publishers' privacy policies (see~\autoref{tab:policy-codebook}), we provide strategies on counteracting deceptive tactics, refining privacy communication and interaction designs, and tailoring privacy policies to users' needs.

\subsubsection{Role of Privacy in VR Self-Expression and Social Interactions}

Avatars in VR are perceived as digital extensions of users' personal identities~\cite{neustaedter2009presenting}, as directly experienced and documented by the lead researcher in ~\autoref{subsubsec:qol-bright} and~\autoref{app-sec:example-diary} during the process of personalizing the avatar to create a strong connection between the virtual self and real-world image that is reinforced by the first-person perspective in VR. The literature suggests that the desire to present an ideal self in VR is driven by the need to align with societal norms~\cite{neustaedter2009presenting}. Our lead researcher's decision to conceal sensitive details, such as default username, indicates that users may also project their privacy concerns onto their avatars. Research has proposed privacy-protecting methods like blurring effects and avatar reconstruction~\cite{wang2021shared,eltanbouly2024avatar}, but these are not ideal for VR, as they unavoidably hinder social interaction and negatively impact user experience. The \pattern{private privacy configuration space} offers a more practical solution, as it allows users to configure their avatars based on privacy concerns in a private setting. This capability of managing their avatar gives users a sense of control over their self-representation and helps them maintain privacy in VR public spaces. However, avatars that closely resemble users can still be used for profiling, as avatar customization choices may reveal personal characteristics, demographic information, and health issues~\cite{eltanbouly2024avatar,symborski2014use,likarish2011demographic,martinovic2014you}. Future research should further explore the phenomenon of projecting personal privacy concerns onto self-representation in virtual worlds as a crucial aspect of examining the privacy implications of VR technology and preserving user privacy.

\subsubsection{VR ``Bright Patterns'' as Countermeasures to Deceptive Design}

We identified several \brightpatterns (See~\autoref{tab:bright-theme}) that offer exemplary privacy-enhancing alternatives to deceptive designs. Literature suggests that designers often rely on established industry conventions and some deceptive patterns were initially created to organize the UI and reduce visual complexity~\cite{zhang2024navigating}. To guide designers toward more ethical practices, we suggest adopting VR-specific \brightpatterns as privacy-friendly standards. ``Bright patterns'' have been explored in 2D websites and mobile environments~\cite{sandhaus2023promoting,truong2022bright,grassl2021dark}. Our observed \brightpatterns in VR reflect and extend those documented in prior literature~\cite{truong2022bright,grassl2021dark,sandhaus2023promoting}. For example, our observed \pattern{privacy-friendly default} resembles the unchecked checkboxes from lab experiments~\cite{truong2022bright} and website observations~\cite{sandhaus2023promoting}, while the \pattern{transparent communication of data practices} reflects the ``dumb it down'' bright pattern~\cite{sandhaus2023promoting}. However, our analysis also identifies novel \brightpatterns, such as \pattern{private privacy configuration space} and \pattern{balanced privacy and functionality} (see~\autoref{tab:bright-theme}), which are not covered in existing research. This gap highlights the need for further research to expand the \brightpatterns beyond websites and develop privacy-friendly bright design standards tailored to immersive environments. 

In addition, current mitigation strategies for deceptive designs, such as browser alerts and warning messages~\cite{schafer2024fighting}, may suffer from issues like software bugs, maintenance failures, and profit motives that undermine its long-term reliability and trust, as seen with digital certificates' failures to alter malicious websites~\cite{hadan2021holistic,johnson2021human}. We encourage future research to expand on \brightpatterns literature~\cite{sandhaus2023promoting,truong2022bright} and our findings by not only identifying deceptive designs but also drawing from ethical design practices to develop a standardized and accessible bright design toolkit for VR designers. 
As designers often prioritize business goals over user convenience due to challenges in justifying privacy-enhancing choices~\cite{zhang2024navigating}, we believe that providing concrete, research-backed examples of \brightpatterns can empower them to make stronger cases for ethical design decisions.

\subsubsection{Towards Effective VR Privacy Communication and Notifications}

Literature suggests that personalized alerts during data collection improve user awareness and comprehension~\cite{frik2022users,bourdoucen2023privacy}. Our analysis in ~\autoref{subsubsec:notification-bright} shows that although just-in-time privacy notifications are not widely adopted in VR, the few examples we observed are useful in informing users about data practices at the moment they occur. This communication approach is particularly useful in VR because users may not fully understand data collection implications from reading a standard privacy policy outside the VR context~\cite{hadan2024privacy}. However, designing notifications in VR can be challenging. Existing approaches, such as head-up displays, haptic feedback, floating displays, and in-situ notifications on virtual objects~\cite{rzayev2019notification}, are often missed by the users due to other rich information and feedback in VR~\cite{ghosh2018notifivr}. On the other hand, notifications that demand attention can disrupt users' VR experience~\cite{george2018intelligent}. Although some studies have explored optimal moments for delivering notifications (e.g.,~\cite{chen2022predicting,zenner2018immersive}), this remains an unsolved issue that requires continuous research. Beyond notifications, \citet{schaub2015design} developed a design space for privacy notices to help system practitioners conceptualize and implement effective privacy notices. ~\citet{feng2021design} later extended this design space by incorporating necessary design dimensions for privacy controls. While some of these design dimensions, such as timing, delivery, and modality, can enhance privacy communication in VR~\cite{schaub2015design,feng2021design}, further research is needed to adapt this space for VR-specific features such as unbounded displays, whole-body interaction, motion sensing, and personalized, context-aware privacy controls enabled by granular sensor data~\cite{krauss2024what,hadan2024privacy,hadan2024deceived}. 

\subsubsection{Independent Privacy Policy Dedicated to VR}
Through our analysis in ~\autoref{subsubsec:policy-ambiguity}, we found that \VRapps from publishers that offer services on multiple platforms, such as computers, mobile, and websites, often provided a general privacy policy that did not clearly distinguish VR-specific privacy concerns. Even when attempts were made to differentiate services using bolded indicator terms, the policy statements often used non-VR examples and failed to clarify privacy implications in VR. Since users generally have limited familiarity with these practices and often rely on their non-VR experiences in making privacy decisions~\cite{hadan2024computer}, we argue that it is crucial to provide clear, accessible information to facilitate users' understanding. The data processing descriptions in \name{Meta Horizon World}'s privacy policy serve as a good example, despite suffering from the \code{plethora of policies} issue. We recommend that \VRapps available across platforms adopt a VR-specific privacy policy to clearly address the unique privacy concerns of immersive technologies, including transparent explanations of how biometric data, such as movement patterns or physiological responses, are collected and used~\cite{pfeuffer2019behavioural,miller2020personal}.

\subsubsection{Accessible Privacy Policies Optimized for VR Environments}
\label{subsubsec:accessible}

Informing users about a system's data practices is crucial for facilitating their informed privacy decisions~\cite{schaub2015design}. However, our selected \VRapps' privacy policies remain complex, text-heavy, and reliant on 2D interfaces (see~\autoref{app-sec:screenshots-deceptive}). Some also include child-specific privacy instructions aimed at guiding both children and parents on protecting children's privacy (see~\autoref{subsubsec:policy-clarity}). However, our analysis shows that these instructions appeared to be buried within the lengthy policies and were difficult to act upon. Although government agencies encourage user-friendly policy designs~\cite{GPEN2024} and mandate special protections for children~\cite{UKICO,CAADCA}, in practice, compliance with legal obligations tends to take priority over user experience due to pressures from compliance teams, senior managers, and security experts, despite designers recognizing the challenges users face~\cite{zhang2024navigating}. However, overloading users with details can be counterproductive~\cite{felt2012android,ramokapane2023skip,momen2020accept}. 

The selected \VRapps often provide a hyperlink to privacy policies on their official websites, but this approach appeared to be insufficient without making the full policy accessible within the VR environment 
(see ~\autoref{subsubsec:consent-interactions}), as users are likely to ignore it~\cite{bourdoucen2023privacy}. Instead, we recommend presenting privacy policies upfront~\cite{chen2022predicting} before users engage with the VR experience, similar to the way they are shown on computers and mobile devices~\cite{bourdoucen2023privacy}, while ensuring the information is accessible within the application for future reference. VR privacy policies may also harness the features of VR technology. For example, previous studies have used AR technology to visualize smart home device data flows (e.g.,~\cite{dasgupta2019user,bermejo2021privacy}). VR privacy policies can also use the technology's features~\cite{hadan2024deceived,krauss2024what} to present data flows through interactive 3D visualizations and sound cues, which have the potential to make privacy information more engaging, intuitive, and usable.

\subsection{Limitations}

As an early exploration of deceptive design in VR, our lead researcher's expertise was important in identifying these patterns, especially since non-expert users often struggle to recognize them~\cite{geronimo2020UI}. However, further research is needed to assess the impact and harm of these patterns across diverse user groups. Our autoethnographic evaluation was guided by an evolving codebook~\cite{hadan2024computer} and supplemented by insights from broader literature (see ~\autoref{subsec:literature-deceptivedesign}). While this codebook incorporated established patterns from foundational deceptive design research~\cite{gray2024ontology,Brignull2010deceptive,gray2018dark,mathur2021makes,bosch2016tales} and privacy regulatory works~\cite{oecd2022dark,cma2022UK,ftc2022bringing}, it does not capture previously unrecognized deceptive design patterns in VR. To address this, the lead researchers' expertise in user-centric privacy research guided the documentation of privacy-related expectations in diary entries and informed the exploration of patterns beyond those identified in the literature. However, blind spots may still persist~\cite{geronimo2020UI,gunawan2021comparative}, and we welcome future research to confirm and expand on our findings.

Our analysis focused on data practices and privacy issues observable in VR design mechanics and privacy policies but left gaps for practices not disclosed in these contexts. While we aimed to identify elements in the information flows, many privacy policy statements were vague or overly broad, which resulted in some barriers in granular classifications (see~\autoref{fig:CI-sankey}). Previous studies have also found discrepancies between actual VR device data transmissions and what privacy policies claim~\cite{egliston2021examining}. We encourage future research to extend our work by conducting network traffic analyses to determine if privacy policies align with actual data transmissions.

Due to ethical concerns about including other users without consent (see~\autoref{subsubsec:ethical_limitations}), our research focused on deceptive patterns that invade user privacy in single-user \VRapps but left multi-user scenarios unexplored. This limitation constrains our understanding of how privacy risks emerge in user-to-user interactions, where social pressure and active data sharing may amplify deceptive practices. Future research could examine how deceptive design exploits privacy in multi-user environments, including (non)collocated interactions and bystander presence, as hypothesized in~\citet{krauss2024what}. 
We also noted non-privacy-related deceptive tactics, such as parasocial pressure and limited-time offers in \name{Beat Saber} that encourage impulsive purchases. Although outside our scope, these observations highlight the need for further research on deceptive design in VR that causes financial, temporal, or psychological harm. 

To deepen our analysis of each VR game and app, we sacrificed the breadth of our study and analyzed fewer total \VRapps than prior work (e.g.,~\cite{geronimo2020UI,gunawan2021comparative}). Although we included at least one game or application from each genre, our limited scope prevents us from assessing the prevalence of deceptive design patterns in VR. We also acknowledge that our analysis was conducted exclusively on a Meta VR headset, which centers our findings around this platform and may not fully capture the deceptive design patterns and information flows present in other VR ecosystems. We encourage future research to explore and compare privacy mechanisms across different VR platforms to obtain a broader perspective and a more comprehensive understanding of the manifestation of deceptive design and information flows.

Lastly, our data was collected from February to early August 2024. We acknowledge that these games and apps can evolve over time. For instance, by October 2024, we observed improvements in permission request notifications in \name{VRChat} and \name{The Room VR}, with added explanations of data usage. We also observed that VR privacy mechanisms were primarily implemented using 2D interfaces, which likely reflects a broader trend of limited integration of VR-specific features in current VR privacy mechanisms (as we discussed in~\autoref{subsubsec:accessible}). We encourage future research to explore how VR privacy mechanisms evolve with greater integration of VR-specific features over time and how deceptive designs emerge.

%% file: 06-Conclusion.tex
In conclusion, our research sheds light on the presence of deceptive design patterns in VR games and applications that hinder users' understanding of VR data practices and impair their ability to make informed privacy decisions. Through a two-phased methodology, we found that while some VR deceptive tactics are migrated from 2D web and mobile environments, and VR's unique features, such as realistic simulation, amplify their impacts on users. We also found that convoluted privacy policies and the absence of explicit consent interactions in VR further diminish users' awareness of the data transmission occurring in VR environments. On the other hand, we also clustered privacy-enhancing design strategies implemented in VR applications that can offer more ethical solutions for VR researchers, designers, and policymakers to create privacy-respecting immersive experiences and establish more transparent privacy policies. As VR continues to evolve, we argue that proactively implementing \brightpatterns that prioritize user privacy is essential for creating a safer and more privacy-friendly VR environment.

%% file: 99-Acknowledge.tex
This project has been funded by the Office of the Privacy Commissioner of Canada (OPC); the views expressed herein are those of the author(s) and do not necessarily reflect those of the OPC, the University of Waterloo, nor the UWaterloo Games Institute. 

L. Zhang-Kennedy also acknowledge support from the Natural Sciences and Engineering Research Council of Canada (NSERC) Discovery Grant (\#RGPIN-2022-03353) and L. Nacke also acknowledge support from the Social Sciences and Humanities Research Council (SSHRC) INSIGHT Grant (\#435-2022-0476), Natural Sciences and Engineering Research Council of Canada (NSERC) Discovery Grant (\#RGPIN-2023-03705), and Canada Foundation for Innovation (CFI) John R. Evans Leaders Fund (\#41844). 

Thank you to the editors for their effort in organizing the peer-review process, and thank you to the reviewers for their insightful feedback that helped us to improve the quality of this manuscript. We also thank post-doctoral researcher Reza Hadi Mogavi and Eugene Kukshinov for their valuable feedback on the manuscript prior to our submission. Screenshots in this manuscript were from the selected games and applications and fall under fair use.

%% file: 98-Appendix.tex
\newpage
\section{Appendix --- Disclosure Statement for Our Use of AI in Writing and Grammar Enhancements}

In accordance with ACM Publications Policy\footnote{Association for Computing Machinery. (July 6, 2023). ACM publications policy guidance for SIGCHI venues. Retrieved October 22, 2024, from ~\url{https://medium.com/sigchi/acm-publications-policy-guidance-for-sigchi-venues-87332173aad1}} and literature suggestion~\cite{hadan2024great}, we acknowledge our use of the Grammarly and Typingmind (GPT-4o model) AI writing assistants. These tools were used to correct grammatical errors and, in a few cases, condense and improve sentence structure using the prompt: ``make the following sentence [our human-written sentence] more concise and fix any grammar errors.'' Our decision to use AI was based on our hope to achieve a concise and succinct writing while preserving the ``human-touch'' in our research writing~\cite{hadan2024great}. We did not use AI for data collection, analysis, or image generation. Analysis was manually conducted by our lead researcher and a graduate research assistant using non-AI-supported version of Dovetail. Figures in this manuscript were captured with the Meta Quest 3 VR headset's screenshot function and formatted using Canva's default templates. Our research team manually reviewed, verified, and copy-edited the full manuscript.


\newpage
\section{Appendix --- Example Diary Entry}
\label{app-sec:example-diary}

\input{Table-DiaryMockup}

\newpage
\clearpage
\section{Appendix --- Example Screenshots Deceptive Design Patterns from VR Games and Playful Applications}
\label{app-sec:screenshots-deceptive}

\begin{figure}[!h]
  \includegraphics[width=\textwidth]{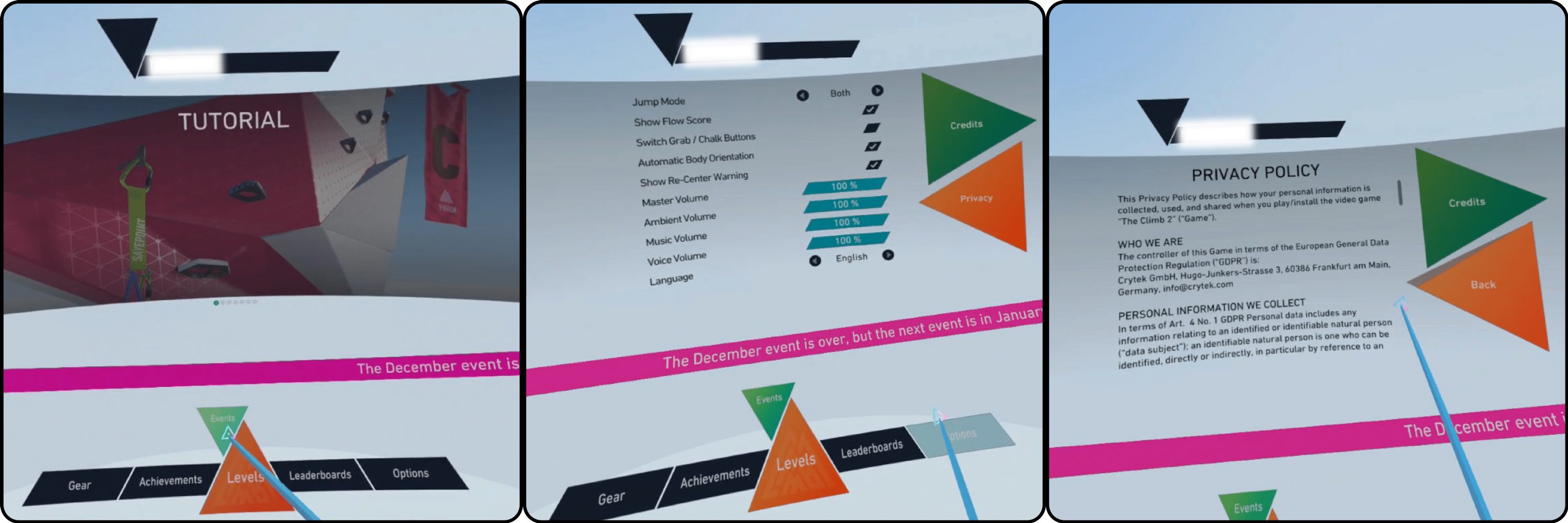}
  \caption{The privacy settings mechanism in \name{The Climb 2} exemplifies the \textbf{\pattern{Privacy Maze}} deceptive design pattern. The main settings screen provides six distinct buttons that each leads to a sub-menu, with privacy-related information hiding within the ``options'' sub-menu without obvious indicator of its availability. To find the privacy-related information, users must navigate to the ``options'' sub-menu (see left screenshot). Inside the ``options'' menu, users encounter two buttons on the right and various application settings in the center (see center screenshot). To access privacy information, users must navigate to the right side and select ``privacy,'' where the privacy policy is located (see right screenshot). \textit{Note. Our researcher's username is concealed in the screenshots for anonymity.}}
  \Description{Example screenshot.}
  \label{fig:screenshot7}
\end{figure}

\begin{figure}[!h]
  \includegraphics[width=0.65\textwidth]{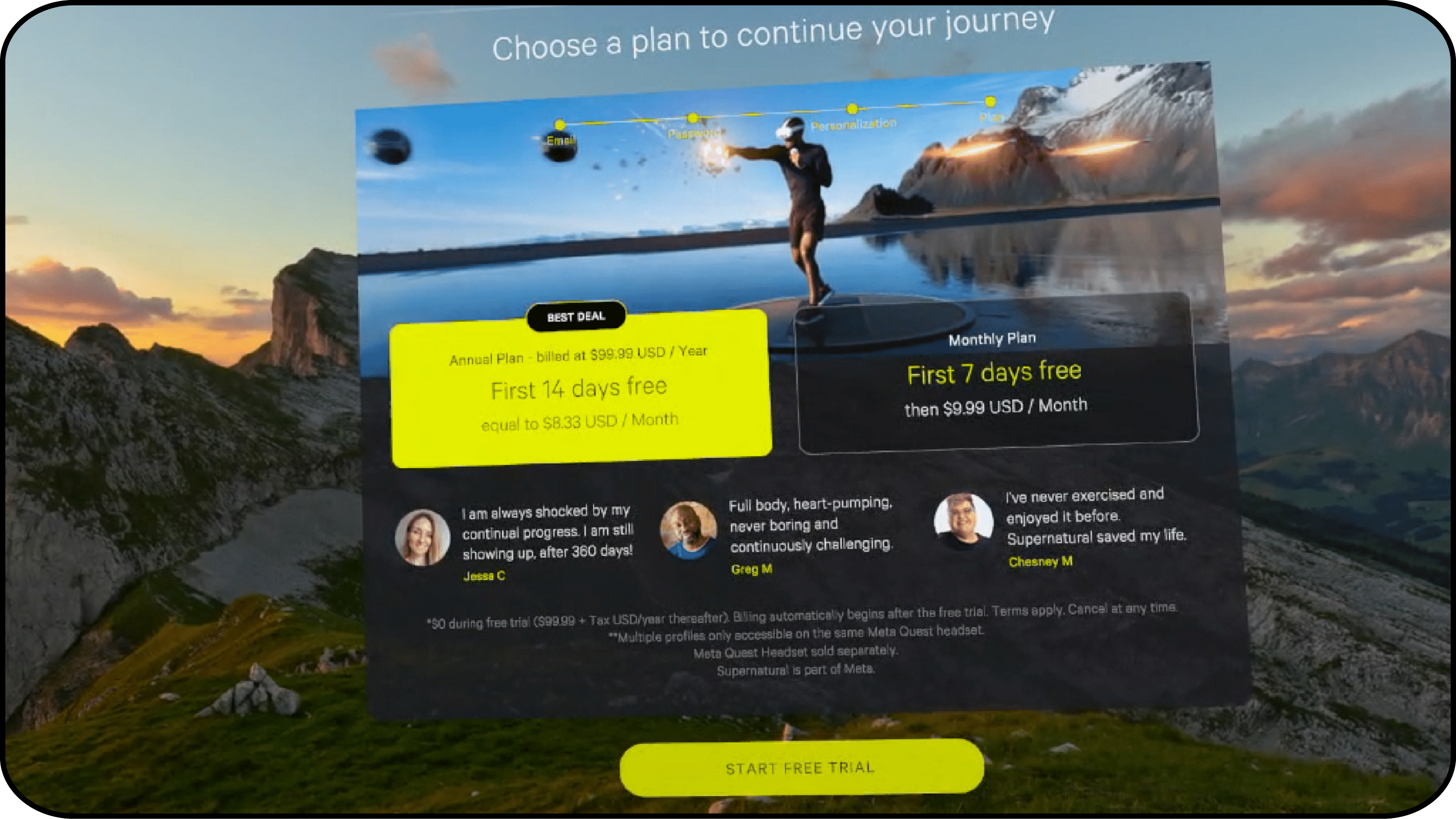}
  \caption{The subscription mechanism in \name{Supernatural} exemplifies the \textbf{\pattern{Endorsements and Testimonials}} deceptive design pattern. The mechanism includes two different subscription plans alongside three testimonials that aim to persuade users to subscribe, which involves providing payment information, by highlighting highly positive experiences. However, the authenticity of these testimonials is questionable, as the professionally polished photos suggest they may have been provided by paid influencers rather than genuine users. This, combined with the realistic virtual environment that demonstrates vivid and lifelike landscapes, create a persuasive setting where users may be misled into believing the testimonials are genuine and be prompted to subscribe under the assumption that they will have similar positive experiences. In addition, the brightly colored and visually emphasized subscription plans and confirmation button also exemplifies \textbf{\pattern{Visual Prominence}} deceptive design. The use of bright green color scheme on an option that involves financial commitment exemplifies \textbf{\pattern{Positive or Negative Framing}} deceptive design, as it suggests emotional safety and subtly encourages users to accept privacy-invasive options without fully understanding their decisions.}
  \Description{Example screenshot.}
  \label{fig:screenshot8}
\end{figure}

\begin{figure}[!h]
  \includegraphics[width=\textwidth]{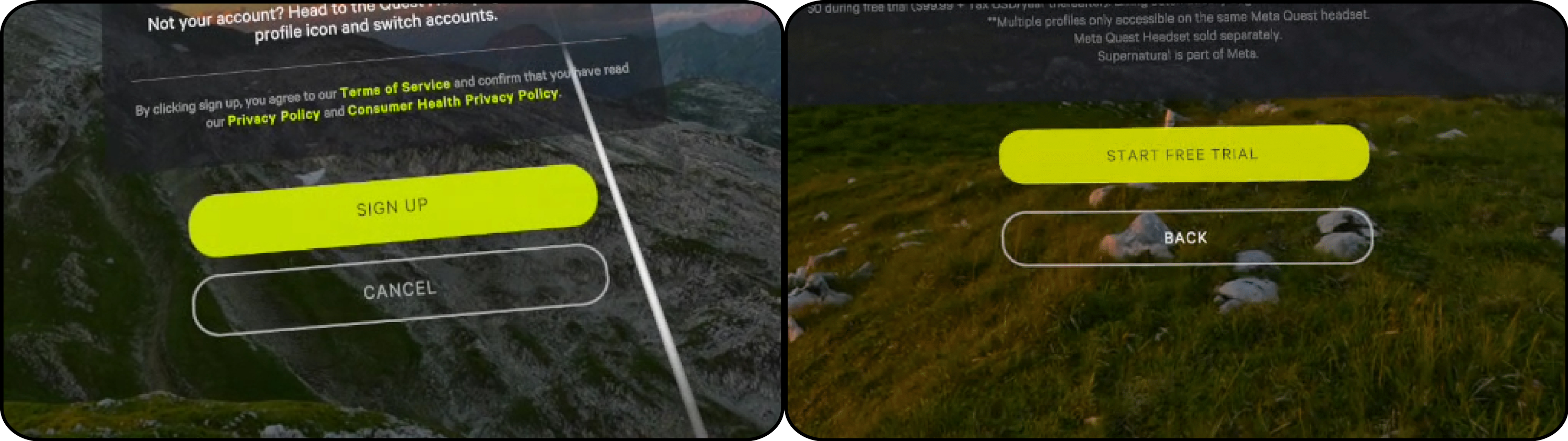}
  \caption{The registration and subscription mechanisms in \name{Supernatural} (see left and right screenshots, respectively) exemplify the \textbf{\pattern{False Hierarchy}}, \textbf{\pattern{Visual Prominence}}, and \textbf{\pattern{Positive or Negative Framing}} deceptive design patterns. Options that involve acceptance are brightly colored, visually emphasized, and ranked above other available choices. The use of a bright green color scheme for options requiring financial commitment exemplifies \pattern{Positive or Negative Framing}, as it conveys a sense of emotional safety and subtly encourage users to opt for privacy-invasive choices. }
  \Description{Example screenshot.}
  \label{fig:screenshot9}
\end{figure}

\begin{figure}[!h]
  \includegraphics[width=0.65\textwidth]{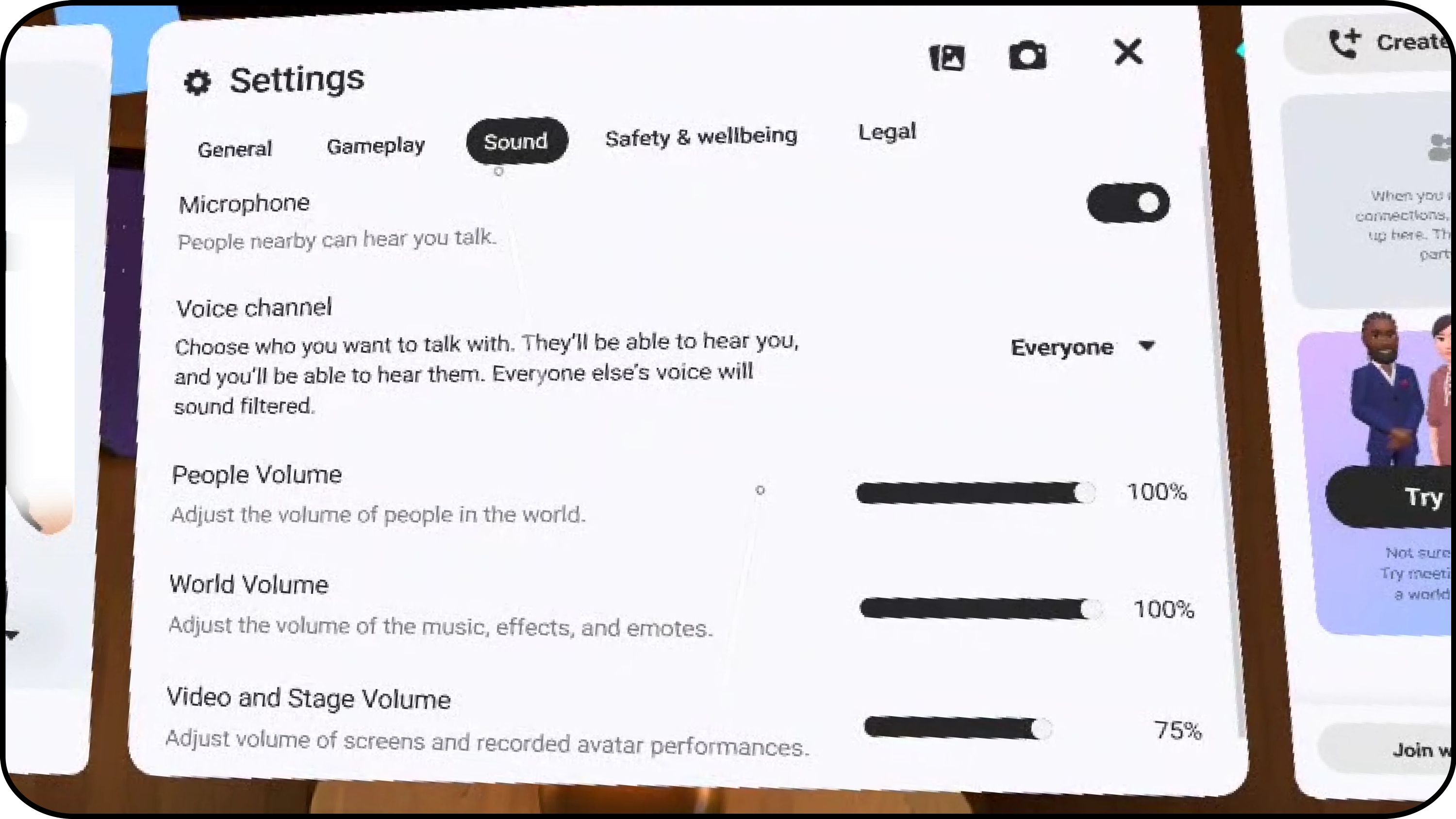}
  \caption{The consent interaction mechanism in \name{Meta Horizon Worlds} exemplifies the \textbf{\pattern{Bad Defaults}} deceptive design pattern. Within the settings menu, several privacy-invasive options are pre-enabled by default. These include automatically activating the microphone (``people nearby can hear you talk''), auto-joining the ``everyone'' voice channel, and automatic leaderboard participation (``allow leaderboard to show your name and score in any world''). While these features are intended to foster social interaction, having them opted-in by default may leave users feeling exposed, especially for those who are more sensitive to personal privacy and boundaries. \textit{Note: Our researcher's username is concealed in the screenshot for anonymity.}}
  \Description{Example screenshot.}
  \label{fig:screenshot10}
\end{figure}

\begin{figure}[!h]
  \includegraphics[width=0.65\textwidth]{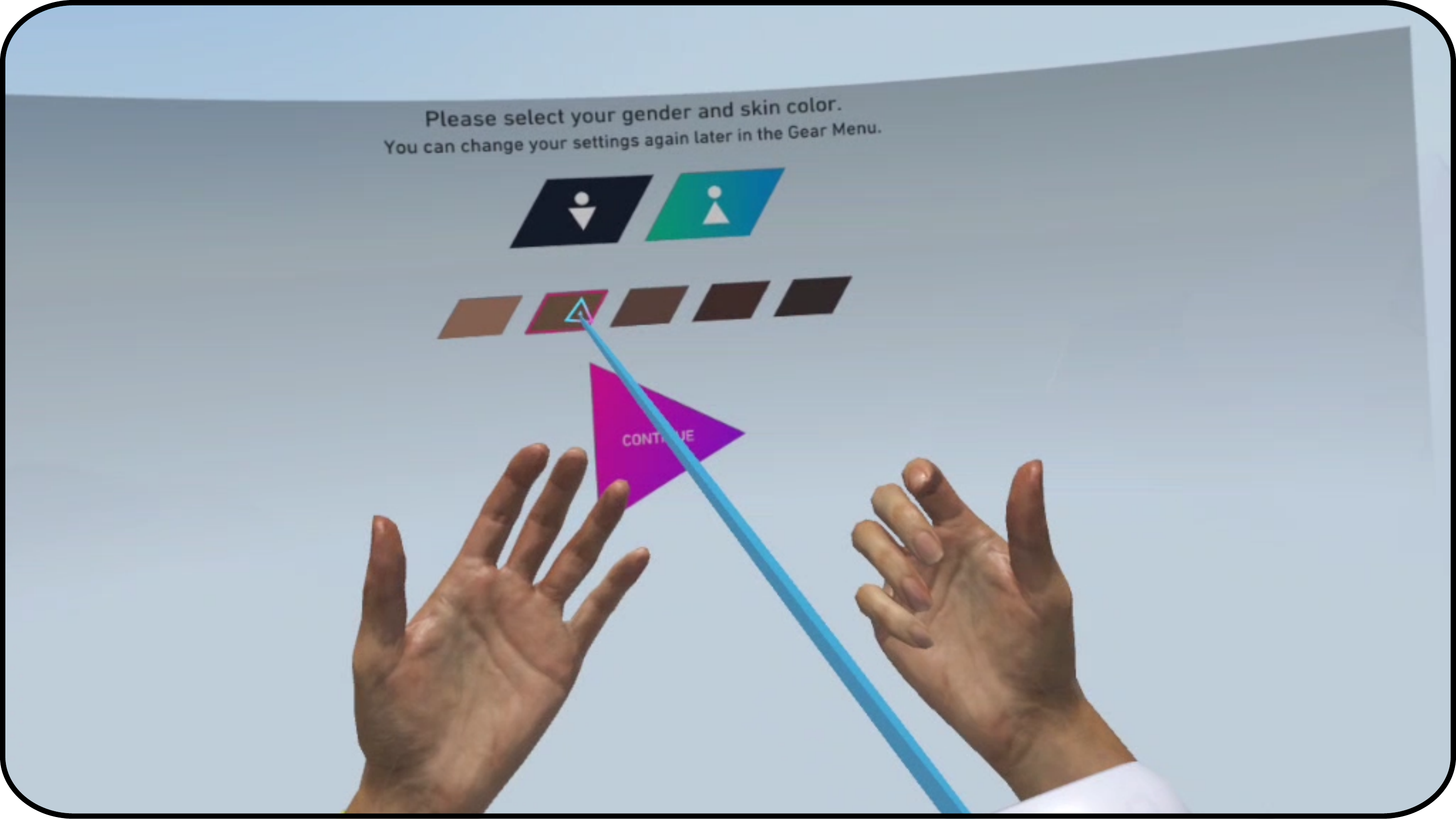}
  \caption{The quality-of-life (QoL) feature that facilitates avatar customization in \name{The Climb 2} exemplifies the \textbf{\pattern{Trick Questions}} deceptive design pattern. In this interface, users are asked to choose their gender and skin color, phrased as ``select your race and gender''. Although this question was intended for avatar customization, the wording exemplifies the misleading nature of Trick Questions. As users see the realistic rendering of their ``hands'' in the avatar creation process from a first-person perspective, they may feel more compelled to create avatars that resemble their real-life appearance. As a result, the data about users' avatar becomes a valuable source for inferring their personal characteristics.}
  \Description{Example screenshot.}
  \label{fig:screenshot11}
\end{figure}

\begin{figure}[!h]
  \includegraphics[width=\textwidth]{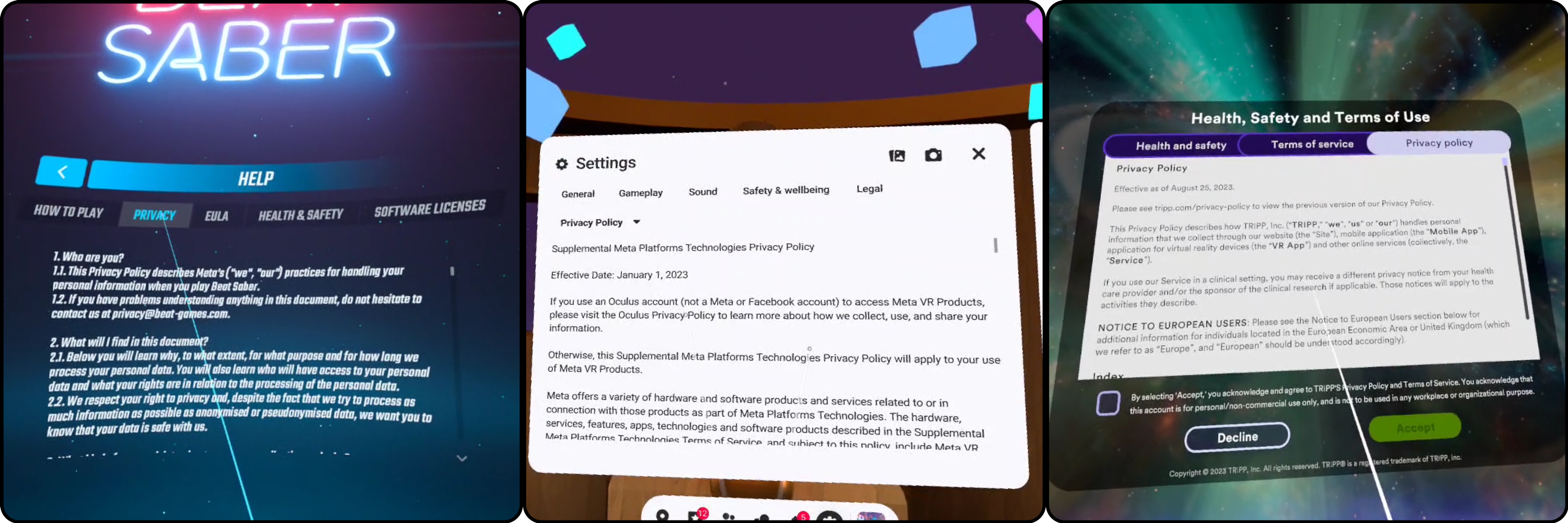}
  \caption{The in-game and in-app privacy policies in \name{Beat Saber} (see left screenshot), \name{Meta Horizon Worlds} (see center screenshot), and \name{TRIPP} (see right screenshot) exemplify \textbf{\pattern{Complex \& Lengthy Language}} deceptive design pattern. Although they are aimed to inform users about data practices upon first entry, these documents are often lengthy and difficult to read. This complexity may discourage users from fully reading the terms and hinders their comprehension of potential privacy risks. Our findings from privacy policy analysis further support this conclusion (see~\autoref{subsubsec:policy-ambiguity} and~\autoref{app-sec:readability}).}
  \Description{Example screenshot.}
  \label{fig:screenshot12}
\end{figure}

\begin{figure}[!h]
  \includegraphics[width=\textwidth]{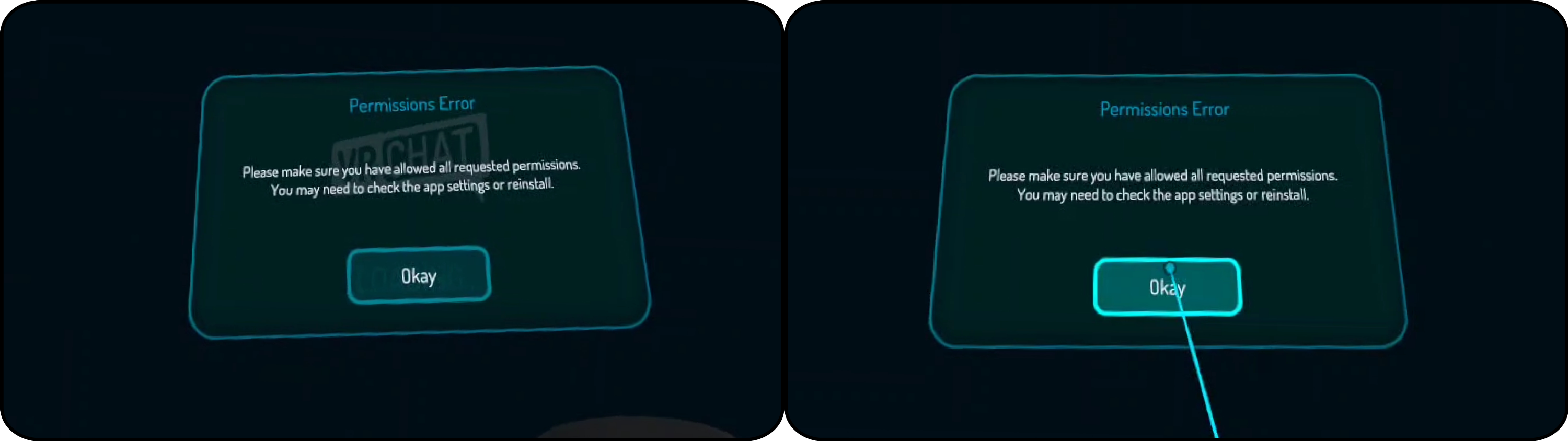}
  \caption{The notification mechanism in \name{VRChat} exemplifies the \textbf{\pattern{Nagging}} deceptive design pattern. When users choose ``Not allow'' during the application's request for access to their photos, media, and files upon their entry, they are repeatedly interrupted by a ``permissions error'' message each time they transfer to a different virtual interaction space (i.e., ``world''). Users must press ``okay'' each time, which can become irritating and may ultimately compel them to re-enter the application and grant the permission to stop the interruptions. Repeatedly prompting users for permissions related to photos, media, and files before those features are even used is excessive. The left screenshot captures the initial error message displayed after denying access, and the right screenshot shows the repeated message at a later time.}
  \Description{Example screenshot.}
  \label{fig:screenshot13}
\end{figure}

\begin{figure}[!h]
  \includegraphics[width=\textwidth]{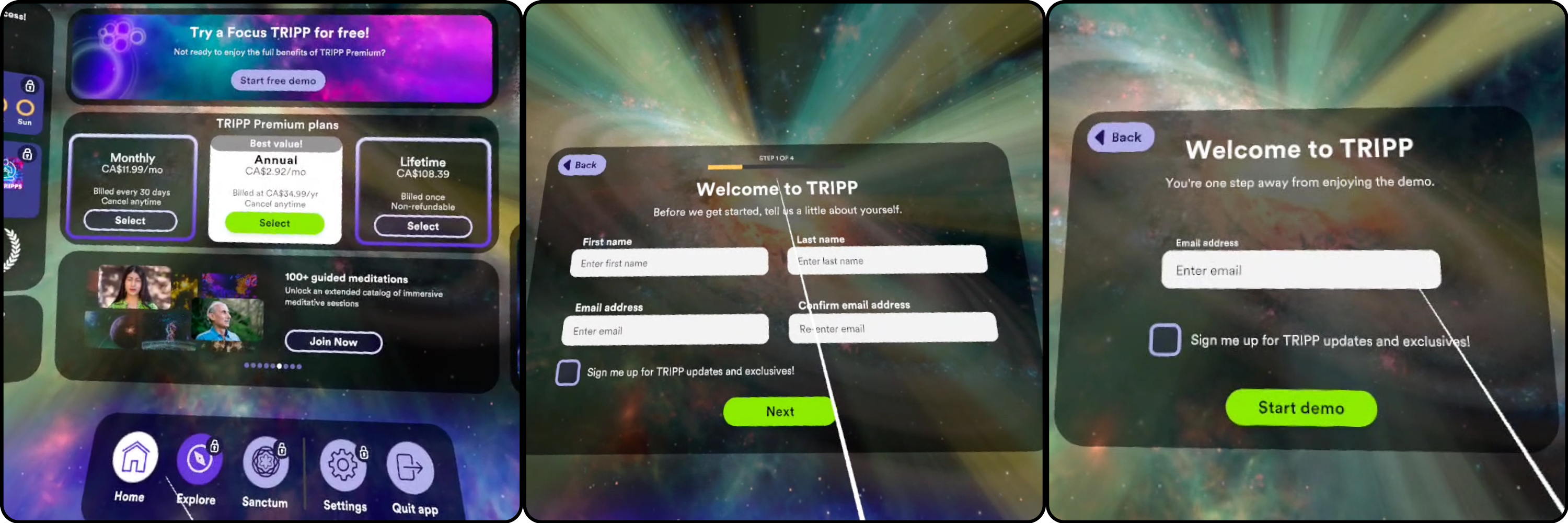}
  \caption{The registration and subscription mechanism in \name{TRIPP} exemplifies the \textbf{\pattern{Forced Registration}} deceptive design pattern. No functions are usable or accessible without registration (see left screenshot). Users are required to provide their email address to even start a demo session (see center screenshot). Users also must register an account and pay a subscription fee with their name and credit card information to unlock all functions, including detailed privacy settings (see right screenshot). While these design mechanisms may support cross-device experiences, they are unnecessary for users intending to use the application solely on VR platforms. The forced disclosure of users' name and email address through the registration process also exemplifies \textbf{\pattern{Privacy Zuckering}}, as this information is not essential for the applications' functionality.}
  \Description{Example screenshot.}
  \label{fig:screenshot14}
\end{figure}

\begin{figure}[!h]
  \includegraphics[width=\textwidth]{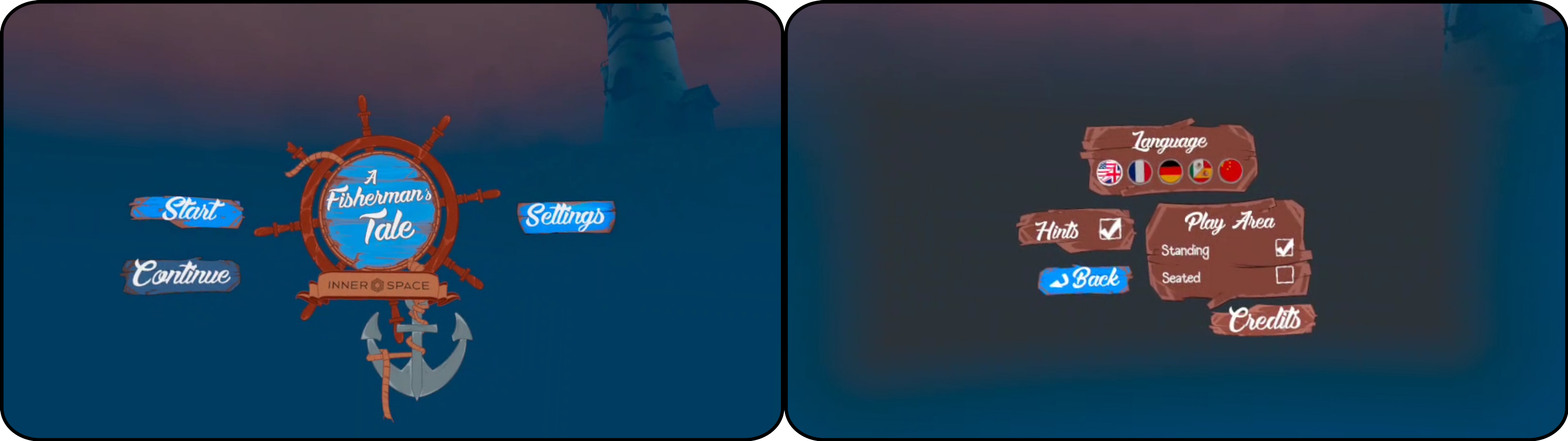}
  \caption{The privacy settings mechanism in \pattern{A Fisherman's Tale} exemplifies the \textbf{\pattern{Mandatory Acceptance}} deceptive design pattern. The settings screen does not provide available settings for users to customize their privacy preferences at all. Therefore, users are forced to consent to all publisher-stated data practices without sufficient awareness, even if these practices are uncomfortable or irrelevant to the game's functions. }
  \Description{Example screenshot.}
  \label{fig:screenshot15}
\end{figure} 

\begin{figure}[!h]
  \includegraphics[width=\textwidth]{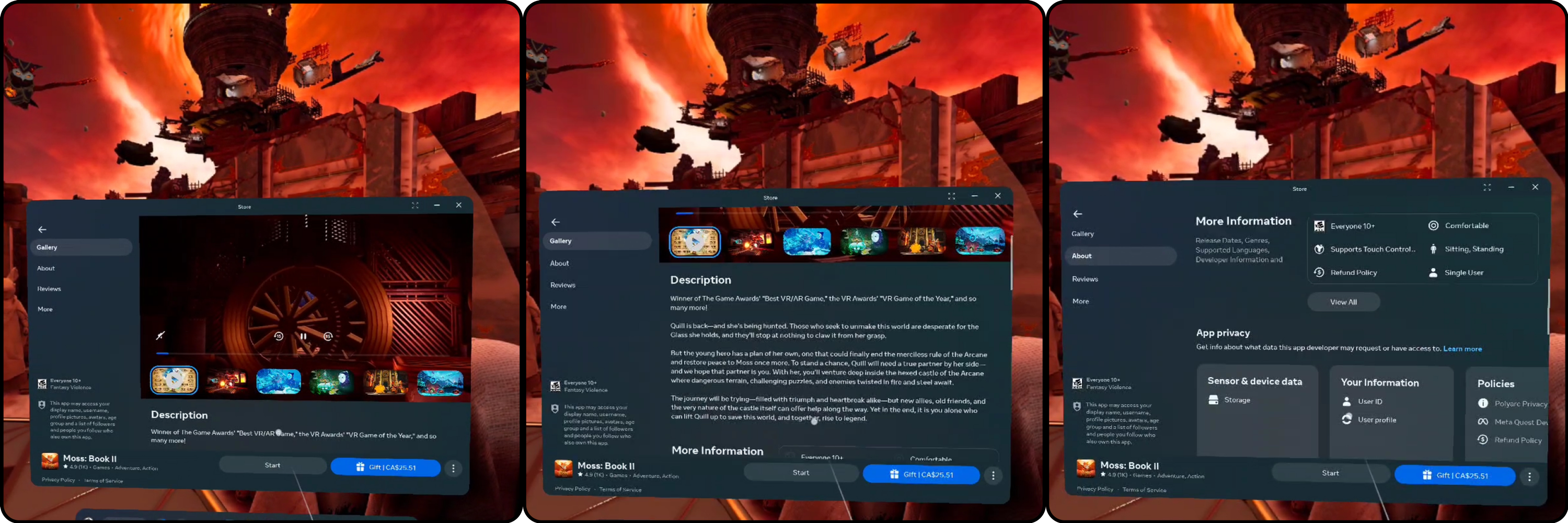}
  \caption{The Meta Quest Store provides a link to the publisher's privacy policy, but it is placed at the very bottom of the page. The store page is primarily dominated by a large and attention-grabbing video preview (see left screenshot). Users are required to scroll down to locate the privacy policy hyperlink at the bottom of the page (see center and right screenshots). As a result, users may not scroll far enough to even notice its presence. This design of the Meta Quest store page can also be seen as an attention-capture tactic~\cite{roffarello2023defining}, as users' focus is naturally drawn to the visually striking VR game and application previews, and leave critical privacy information overlooked.}
  \Description{Example screenshot.}
  \label{fig:screenshot16}
\end{figure}

\newpage
\clearpage
\section{Appendix --- Example Screenshots Bright Design Patterns from VR Games and Playful Applications}
\label{app-sec:screenshots-bright}

\begin{figure}[!h]
  \includegraphics[width=\textwidth]{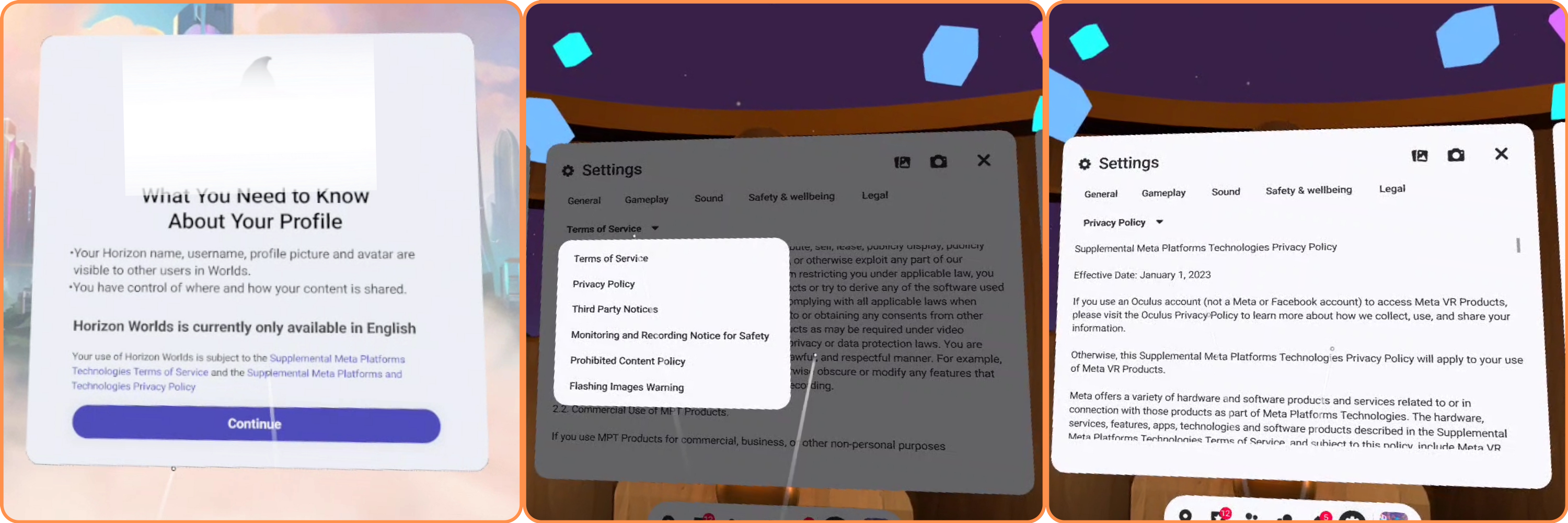}
  \caption{The in-app privacy policy in \name{Meta Horizon Worlds} exemplifies the \textbf{\pattern{Accessible privacy policy}} bright design pattern. Upon entry, the application provides users with a brief description of its data practices, along with hyperlinks to more detailed information available on external webpages (see left screenshot). This privacy information can also be accessed at any time through the settings menu (see center and right screenshots). This design ensures users can easily revisit it as needed. \textit{Note: Our researcher's username and avatar image are concealed in the screenshot for anonymity.}}
  \Description{Example screenshot.}
  \label{fig:screenshot17}
\end{figure}

\begin{figure}[!h]
  \includegraphics[width=\textwidth]{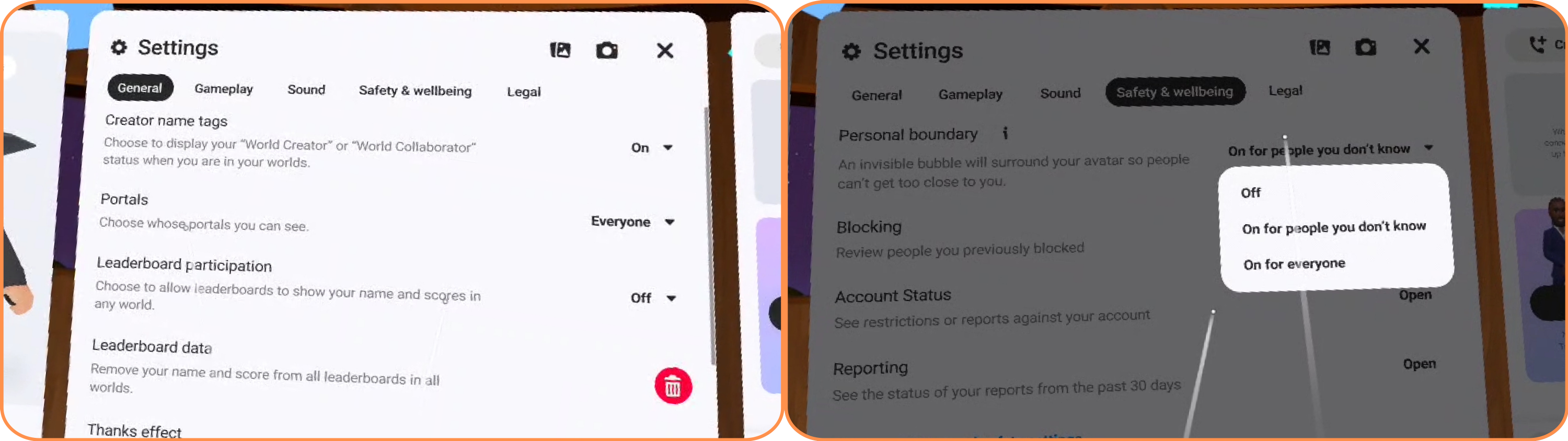}
  \caption{The settings mechanism in \name{Meta Horizon Worlds} exemplifies the  \textbf{\pattern{unbiased presentation of choices}} bright design pattern. All options in the menu are in black and white, with the data deletion option distinctively highlighted in red to make it stand out (see left screenshot). Even within the drop-down menus, the options maintain a consistent color scheme, with the most privacy-preserving choice (``off'') placed at the top (see right screenshot).}
  \Description{Example screenshot.}
  \label{fig:screenshot18}
\end{figure}

\begin{figure}[!h]
  \includegraphics[width=\textwidth]{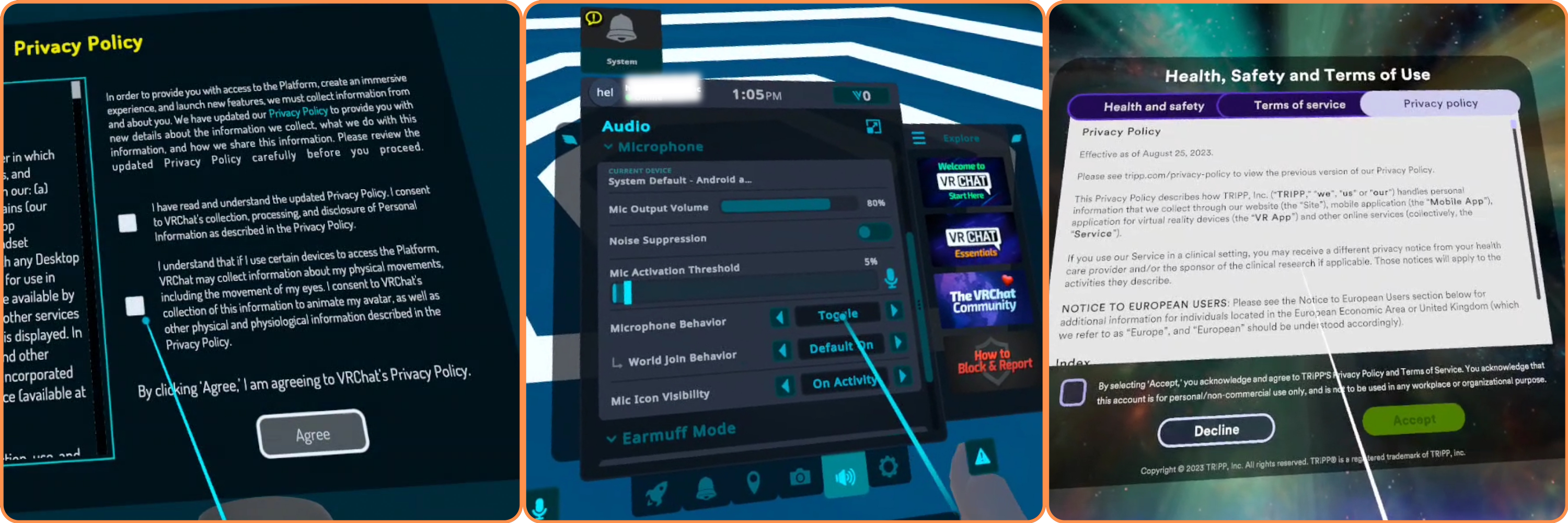}
  \caption{The consent interactions and settings mechanisms in \name{VRChat} (see the left and center screenshots) and the consent interactions mechanism in \name{TRIPP} (see right screenshot) exemplify the \textbf{\pattern{privacy-friendly default}} bright design pattern. These mechanisms ensure that privacy-related options are unchecked by default and require users to decide to opt-in. In \name{VRChat}, the microphone is set to ``toggle'' mode by default to prevent unintended listening by others. We note that although \name{VRChat} also enables auto-joining the world voice channel by default, we do not consider this a privacy-invasive bad default since the microphone remains in ``toggle'' mode and must be manually activated by the user. \textit{Note: Our researcher's username is concealed in the right screenshot for anonymity.}}
  \Description{Example screenshot.}
  \label{fig:screenshot19}
\end{figure}

\begin{figure}[!h]
  \includegraphics[width=\textwidth]{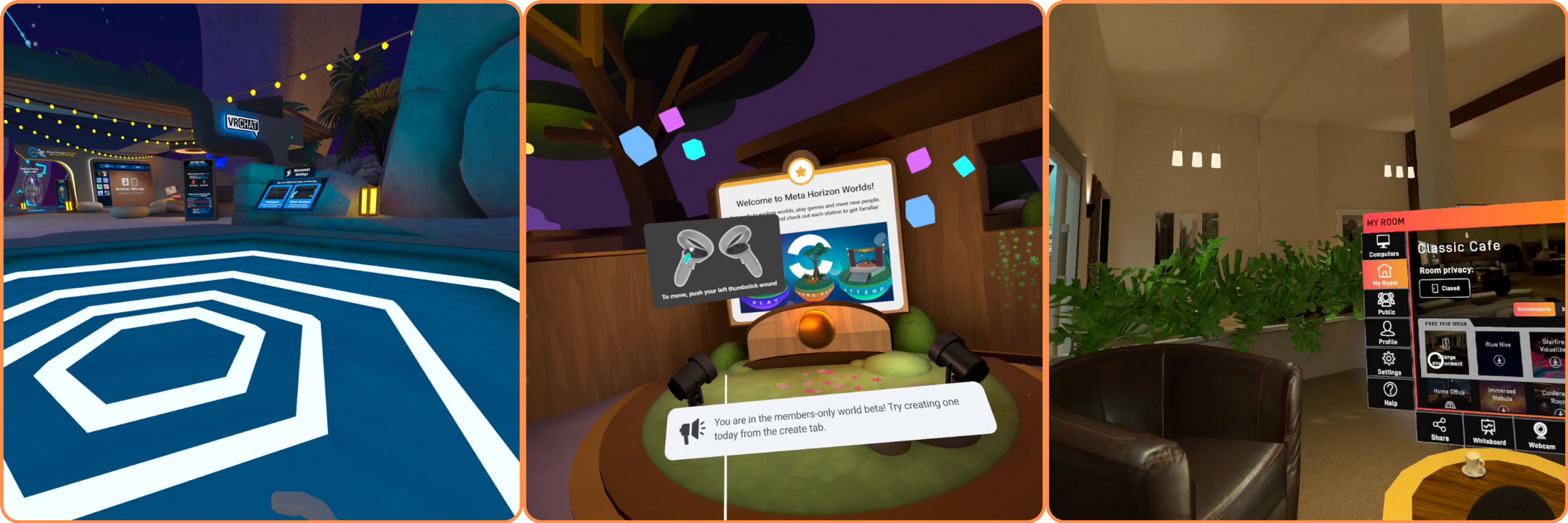}
  \caption{The quality-of-life (QoL) features, such as the private beginner spaces in \name{VRChat} (see left screenshot) and \name{Meta Horizon Worlds} (see center screenshot) and private workspace in \name{Immersed} (see right screenshot), exemplify the \textbf{\pattern{private privacy configuration space}} bright design pattern. Upon entry, these applications place users in private spaces where they can explore functionalities, adjust settings, and customize avatars before interacting in public spaces. This design allows users to familiarize themselves with privacy settings, review policy information, configure privacy preferences, and customize their avatar and username tag to conceal personal information before joining public virtual worlds.}
  \Description{Example screenshot.}
  \label{fig:screenshot20}
\end{figure}

\begin{figure}[!h]
  \includegraphics[width=\textwidth]{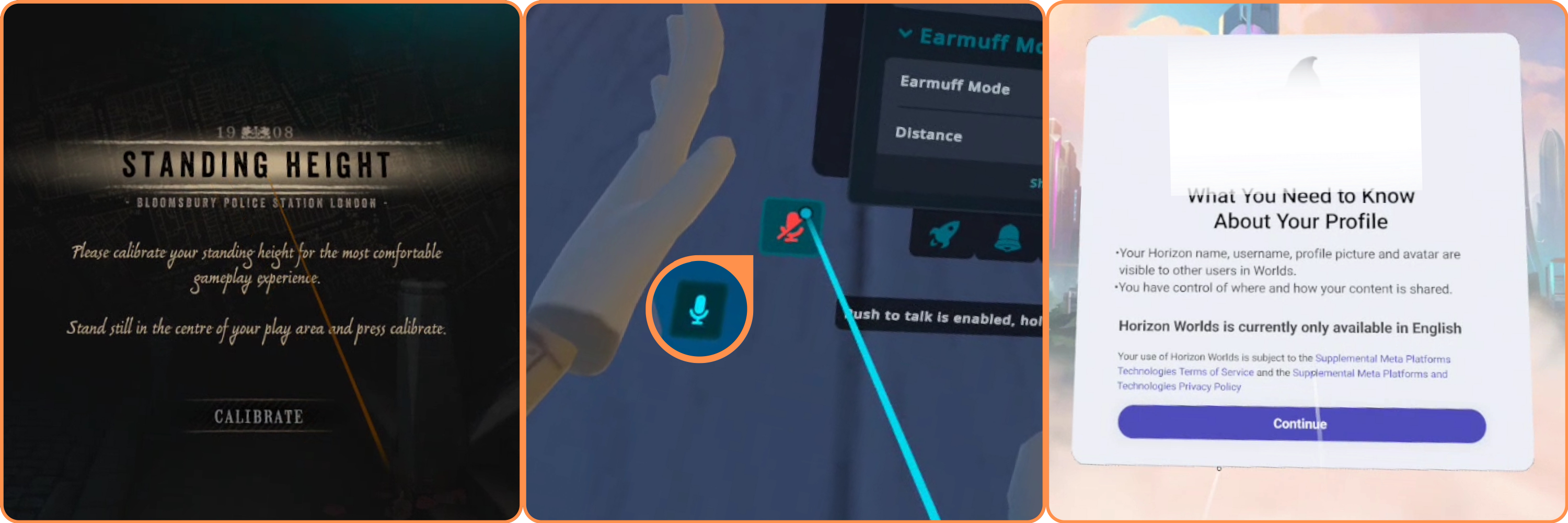}
  \caption{The quality-of-life (QoL) feature, specifically the microphone icon in \name{VRChat} (see center screenshot), and the notification mechanisms in \name{The Room VR} (see left screenshot) and \name{Meta Horizon Worlds} (see right screenshot) exemplify the \textbf{\pattern{transparent communication of data practices}} bright design pattern. The \name{VR Chat} implemented a microphone icon on the bottom of the screen to keep users aware of their microphone status. \name{The Room VR} notifies users when their height is calibrated specifically for gameplay purposes, and \name{Meta Horizon Worlds} provides a brief description of data practices upon users' entry that clearly states what data is collected and for what purposes.}
  \Description{Example screenshot.}
  \label{fig:screenshot21}
\end{figure}

\begin{figure}[!h]
  \includegraphics[width=\textwidth]{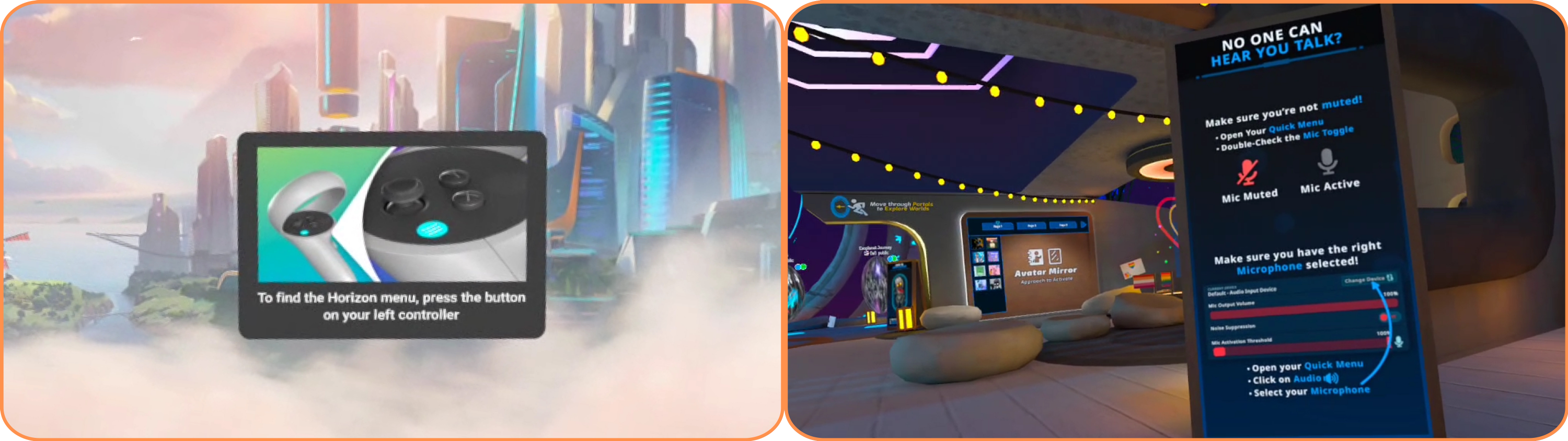}
  \caption{The privacy resource mechanisms in \name{Meta Horizon Worlds} (see left screenshot) and \name{VR Chat} (see right screenshot) exemplify the \textbf{\pattern{clear control mechanisms}} bright design pattern. \name{Meta Horizon Worlds} uses an illustrative image to show users how to access the settings menu, and \name{VR Chat} provides sign boards to guide users to ensure that privacy controls can easily be located. }
  \Description{Example screenshot.}
  \label{fig:screenshot22}
\end{figure}

\begin{figure}[!h]
  \includegraphics[width=\textwidth]{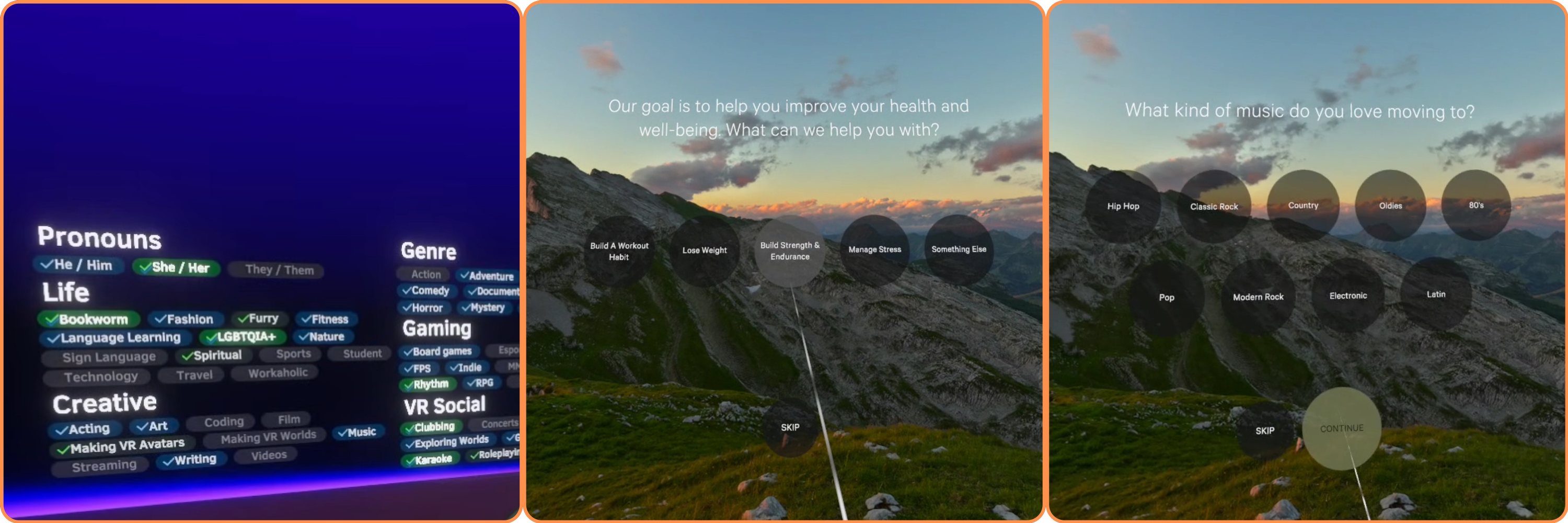}
  \caption{The quality-of-life (QoL) features in \name{VRChat} (see left screenshot) and \name{Supernatural} (see center and right screenshots) exemplify the \textbf{\pattern{balanced privacy and functionality}} bright design pattern. In \name{VRChat}, users are invited to provide personal details like lifestyle and social preferences to help find matching friends, but this step is optional. The friend-matching can still be made without sharing this information. In \name{Supernatural}, users are asked to select their fitness goals and preferred music genre for sessions, with the option to skip these selections if preferred.}
  \Description{Example screenshot.}
  \label{fig:screenshot23}
\end{figure}

\newpage
\clearpage
\section{Appendix --- Privacy Policy Readability Assessment}
\label{app-sec:readability}

Following GPEN's approach~\cite{GPEN2024}, we assessed the readability of the 13 selected privacy policies using the Flesch Reading Ease Score in Microsoft 365.\footnote{Microsoft Support. (n.d.). Get your document's readability and level statistics. Retrieved August 27, 2024, from \url{https://support.microsoft.com/en-us/office/get-your-document-s-readability-and-level-statistics-85b4969e-e80a-4777-8dd3-f7fc3c8b3fd2}} We copied the content of each privacy policy into a Word document and evaluated its readability using the \textit{Flesch Reading Ease} option under the \textit{Editor} menu. ~\autoref{tab:readbility} presents the results, where lower scores indicate more difficult contents requiring a higher education level to understand.

\input{Table-PolicyReadabilityAssessment}

\newpage
\section{Appendix --- Information Flow by Senders}
\label{app-sec:flow-by-sender}

\begin{figure}[!h]  
  \includegraphics[width=\textwidth]{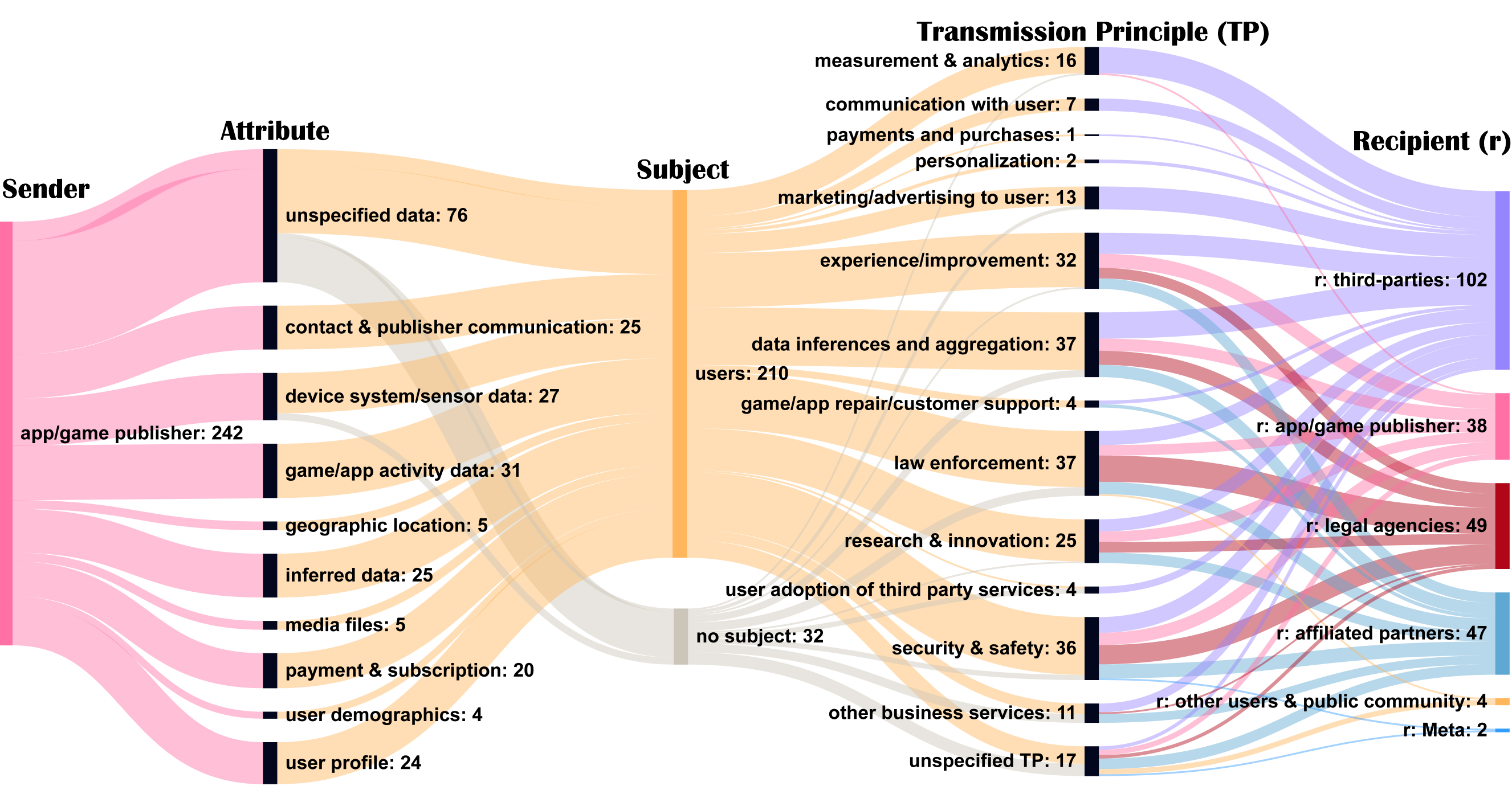}
  \caption{This Sankey Diagram visualizes the 242 information flows sent from the \code{app/game publisher} that we identified from 21 privacy policy statements through our thematic analysis of 12 privacy policies. Multiple information flows may be derived from a single policy statement if it involves multiple instances of the same CI factor. For example, if an \textit{attribute} is transmitted to various \textit{recipients}, the transmission to each \textit{recipient} represents a separate information flow. The diagram moves from left to right, showing the codes in the information flows across the five \cif themes: (1) sender, (2) attribute, (3) subject, (4) transmission principle, and (5) recipient. For example, \code{device system/sensor data} appears as the data attribute in 27 flows. Color schemes distinguish different information flows. For example, colors flowing from \textit{attribute} to \textit{subject} identify the subject of which \textit{attribute} relates to. \textit{Note: despite our efforts to accurately classify and differentiate these information flows in the thematic analysis, ambiguous language in privacy policies may affect the accuracy of this figure, particularly when multiple CI factors are grouped together in a same policy statement without clear distinctions between their relationships. Nonetheless, the classification of these information flows still provides valuable insights into general patterns of data collection and sharing in VR applications.}}
  \Description{Sankey Diagram that demonstrates the user information flow based on our analysis of privacy policies.}
  \label{fig:CI-sankey-publisher}
\end{figure}

\newpage
\begin{figure}[!h]  
  \includegraphics[width=\textwidth]{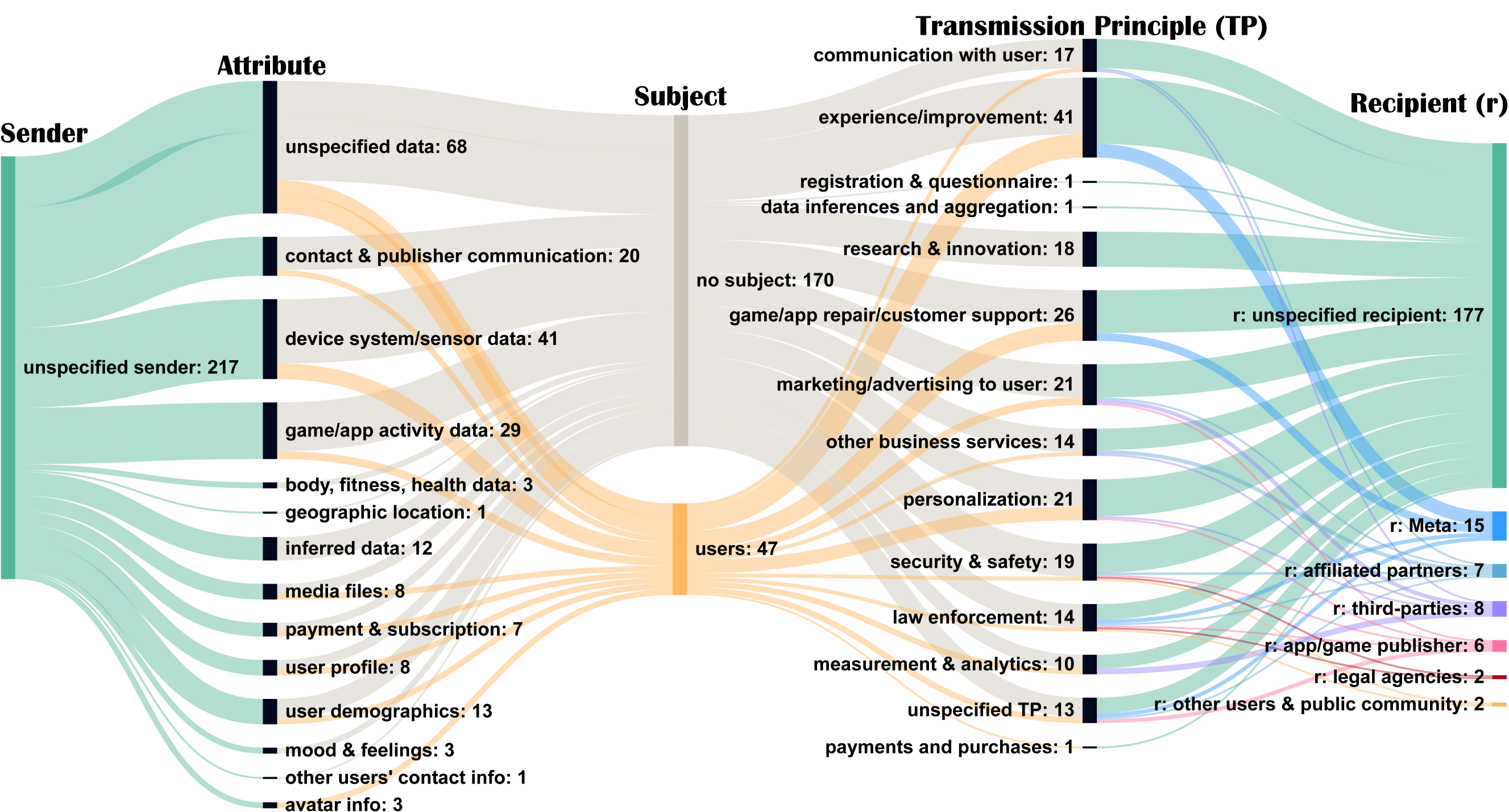}
  \caption{This Sankey Diagram visualizes the 217 information flows sent from the \code{unspecified sender} that we identified from 87 privacy policy statements through our thematic analysis of 12 privacy policies. Multiple information flows may be derived from a single policy statement if it involves multiple instances of the same CI factor. For example, if an \textit{attribute} is transmitted to various \textit{recipients}, the transmission to each \textit{recipient} represents a separate information flow. The diagram moves from left to right, showing the codes in the information flows across the five \cif themes: (1) sender, (2) attribute, (3) subject, (4) transmission principle, and (5) recipient. For example, \code{device system/sensor data} appears as the data attribute in 41 flows. Color schemes distinguish different information flows. For example, colors flowing from \textit{attribute} to \textit{subject} identify the subject of which \textit{attribute} relates to. \textit{Note: despite our efforts to accurately classify and differentiate these information flows in the thematic analysis, ambiguous language in privacy policies may affect the accuracy of this figure, particularly when multiple CI factors are grouped together in a same policy statement without clear distinctions between their relationships. Nonetheless, the classification of these information flows still provides valuable insights into general patterns of data collection and sharing in VR applications.}}
  \Description{Sankey Diagram that demonstrates the user information flow based on our analysis of privacy policies.}
  \label{fig:CI-sankey-unspecified}
\end{figure}

\newpage
\begin{figure}[!h]  
  \includegraphics[width=\textwidth]{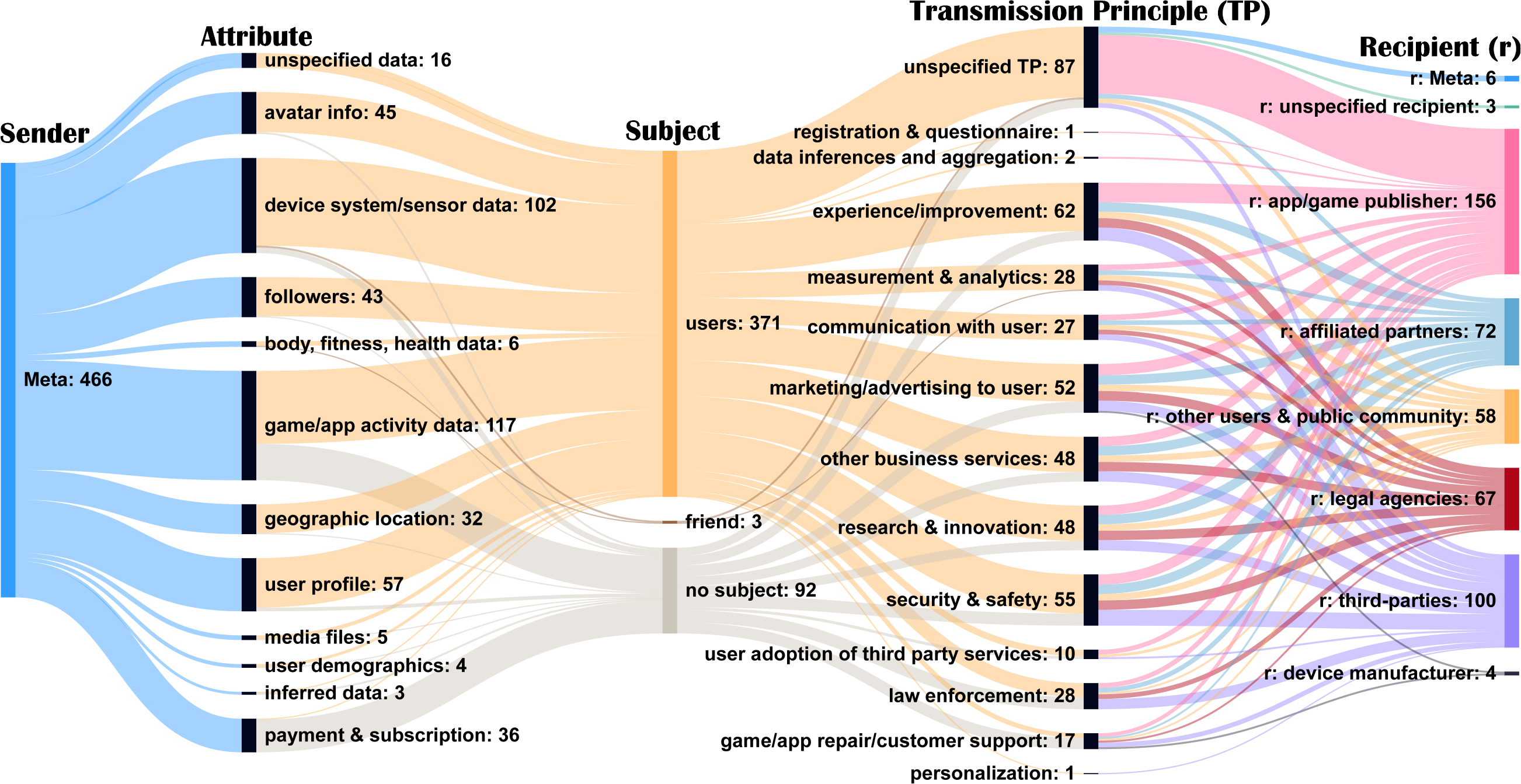}
  \caption{This Sankey Diagram visualizes the 466 information flows sent from the \code{Meta} Quest VR platform that we identified from 53 privacy policy statements through our thematic analysis of 12 privacy policies. Multiple information flows may be derived from a single policy statement if it involves multiple instances of the same CI factor. For example, if an \textit{attribute} is transmitted to various \textit{recipients}, the transmission to each \textit{recipient} represents a separate information flow. The diagram moves from left to right, showing the codes in the information flows across the five \cif themes: (1) sender, (2) attribute, (3) subject, (4) transmission principle, and (5) recipient. For example, \code{device system/sensor data} appears as the data attribute in 41 flows. Color schemes distinguish different information flows. For example, colors flowing from \textit{attribute} to \textit{subject} identify the subject of which \textit{attribute} relates to. \textit{Note: despite our efforts to accurately classify and differentiate these information flows in the thematic analysis, ambiguous language in privacy policies may affect the accuracy of this figure, particularly when multiple CI factors are grouped together in a same policy statement without clear distinctions between their relationships. Nonetheless, the classification of these information flows still provides valuable insights into general patterns of data collection and sharing in VR applications.}}
  \Description{Sankey Diagram that demonstrates the user information flow based on our analysis of privacy policies.}
  \label{fig:CI-sankey-meta}
\end{figure}

\newpage
\begin{figure}[!h]  
  \includegraphics[width=\textwidth]{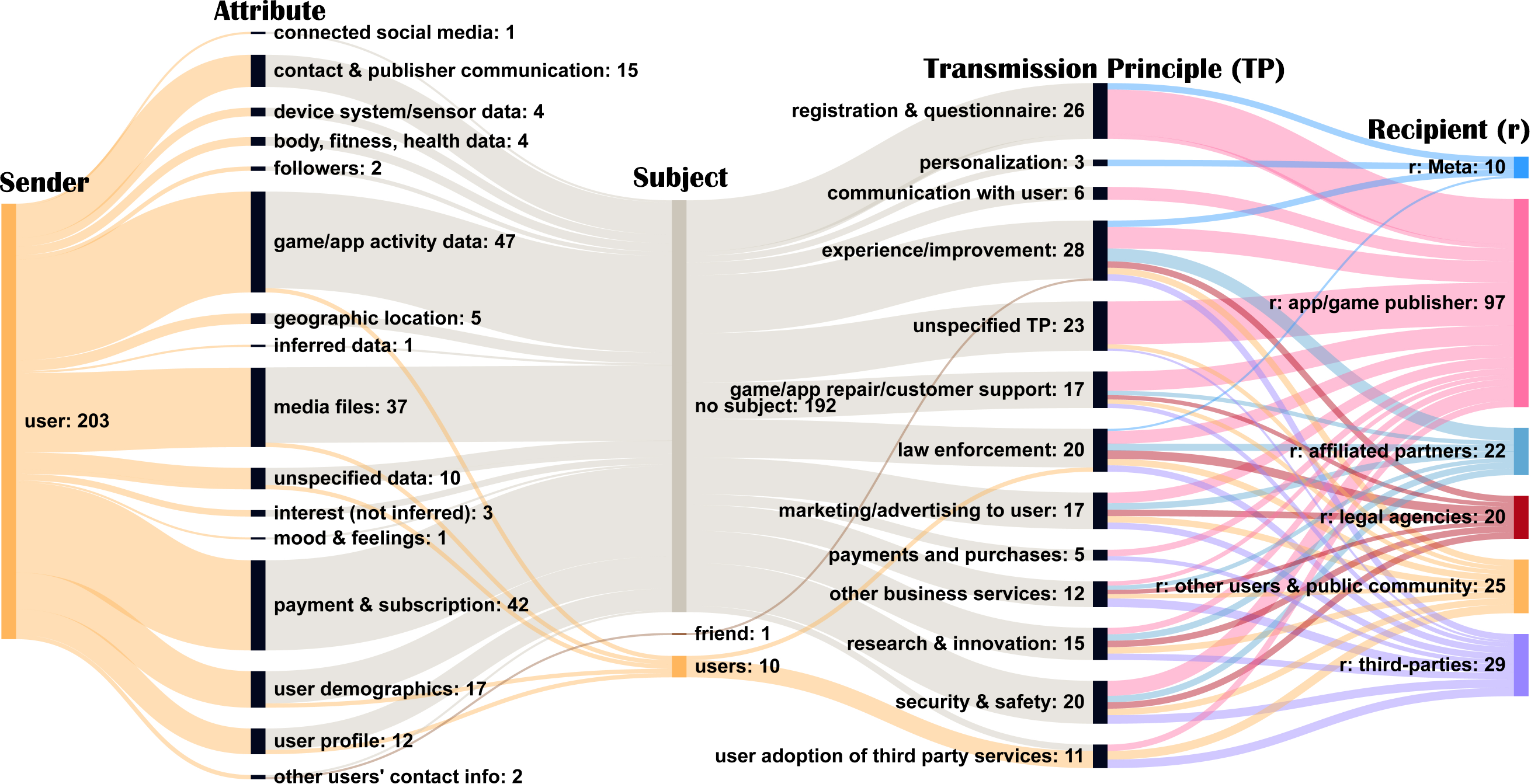}
  \caption{This Sankey Diagram visualizes the 203 information flows sent from the \code{User} directly that we identified from 42 privacy policy statements through our thematic analysis of 12 privacy policies. Multiple information flows may be derived from a single policy statement if it involves multiple instances of the same CI factor. For example, if an \textit{attribute} is transmitted to various \textit{recipients}, the transmission to each \textit{recipient} represents a separate information flow. The diagram moves from left to right, showing the codes in the information flows across the five \cif themes: (1) sender, (2) attribute, (3) subject, (4) transmission principle, and (5) recipient. For example, \code{device system/sensor data} appears as the data attribute in 41 flows. Color schemes distinguish different information flows. For example, colors flowing from \textit{attribute} to \textit{subject} identify the subject of which \textit{attribute} relates to. \textit{Note: despite our efforts to accurately classify and differentiate these information flows in the thematic analysis, ambiguous language in privacy policies may affect the accuracy of this figure, particularly when multiple CI factors are grouped together in a same policy statement without clear distinctions between their relationships. Nonetheless, the classification of these information flows still provides valuable insights into general patterns of data collection and sharing in VR applications.}}
  \Description{Sankey Diagram that demonstrates the user information flow based on our analysis of privacy policies.}
  \label{fig:CI-sankey-user}
\end{figure}

\newpage
\begin{figure}[!h]  
  \includegraphics[width=\textwidth]{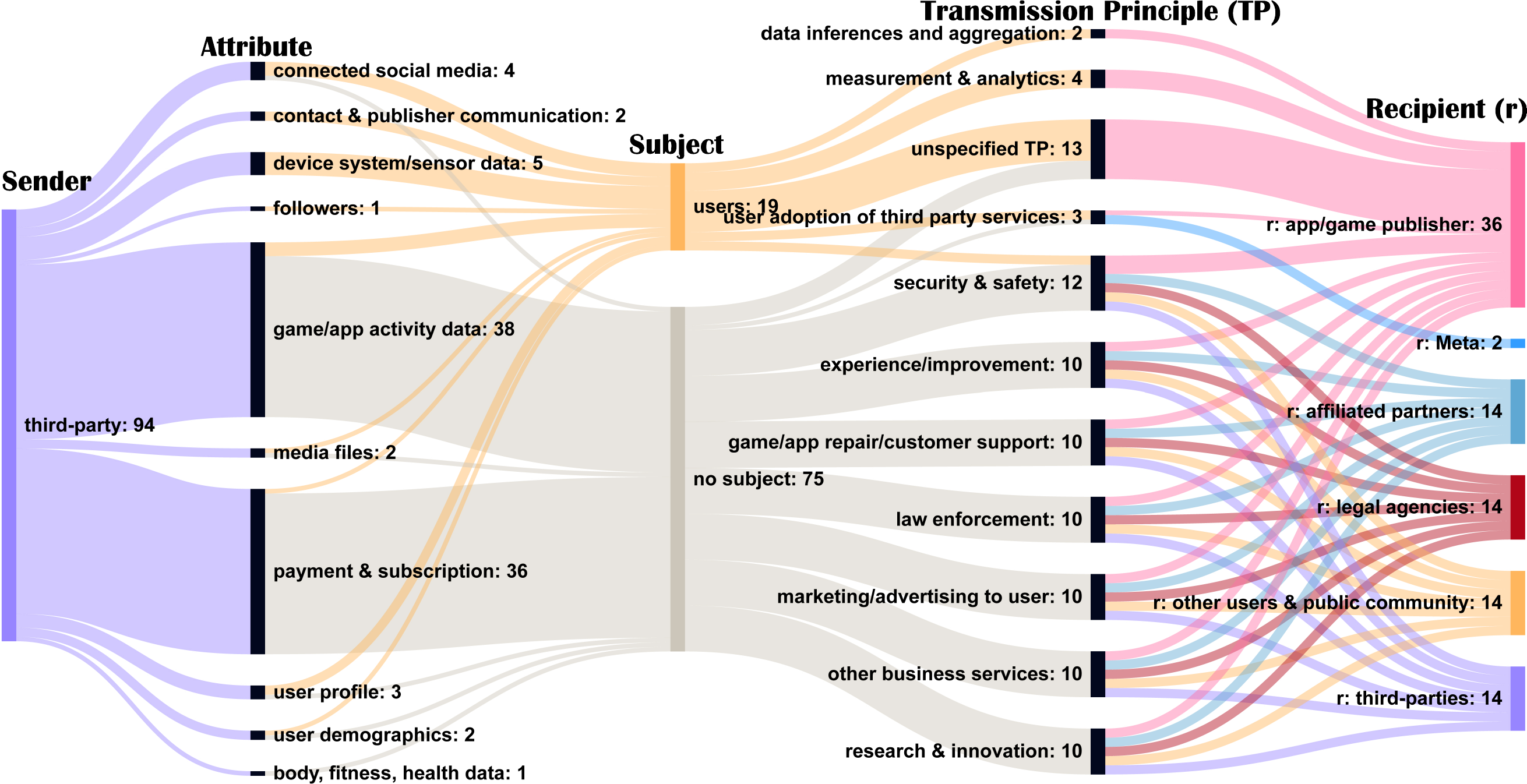}
  \caption{This Sankey Diagram visualizes the 94 information flows sent from the \code{third parties} that we identified from 9 privacy policy statements through our thematic analysis of 12 privacy policies. Multiple information flows may be derived from a single policy statement if it involves multiple instances of the same CI factor. For example, if an \textit{attribute} is transmitted to various \textit{recipients}, the transmission to each \textit{recipient} represents a separate information flow. The diagram moves from left to right, showing the codes in the information flows across the five \cif themes: (1) sender, (2) attribute, (3) subject, (4) transmission principle, and (5) recipient. For example, \code{device system/sensor data} appears as the data attribute in 41 flows. Color schemes distinguish different information flows. For example, colors flowing from \textit{attribute} to \textit{subject} identify the subject of which \textit{attribute} relates to. \textit{Note: despite our efforts to accurately classify and differentiate these information flows in the thematic analysis, ambiguous language in privacy policies may affect the accuracy of this figure, particularly when multiple CI factors are grouped together in a same policy statement without clear distinctions between their relationships. Nonetheless, the classification of these information flows still provides valuable insights into general patterns of data collection and sharing in VR applications.}}
  \Description{Sankey Diagram that demonstrates the user information flow based on our analysis of privacy policies.}
  \label{fig:CI-sankey-thirdparties}
\end{figure}

%% file: Table-DiaryMockup.tex
\begin{table*}[!h]
\renewcommand{\arraystretch}{1.5}  
\centering
\caption{Structural outline of a sample diary entry from a \name{VRChat} autoethnography exploration session on July 22, 2024. The diary was documented in Miro given its feasibility in linking videos, screenshots, and diary texts. From left to right: (1) detailed observations of privacy communication and interaction design mechanics that incorporated deceptive designs, and (2) reflections on experiences, reactions, thoughts, and emotional responses when the design mechanics were encountered. We focused on the design mechanisms that allow users to obtain privacy information, make privacy choices, and express privacy preference.}
\label{tab:diary-mockup}
\resizebox{0.85\textwidth}{!}{%
\begin{tabular}{@{}p{0.55\textwidth}|p{0.55\textwidth}@{}}
\toprule
\multicolumn{2}{l}{\textbf{Date:} July 22, 2024}   \\ 
\multicolumn{2}{l}{\textbf{Estimated Time:} $\sim$1.5 hours including diary writing} \\ \midrule
\multicolumn{2}{l}{\textbf{{[}Imported Video Recording Available for Re-watch{]}} } \\ \midrule
\textbf{{[}\textit{Diary:}{]}} & \textbf{{[}Reflection Note:{]}}\\
\textit{Today, I had my first experience with VR Chat. The initial on-boarding process was straightforward, and I was placed in a beginner world. While I'm unsure if this was intentional or simply a result of joining at a less populated time, I appreciated the opportunity of no immediate social interaction. I was able to explore the environment, customize my avatar, and familiarize myself with the platform's features and adjust my privacy settings without the pressure of being observed by other users.} & I was very satisfied that I was placed in a private world upon joining. As a newcomer to social VR experiences, this private world gave me a safe and controlled space to get familiarized with the platform's features and configure my virtual identity without worrying about unwanted judgment or exposure. It gave me a safe adoption of social VR experiences and a smooth transfer from private to public spaces. Had I been immediately placed in a public world, I may have accidentally revealed sensitive information or made mistakes that could have compromised my privacy. \\
\multicolumn{1}{r|}{ \textbf{{[}Screenshots go here{]}} }& \textbf{{[}Screenshots go here{]}} \\
\textit{One interesting aspect of VR Chat was the way it made me feel more immersed in the virtual world compared to traditional PC games. The ability to view my avatar from a first-person perspective and interact with the environment using my full body made it feel like a more personal and realistic experience. While my avatar was entirely fictional and doesn't mirror my actual physical appearance, I still found myself carefully selecting clothing and accessories for my avatar and be mindful of what I wear or say in a public space, as if curating a public image of myself.} & Overall, I experienced a heightened sense of self-awareness and concern for personal boundary in VRChat. I also wondered, can people infer my interests or preferences based on my avatar's clothing choices? Can they guess my personality and gender from my name tag? The possibility of being deanonymized through these cues made me feel vulnerable but having access to a private beginner world really eased my concerns.
\\
 \bottomrule
\end{tabular}%
}
\end{table*}

%% file: Table-PolicyReadabilityAssessment.tex
\begin{table}[!ht]
\centering
\caption{Readability of Privacy Policies based on Flesch Reading Ease (FRE) Assessment}
\label{tab:readbility}
\resizebox{0.8\columnwidth}{!}{%
\begin{tabular}{@{}lllll@{}}
\toprule
Privacy Policy & Date Published & Date Retrieved & Language & FRE Score* \\ \midrule
Beat Saber & June 2023 & March 2024 & English & 42.4\\
Moss: Book II & September 2021 & February 2024 & English & 36.5\\
The Room VR & February 2022 & June 2024 & English & 46.4\\
A Fisherman's Tale & April 2023 & June 2024 & English & 50.6 \\
Down the Rabbit Hole & March 2020 & June 2024 & English & 34.8  \\
LEGO\textregistered Bricktales & November 2023 & June 2024 & English & 39.2 \\
The Climb 2 & n.d. & July 2024 & English & 38.7  \\
VRChat & November 2023 & June 2024 & English & 35.9 \\
TRIPP & November 2023 & July 2024 & English & 36.1  \\
Supernatural & July 2024 & July 2024 & English & 35.6  \\
Immersed & August 2024 & August 2024 & English & 41.3  \\
Meta Horizon Worlds & \multirow{2}{*}{June 2024} & \multirow{2}{*}{August 2024} & \multirow{2}{*}{English} & \multirow{2}{*}{42.5}  \\
Meta Platforms Technologies Privacy Policy & & & &  \\
 \bottomrule
 \multicolumn{5}{l}{\begin{tabular}[c]{@{}l@{}}\textit{Note.} *All privacy policies have a Flesch Reading Ease (FRE) score ranging from 30 to 50, indicating they \\ are ``difficult to read'' and require at least ``college'' level knowledge. According to: \\~\url{https://en.wikipedia.org/wiki/Flesch\%E2\%80\%93Kincaid\_readability\_tests}\\ 
\end{tabular}}
\end{tabular}%
}
\end{table}